\documentclass{jfm}
\usepackage{amscd} 

\usepackage{psfrag}
\usepackage{epstopdf}
\usepackage{color}

\usepackage{mathrsfs}
\usepackage{natbib}
\usepackage[toc,page]{appendix}
\usepackage{bm}
\usepackage{psfrag}
\usepackage{epsfig}
\usepackage{amsmath, amsfonts, amsbsy, amssymb, array, float}
\usepackage{graphicx,psfrag,subfigure,overpic}
\usepackage[toc,page]{appendix}
\usepackage[dvipsnames]{xcolor}
\usepackage{bm}
\usepackage{wasysym}
\ifCUPmtlplainloaded \else
  \checkfont{eurm10}
  \iffontfound
    \IfFileExists{upmath.sty}
      {\typeout{^^JFound AMS Euler Roman fonts on the system,
                   using the 'upmath' package.^^J}%
       \usepackage{upmath}}
      {\typeout{^^JFound AMS Euler Roman fonts on the system, but you
                   dont seem to have the}%
       \typeout{'upmath' package installed. JFM.cls can take advantage
                 of these fonts,^^Jif you use 'upmath' package.^^J}%
      }
  \else
  \fi
\fi
\ifCUPmtlplainloaded \else
  \checkfont{msam10}
  \iffontfound
    \IfFileExists{amssymb.sty}
      {\typeout{^^JFound AMS Symbol fonts on the system, using the
                'amssymb' package.^^J}%
       \usepackage{amssymb}%

      }{}
  \fi
\fi
\ifCUPmtlplainloaded \else
  \IfFileExists{amsbsy.sty}
    {\typeout{^^JFound the 'amsbsy' package on the system, using it.^^J}%
     \usepackage{amsbsy}}
    {\providecommand\boldsymbol[1]{\mbox{\boldmath $##1$}}}
\fi

\newcommand{\EQ}{\begin{equation}}
\newcommand{\EN}{\end{equation}}
\newcommand{\EQA}{\begin{eqnarray}}
\newcommand{\ENA}{\end{eqnarray}}
\definecolor{2deg}{RGB}{128, 0, 255}
\definecolor{3deg}{RGB}{25, 150, 243}
\definecolor{4deg}{RGB}{77, 243, 206}
\definecolor{5deg}{RGB}{179, 243, 150}
\definecolor{6deg}{RGB}{255, 150, 79}
\definecolor{7deg}{RGB}{255, 0, 0}
\author{W. Mostert}
\affiliation{Mechanical and Aerospace Engineering, Princeton University, Princeton, NJ 08544, USA}
\author{S. Popinet}
\affiliation{Institut Jean Le Rond d'Alembert, CNRS UMR 7190, Sorbonne Universit\'{e}, Paris, France}
\title[High-resolution direct simulation of deep water breaking waves]
{High-resolution direct simulation of deep water breaking waves: transition to turbulence, bubbles and droplets production}
\author[ ]%
{W.\ns M\ls O\ls S\ls T\ls E\ls R\ls T$^{1,2}$,\ns%
  \ns
  S. \ns P\ls O\ls P\ls I\ls N\ls E\ls T$^3$\ns%
  \and \ns
 L.\ns D\ls E\ls I\ls K\ls E$^{1,4}$%
  \break
 }
\affiliation{
  $^1$Mechanical \& Aerospace Engineering, Princeton University, Princeton NJ 08544, USA\\[\affilskip]
  $^2$Department of Mechanical and Aerospace Engineering, Missouri University of Science and Technology, Rolla MO 65401, USA\\[\affilskip]
  $^3$Institut Jean Le Rond d'Alembert, CNRS UMR 7190, Sorbonne Universit\'{e}, Paris, France\\[\affilskip]
  $^4$High Meadows Environmental Institute, Princeton University, Princeton NJ 08544, USA
}
\pubyear{2018}
\volume{xxx}
\pagerange{xxx-yyy}
\date{?? and in revised form ??}
\begin{document}
\large
\maketitle
\begin{abstract}
We present high-resolution three-dimensional direct numerical simulations of breaking waves solving for the two-phase Navier-Stokes equations. We investigate the role of the Reynolds (wave inertia relative to viscous effects) and Bond numbers (wave scale over the capillary length) on the energy, bubble and droplet statistics of strong plunging breakers. We explore the asymptotic regimes at high Reynolds and Bond numbers and compare with laboratory breaking waves. Energetically, the breaking wave transitions from laminar to three-dimensional turbulent flow on a timescale that depends on the turbulent Reynolds number up to a limiting value of $Re_\lambda \sim 100$, consistent with the mixing transition in other canonical turbulent flows. We characterize the role of capillary effects on the impacting jet and ingested main cavity shape and subsequent fragmentation process, and extend the buoyant-energetic scaling from \cite{Deike2016} to account for the cavity shape and its scale separation from the Hinze scale, $r_H$. We confirm two regimes in the bubble size distribution, $N(r/r_H)\propto (r/r_H)^{-10/3}$ for bubbles above $r_H$, and $N(r/r_H)\propto (r/r_H)^{-3/2}$ below it. We show resolved bubbles up to one order of magnitude below the Hinze scale and observe a good collapse of the numerical data compared to laboratory breaking waves \citep{Deane2002}. We resolve droplet statistics at high Bond number in good agreement with recent experiments \citep{Erinin2019}, with a distribution shape close to $N_d(r_d)\propto r_d^{-2}$. The evolution of the droplet statistics appears controlled by the details of the impact process and subsequent splash-up. We discuss velocity distributions for the droplets, finding ejection velocities up to four times the phase speed of the wave, which are produced during the most intense splashing events of the breaking process.

\end{abstract}
\section{Introduction}
\label{sec:introduction}

\subsection{The broader context}
The action of breaking waves on the ocean surface has a large and incompletely understood effect on the dynamics of mass, momentum and energy transfer between the ocean and the atmosphere, converting much of the wave energy into heat in a complex process which spans a wide range of scales \citep{Melville1996}. Breaking also marks a transition at the ocean surface from laminar flow to two-phase turbulent mixing at small scales, modulating the dynamics of the upper ocean sub-mesoscales, particularly via Langmuir turbulence and fronts \citep{Mcwilliams2016}, and affects the transport of particles with implications for the fate of oil spills and plastic pollutants \citep{Deike2017,Pizzo2019}. Furthermore, surface breaking injects a large amount of gas into the ocean via the entrainment of bubbles, including approximately $30\%$ of the $\textrm{CO}_2$ which has been released into the atmosphere \citep{Deike2018,Reichl2020}; breaking also ejects spray into the atmosphere, where it can convect and evaporate to leave salt crystals which may serve as cloud condensation nuclei \citep{deLeeuw2011,Veron2015}.

Wave breaking involves transition from two-dimensional (2D) laminar wave flow to three-dimensional (3D) turbulence. As wave energy focuses through linear or nonlinear processes, local conditions on a wave surface become unstable and cause breaking, which transfers energy and momentum to the water column. The geometry and kinematics of the breaking waves have been extensively studied \citep{Longuet-Higgins1976,Perlin2013,Schwendeman2017,Fedele2020}, and the identification of a breaking threshold with approaches based on the wave kinematics, dynamics or geometry remains a longstanding issue \citep{Melville1982,Banner2007,Perlin2013} with recent work discussing the link between the breaker kinematics and dynamics \citep{Saket2017,Derakhti2020,Pizzo2020}. 

While the initiation of the breaking phenomenon and the turbulence generated by it have been characterized \citep{Rapp1990,Duncan1999,Tulin1999,Melville2002,Banner2007,Drazen2008,Drazen2009}, the time and length scales of the transition process remain to be explored. During this transition to turbulence, air is entrained, and bubbles are formed  \citep{Lamarre1991,Deane2002} and spray droplets are ejected \citep{Erinin2019}. The measurements of 3D two-phase turbulence in the laboratory and in the field present many technical challenges in terms of successfully accessing the turbulent flow field and the size distributions of drops and bubbles during the active time of breaking.

Direct numerical simulations (DNS) therefore appear as an appealing tool. Owing to the computational difficulty and expense of modelling 3D multiphase flows, numerical studies began by using 2D breakers as analogues for the full 3D processes \citep{Chen1999,Song2004,Iafrati2009,Iafrati2011,Deike2015}. Early development of nonlinear potential flow models has shed light on the breaking process up to the moment of impact \citep{Longuet-Higgins1976,Dommermuth1988}, while 3D simulations have used reduced models such as large-eddy simulation to capture the breaking process itself \citep{Watanabe2005,Lubin2015,Hao2019}, but the complete resolution of the breaker in DNS in 3D has only recently become feasible \citep{Fuster2009,Deike2016,Wang2016,Yang2018}. Surprisingly, despite the essentially 3D nature of the turbulence resulting from breaking, 2D breakers at the tested conditions have provided a reasonable estimate of the dissipation rates obtained from experiments and 3D computation (discussed further below). In contrast, the turbulent dissipation in internal wave breaking has been shown to be a clear 3D process \citep{Gayen2010}.

\subsection{Laboratory experiments and direct numerical simulations of canonical breaking waves}
Canonical breaking waves have been studied using a variety of different approaches both experimental and numerical \citep{Duncan1981,Melville1982,Rapp1990,Melville1994,Duncan1999,Banner2007,Tian2010,Drazen2008,Erinin2019}. Studies such as these have identified the main controlling parameters of breaking waves, namely the breaking speed and the wave slope at breaking $S=ak$, where $a$ is the wave amplitude and $k$ the wavenumber. The bandwidth of the wave packet is also important, and the detailed kinematics before breaking, in particular a significant slowdown of the wave crest, have been discussed in order to propose breaking threshold criteria \citep{banner2014linking,saket2017threshold,pizzo2019focusing,Derakhti2020,Fedele2020}, although we will neglect its influence from hereon in.


It follows that DNS of breaking waves can be framed in terms of a set of non-dimensional numbers. The relevant parameters are the air-water density and viscosity ratio, the wave speed and wavenumber, and amplitude. These define a wave Reynolds number, and the wave slope
\begin{equation}
    \textrm{Re} = \frac{\sqrt{g\lambda_0^3}}{\nu}, \quad S = ak
\end{equation}
where $\lambda_0 = 2\pi/k$ is the wavelength and $\nu$ is the kinematic viscosity of the water. Similarly to turbulent DNS, numerical simulations of breaking waves are typically confined to the highest Re accessible to available computation effort, which has grown over time. \citet{Iafrati2009,Deike2015,Deike2016,Vita2018} have typically used $Re=40\times 10^3$. 

To consider bubble and droplet generation, the Bond number is needed,
\begin{equation}
  \textrm{Bo} = \frac{\Delta \rho g}{\sigma k^2}.
\end{equation} 
$\Delta \rho$ is the density difference between air and water and $\sigma$ is the surface tension. Bo corresponds to the ratio between the wavelength and the capillary length scale.

\citet{Deike2015,Deike2016} used the Bond number to compare the numerical wavelength to experimental results. \citet{Deike2015} describes the wave patterns for a large range of Bo and $S$, discussing the energetics of parasitic capillary waves, spilling breakers and plunging breakers. As discussed in \citet{Iafrati2009,Deike2015,Deike2016}, the breaking waves in laboratory would approach $Re=10^6$. Despite this difference in Re, DNS \citep{Iafrati2009,Deike2015,Deike2016} and LES \citep{Derakhti2014,Derakhti2016} found good agreement between experiments and simulations for the non-dimensional energy dissipation due to breaking as a function of the breaker slope (see figure \ref{fig:Deike2016-5b})){. Nevertheless, an outstanding challenge in direct numerical simulation (DNS) is the correct numerical resolution of processes whose separation of scales increases with $Re$ and $Bo$. For such simulations to capture the physics of breaking waves correctly, they must resolve all scales between and including those of energy dissipation and the formation and breakup of bubbles and droplets in a two-phase turbulent environment. This requires capturing the full physics of the problem, while retaining a qualitatively faithful representation of the breaking process in comparison with experiment. These very difficult challenges have historically limited the scope of DNS investigations, the details of whose approaches are discussed in more detail below.

Both the wave Reynolds number and the Bond number characterize the overall scale of the wave through its wavelength and phase speed, compared with viscous and capillary effects. Once the wave breaks, the turbulence it generates is controlled by the breaking slope together with the speed of the breaker, and is itself characterized by a turbulence Reynolds number, typically defined using the Taylor micro-scale $Re_{\lambda}$, with \citet{Drazen2009} typically finding values around $Re_{\lambda}\approx 500$. Similarly, the fragmentation processes and generation of drops and bubbles in a turbulent flow is usually analysed in terms of a Weber number, comparing the inertial stresses due to the turbulence to the surface tension.


\subsection{Energetics and dimensionality of breaking waves}
\label{intro-energy}

Breaking waves dissipate energy, generating a turbulent two-phase flow, with properties that can be related to the local breaking properties \citep{Duncan1981}. The local turbulent dissipation rate due to breaking can be described by an inertial scaling \citep{Drazen2008}, 
\begin{equation}
\varepsilon=\sqrt{gh}^3/h
\end{equation}
where $h$ is the breaking height{\color{black}, here consistently defined as half the distance between wave crest and trough}, $\sqrt{gh}$ the ballistic velocity of the plunging breaker where $g$ is the acceleration due to gravity. The turbulence is confined to a volume {\color{black}$\mathcal{V}_0=AL_c$}, of cross section which is generally assumed to be $A\simeq \pi h^2/4$ \citep{Duncan1981,Drazen2008}, and length of breaking crest $L_c$, leading to an integrated dissipation rate per unit length of breaking crest,
\begin{equation}
  \epsilon_l = \rho A \varepsilon.
\end{equation}
This scaling can be related to the initial slope, bandwidth, and speed of the wave packet in controlled laboratory experiments  \citep{Duncan1981,Rapp1990,Banner2007,Drazen2008,Tian2010,Grare2013} and numerical simulations \citep{Deike2015,Deike2016,Iafrati2009,Derakhti2016}.  The breaking parameter $b$ is a non-dimensional measure of the dissipation that was introduced by \citet{Duncan1981,Phillips1985}, and relates to $\epsilon_l$
\begin{equation}
\epsilon_l=b\rho c^5/g,
\end{equation}
which combined with the local dissipation rate argument above and assuming the breaking speed is related to the wavenumber by the dispersion relation $c = \sqrt{g/k}$ leads to $b \propto S^{5/2}$ \citep{Drazen2008}. Introducing a slope-based breaking threshold $S_0$, this formulation for the breaking parameter reads,
\begin{equation}
b=\chi_0(S-S_0)^{5/2}. 
\end{equation}
Extensive laboratory experiments have demonstrated the accuracy of the physics-based model, with $\chi_0\approx0.4$ and $S_0\approx0.08$ used as fitting parameters by \citet{Romero2012} and allowing to account for numerous laboratory data \citep{Duncan1981,Rapp1990,Banner2007,Drazen2008,Tian2010,Grare2013}. Several numerical studies have confirmed this scaling and validated their approaches against this result \citep{Deike2015,Deike2016,Deike2017,Derakhti2014,Derakhti2016,Vita2018}. Figure \ref{fig:Deike2016-5b} shows $b$ as a function of $S$ for a variety of experimental and numerical data, including from the present study. 
We note that experimental work using the linear focusing technique typically considers the linearly predicted wave slope, summed over all components, while numerical work using compact wave initialization have considered the initial slope. In all cases, the slope being used is proportional to the breaking slope, as discussed in \citet{Drazen2008} for experimental data and \citet{Deike2015,Deike2016} for numerical data, which allows comparison between the experimental and numerical work. The differences in definitions and estimations may therefore be responsible for some of the scatter in figure \ref{fig:Deike2016-5b} between the various data sets, and uncertainties in the fitting coefficients are indicated by the shaded area. Note that the scaling $b\propto S^{5/2}$ is observed at high slopes for both the experiments and DNS. Moreover, the proportion of energy dissipated by breaking for a given slope in is similar between experiments and simulations. This fundamental model for the turbulent dissipation rate has been successfully used as the physical basis of larger-scale spectral wave models \citep{Romero2012,Romero2019}. Moreover, we recently proposed an extension of the inertial argument to certain types of shallow-water breakers \citep{Mostert2020}.


\begin{figure}
  \centering
    \includegraphics[width=0.65\linewidth]{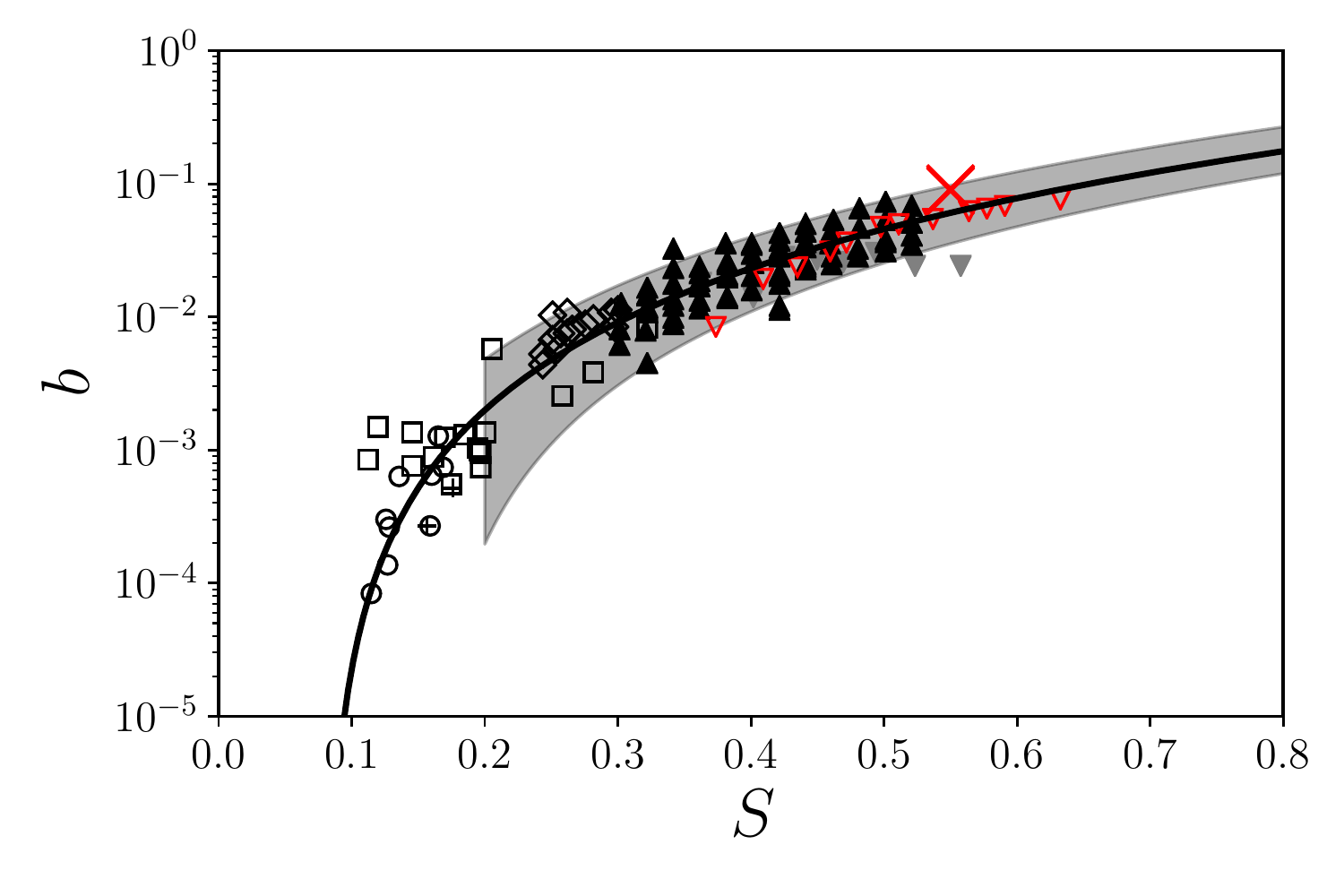}
   \caption{Breaking parameters $b$ as a function of wave slope $S$. Red X: Present DNS data. Red circles: DNS data of \cite{Deike2016}, blue circles are DNS data from \cite{Deike2015}. Black and grey data: Experimental data due to \cite{Drazen2008,Banner2007,Grare2013}. Solid line: $b=0.4(S-0.08)^{5/2}$, semi-empirical result of \cite{Romero2012}. Shaded area indicates the uncertainties on the scaling for b.}
\label{fig:Deike2016-5b}
\end{figure}

It remains to determine the particular transition characteristics of the fully three-dimensional (3D) flow, and to investigate the dependence of these characteristics on the flow Reynolds number, as well as on the evolution of the ingested bubble plume. Furthermore, even aside from limitations on the maximum values of $Re,Bo$ attainable in computation, many numerical studies have investigated two-dimensional (2D) breakers as computationally feasible analogues for the full 3D processes \cite[]{Song2004,Hendrickson2006,Iafrati2009,Deike2015}. Surprisingly, despite the essentially 3D nature of the turbulence resulting from the breaking process, 2D breakers at the tested conditions provided a reasonable estimate of the dissipation rates for 3D breakers obtained from computation and experiment, with discrepancies sometimes as small as $5\%$ \cite[]{Lubin2006,Iafrati2009}. Favourable comparison with semi-empirical models as discussed above also suggests the usefulness of 2D computations for the dissipation rate \cite[]{Deike2015}. Nonetheless, the details of the 2D-3D transition physics in breaking waves constitute an open question. The present study will go some way to addressing these questions, with suggestion of a possible transition to turbulence with an associated turbulence Reynolds number.


\subsection{Bubble size distributions in breaking waves}
A breaking wave entrains air, which is characterized by a broad size distribution of bubbles. Direct investigation of the bubble distribution, obviously not available within a 2D study, is important to inform subgrid scale models used in LES \citep{Shi2010,Liang2011,Liang2012,Derakhti2014} and gas transfer models \citep{Liang2011,Deike2018}. \citet{Garrett2000} proposed a turbulent break-up cascade model for the size distribution per unit volume $\mathcal{N}(r)$, where $r$ is the bubble radius, as a function of the local dissipation rate $\bar{\varepsilon}$ with constant volumetric air flow-rate $Q$, with a dimensional analysis yielding,
\begin{equation}
  \mathcal{N}(r) \propto Q \bar{\varepsilon}^{-1/3} r^{-10/3}.
\end{equation}
We note that a time-averaged dissipation rate $\bar{\varepsilon}$ over the breaking time has been considered when analyzing and scaling various data sets in \citet{Deane2002,Deike2016}. The corresponding breakup model assumes a turbulent inertial subrange with a direct cascade, with large bubbles injected at one end of the cascade by a notional entrainment process and turbulent fluctuations then breaking these into smaller bubbles. The lower end of the cascade is set by the Hinze scale \citep{Hinze,Deane2002,Perrard2020},

\begin{equation}
  r_H = \mathcal{C}_0 \left( \frac{\sigma}{\rho}\right)^{3/5} \bar{\varepsilon}^{-2/5}. \label{Hinze}
\end{equation}
Here $\mathcal{C}_0\simeq 0.4$ \citep{Deane2002} is a dimensionless constant. Its value is related to the critical Weber number defining bubble break-up which ranges typically from 1 to 5 \citep{risso1998oscillations,Martinez-Bazan1999,Deane2002,Vejravzka2018,Perrard2020,Riviere2021}, with estimations of $\mathcal{C}_0$ varying by about a factor of 2. These differences are related to variations in the experimental protocols and the large scale structure of the turbulent flow. Note also that the breaking wave problem is transient in nature, so that the Hinze scale might present variations in time, and estimations of the Hinze scale based on the averaged turbulence dissipation rate presents an added uncertainty. For all these reasons, it should be considered a soft limit. The size distribution below the Hinze scale is not addressed by \citet{Garrett2000}.

Laboratory experiments have reported measurements of the bubble size distribution under a breaking wave using various optical and acoustic techniques \citep{Loewen1996,Terrill2001,Deane2002,Leifer2006,Rojas2007,Blenkinsopp2010} in general agreement with the model from \citet{Garrett2000}. Theoretical and numerical investigation has further strengthened understanding of the turbulent bubble cascade above the Hinze scale \citep{Chan2020,Chan2020a}.  \citet{Deike2016} demonstrated the ability of numerical methods to reproduce the size distribution observed experimentally and described theoretically, with an extension of the theory to constrain the mean air flow rate for increasing wave slopes. That study also noted a correspondence between the development of the entrained bubble population and the wave's energy dissipation rate. For bubbles below the Hinze scale, however, there is significant scatter between existing datasets, although \citet{Deane2002} suggests a relationship of $\propto r^{-3/2}$.

The numerical studies from \citet{Deike2016,Wang2016} had limited resolution of sub-Hinze scale bubbles and were performed at $Re=40\times 10^3$, $Bo=200$ with the assumption that the bubble size distributions were independent of Re, Bo, like the dissipation rate (see \S\ref{intro-energy}). The present DNS study brings to bear sophisticated methods and computational resources to test the dependence in Re, Bo, of the bubble size distribution, and to resolve the sub-Hinze bubble statistics. These constitute two of the main objectives of the present study.


\subsection{Droplet size distributions in breaking waves}
The mechanisms of spray generation by breaking waves have been recently reviewed by \citet{Veron2015}. Droplet size distributions have been explored experimentally in the presence of wind \citep{Wu1979,Veron2012,ortiz2016sea,troitskaya2018bag} as well as for deep water breaking waves generated by linear focusing \citep{Erinin2019}, while numerical investigations have been made of Lagrangian transport of spume droplets in the air \citep{Richter2013,Druzhinin2017,Tang2017}. However, a general theoretical model for the droplet size distribution has not been formulated.



In the context of breaking waves, spray is not created in the same manner as bubbles in the flow, being instead more analogous to atomization and fragmentation droplets \citep{Veron2012,troitskaya2018bag,Villermaux2020}. They are generated by two main mechanisms: direct ejection from wave impact and the related dynamic interface evolution, and indirect jet ejection resulting from the bursting of bubbles that were initially entrained by the breaker \citep{Lhuissier2012,Deike2018b,Berny2020}. The latter population is typically much smaller than the former \citep{Veron2015}, and hence even more challenging to resolve numerically within the breaking wave event, but can be studied separately \citep{Deike2018b,Berny2020}. Separately, a major complicating factor is that spray droplet populations are typically significantly smaller than bubble populations for a given breaking wave, leading to challenges in statistical convergence of the data. For these reasons, experimental and numerical studies of droplet production by breaking waves are limited \citep{Wang2016,Erinin2019}. In this study, droplet populations are resolved over a sufficient range of length scales to allow comparison with experiment, showing good agreement in the shape of the resolved size distribution. Velocity and joint velocity-size distributions are also shown, which will aid future studies.

\subsection{Outline}

In this paper, we present high-resolution DNS, which mobilizes sophisticated tools and computational resources to advance the following challenges; we will show statistics spanning multiple scales of fluid behaviour for full 3D simulations which capture breaking physics as seen in laboratory experiments regarding energy dissipation, bubble and droplets size distribution. The setup is similar to \citet{Deike2016} and is analogous to deep-water breaking waves in the laboratory obtained by focusing packets \citep{Drazen2008,Deane2002} as demonstrated by \citet{Deike2016}, but increased resolution of the interfacial processes allows access to higher Reynolds and Bond number to describe the transition to 3D turbulence and the formation of droplets and bubbles down to scales comparable to state-of-the-art laboratory experiments. These simulations represent the current state of the art in multiphase simulations of breaking waves and further confirm that the physics of breaking waves can be profitably investigated through these high fidelity numerical data. We analyse the role of these parameters in interfacial processes, including air entrainment, bubble statistics and droplet statistics. We discuss how energy dissipation, bubble and droplet statistics seem independent of the Reynolds number above a certain value, for the strong plunging breakers, confirming the results obtained previously at lower Reynolds numbers by comparison with experimental data. Next, we investigate the role of the capillary length and other flow scales on the air entrainment and spray production, which are most likely to mediate the development of transverse instabilities in the breaking process. We emphasize that such a study is only possible thanks to improvement in adaptive mesh refinement (AMR) techniques, along with increasing computational power, which has enabled sufficiently high resolution.

The paper proceeds as follows. In \S\ref{sec:formulation}, we describe the numerical methods and the formulation of the physical problem, the transition from the initial planar configuration to fully-developed 3D flows, and the general processes which produce entrained bubbles and ejected spray. In \S\ref{sec:energetics}, we investigate the development of the 3D flow in direct comparisons with 2D computations as well as the role of transverse instabilities and their influence on the dissipation rate. We study the transition time and length scale of the breaking flow, from its initial 2D configuration, to final 3D turbulent one. Then, in \S\ref{sec:bubbles}, we present bubble size distribution at higher $Re, Bo$ and numerical resolutions than those found in the numerical literature, and extend below the Hinze scale at lower $Re, Bo$. Droplet size and velocity distributions are presented in \S\ref{sec:droplets}, before concluding in \S\ref{sec:conclusion}.

\begin{figure}
\begin{minipage}[t]{0.4cm}
\raisebox{1.7cm}{(a)}
\end{minipage}
\begin{minipage}{6cm}
\centering
\includegraphics[width=1.0\textwidth]{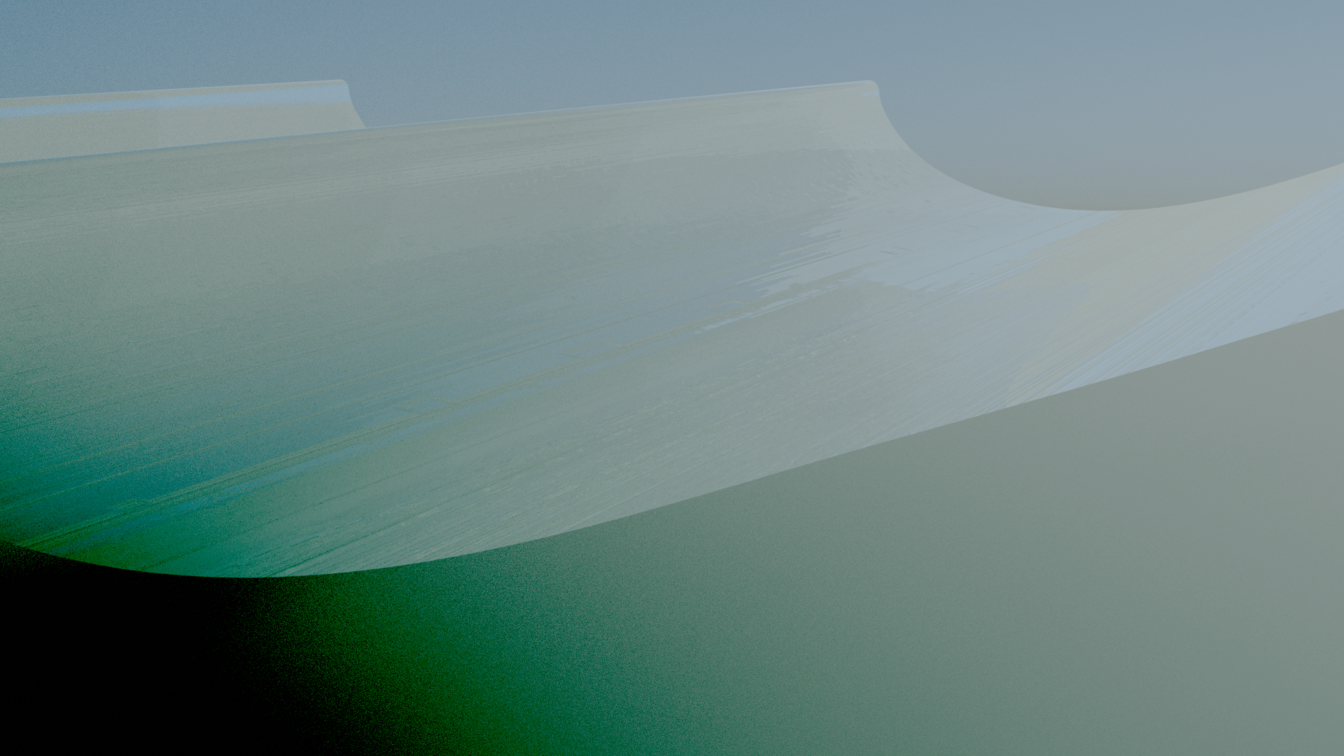}
\end{minipage}
\begin{minipage}[t]{0.4cm}
\raisebox{1.7cm}{(b)}
\end{minipage}
\begin{minipage}{6cm}
\centering
\includegraphics[width=1.0\textwidth]{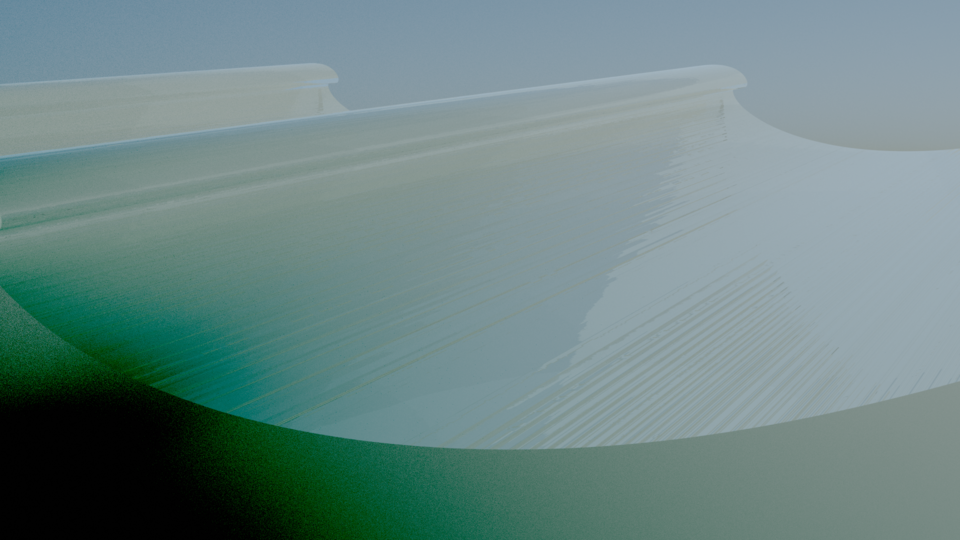}
\end{minipage}\hfill\\
\begin{minipage}[t]{0.4cm}
\raisebox{1.7cm}{(c)}
\end{minipage}
\begin{minipage}{6cm}
\centering
\includegraphics[width=1.0\textwidth]{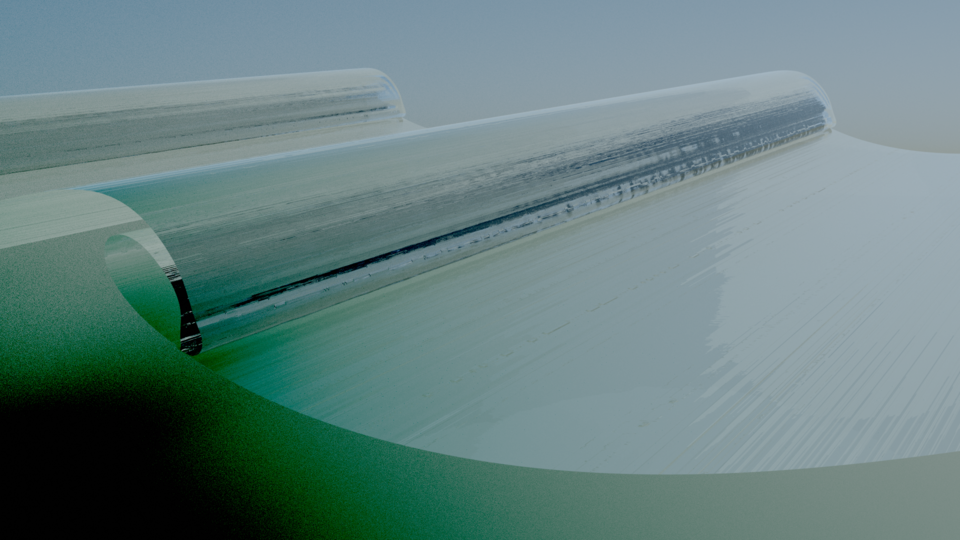}
\end{minipage}
\begin{minipage}[t]{0.4cm}
\raisebox{1.7cm}{(d)}
\end{minipage}
\begin{minipage}{6cm}
\centering
\includegraphics[width=1.0\textwidth]{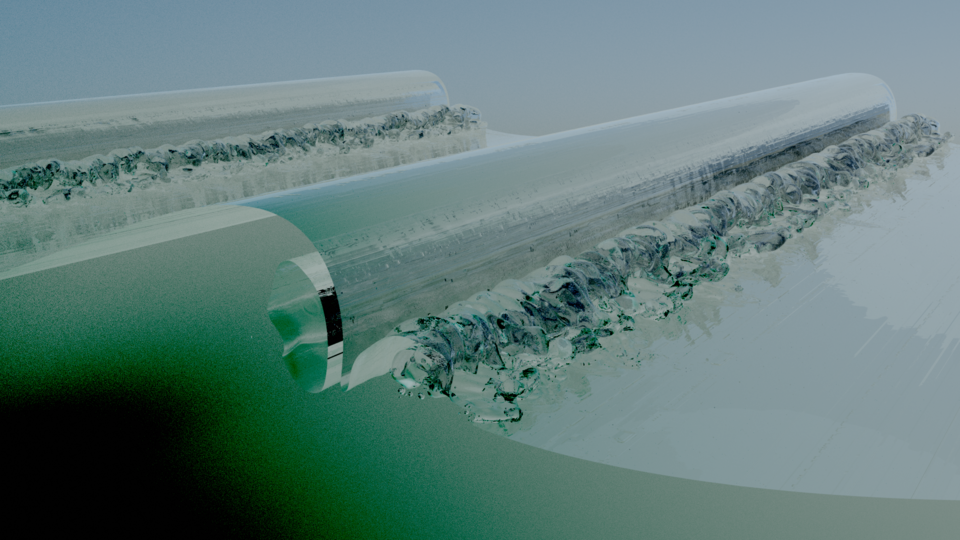}
\end{minipage}\hfill\\
\begin{minipage}[t]{0.4cm}
\raisebox{1.7cm}{(e)}
\end{minipage}
\begin{minipage}{6cm}
\centering
\includegraphics[width=1.0\textwidth]{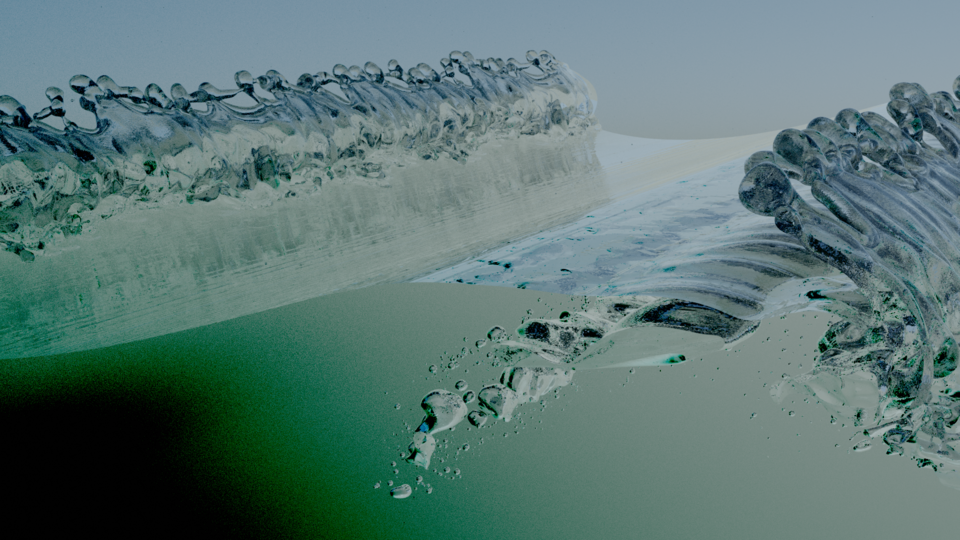}
\end{minipage}
\begin{minipage}[t]{0.4cm}
\raisebox{1.7cm}{(f)}
\end{minipage}
\begin{minipage}{6cm}
\centering
\includegraphics[width=1.0\textwidth]{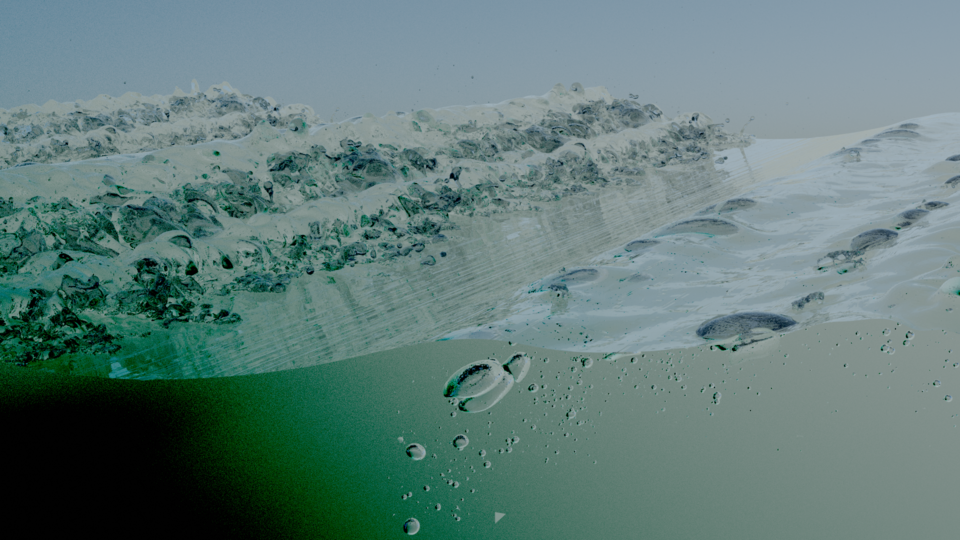}
\end{minipage}\hfill\\
\begin{minipage}[t]{0.4cm}
\raisebox{1.7cm}{(g)}
\end{minipage}
\begin{minipage}{6cm}
\centering
\includegraphics[width=1.0\textwidth]{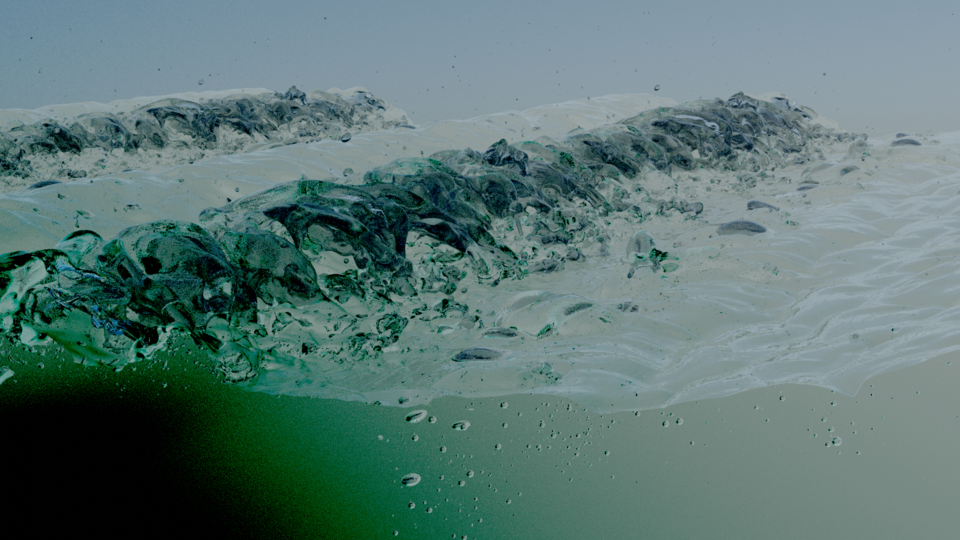}
\end{minipage}
\begin{minipage}[t]{0.4cm}
\raisebox{1.7cm}{(g)}
\end{minipage}
\begin{minipage}{6cm}
\centering
\includegraphics[width=1.0\textwidth]{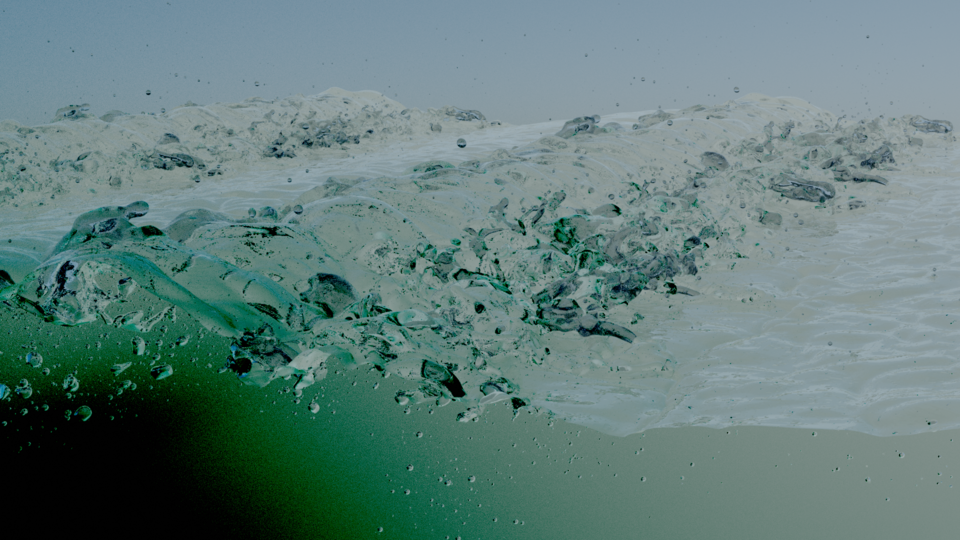}
\end{minipage}
\caption{Snapshot renderings of the three-dimensional (3D) breaking wave water-air interface at different times, for the case $Bo=500, Re=100\times 10^3$, at resolution $L=11$. (a) $t/T=0.37$, nonlinear steepening and initial overturning, (b) $t/T=0.56$, jet formation (c,d) $t/T=0.67, 0.8$, impact and ingestion of main cavity (e) $t/T=1.04$, splash-up of main wave and rupture of main cavity (f, g, h) $t/T=1.2, 1.36, 1.52$ continuation and slowdown of main breaking process.}
\label{fig:breakersequence}
\end{figure}

\section{Problem formulation and numerical method}
\label{sec:formulation}
\subsection{Basilisk library}
{\color{black}
  We use the Basilisk library to solve the two-phase incompressible Navier-Stokes equations with surface tension, in 2D and 3D. The successor of the Gerris flow solver \citep{Popinet2003,Popinet2009}, Basilisk is able to solve a diversity of partial differential equation systems in an adaptive mesh refinement (AMR) framework which significantly decreases the cost of high-resolution computations, allowing an efficient representation of multiscale processes. Flow advection is approximated using the Bell-Collela-Glaz method \citep{Bell1989} and the viscous terms are solved implicitly. The interface between distinct gas and liquid is described by a geometric volume-of-fluid (VOF) advection scheme, with a well-balanced surface tension treatment which mitigates the generation of parasitic currents \citep{Popinet2018}. A momentum-conserving implementation allows to avoid artefacts due to momentum ``leaking'' between the dense and light phases \citep{Fuster2018,Zhang2020}. }
{\color{black}
  The governing equations can be written,
  \begin{align}
    \frac{\partial \rho}{\partial t} + \nabla\cdot\left(\rho\mathbf{u} \right) &= 0, \label{momentum}\\
    \rho \left( \frac{\partial \mathbf{u}}{\partial t } + \mathbf{u} \cdot \nabla \mathbf{u} \right) &= -\nabla p + \nabla \cdot \left(2\mu \mathbf{D}\right) + \rho\mathbf{g} + \sigma \kappa \delta_s \mathbf{n},\label{mass}\\
    \nabla \cdot \mathbf{u} &= 0 \label{divergence},
  \end{align}
  where $\rho, \mathbf{u}, \mu, \sigma, \mathbf{D}, \mathbf{g}$ are the fluid density, velocity vector, dynamic viscosity, surface tension, deformation tensor, and gravitational acceleration vector respectively. The density and viscosity are allowed to vary according to a volume fraction field $c(\mathbf{x}, t)$ which in these simulations takes the value of zero in the gas phase and unity in the liquid phase. The variable $\delta_s$ is a Dirac delta which concentrates surface tension effects into the liquid-gas interface; $\kappa$ is the curvature of the interface, and $\mathbf{n}$ is its unit normal vector.
  }

\subsection{Wave initialization}
We consider breaking waves in deep water. The relevant physical parameters are: the liquid and gas $\rho_w, \rho_a$ respectively, the respective dynamic viscosities $\mu_w, \mu_a$, the surface tension $\sigma$, the wavelength $\lambda_0$, initial wave amplitude $a$, and gravitational acceleration $g$. The water depth $h_0$, while finite, is assumed sufficiently large so that it does not significantly affect the breaking physics. The eight significant parameters, which are expressed in three physical dimensions, can thus be reduced into five dimensionless groups according to Buckingham's theorem; these are the density ratio $\rho_a/\rho_w$, viscosity ratio $\mu_a/\mu_w$, wave slope $S = a k$ where $k=2\pi/\lambda_0$ is the wavenumber, and the Bond and Reynolds numbers, as defined previously, {\color{black} $\textrm{Bo} = \Delta \rho g/\sigma k^2$, $\textrm{Re} = \sqrt{g\lambda_0^3}/\nu_w$,
where $\Delta \rho = \rho_w - \rho_a \simeq \rho_w$, and $\nu_w = \mu_w/\rho_w$ is the kinematic viscosity. The wave period is $T = \lambda_0/c = 2\pi/\sqrt{gk}$, where $c=\sqrt{g/k}$ is the linear phase speed for deep water gravity waves.} {\color{black} The governing equations \eqref{momentum}-\eqref{divergence} can be non-dimensionalized in terms of these groups.} These definitions follow the literature, see \citet{Chen1999,Iafrati2009,Deike2015,Deike2016}. 

The numerical resolution is indicated by the smallest cell size attained in the simulation, given by $\Delta = \lambda_0/2^L$, where $L$ is the maximum level of refinement used in the AMR scheme. The refinement criterion is based on both the velocity field and the VOF tracer field. The maximum resolution used in this study is $L=11$, corresponding to a conventional grid of $(2^{11})^3$, or approximately 8.6 billion total cells. Under the AMR scheme, the grid size reduces to the order of 150 million cells at $L=11$.

We initialize the breaking wave following \citet{Chen1999,Iafrati2011,Deike2015,Deike2016,Wang2016,Chan2020,Chan2020a}, based on an unstable third-order Stokes wave {\color{black} for the water velocity and zero velocity in the air. The flow is regularized in the first time step. We note that the Stokes wave solution has been derived for an irrotational, inviscid, free surface wave, hence remains an imperfect initial condition for the full two-phase flow problem, accounting for viscosity and surface tension. However, numerous studies have demonstrated that it provides an efficient and compact initialization to study the post-breaking processes.} Both 2D and 3D simulations are conducted in order to investigate the transition from the laminar, planar, and essentially two-dimensional initial evolution to the final, turbulent, three-dimensional flow. Besides the dimensional difference, the 2D simulations are initialized identically to the 3D simulations. {\color{black} In the 3D simulations, no perturbation is used to seed the transition from planar to non-planar evolutioin of the wave; this transition is brought about by numerical noise during the breaking process.}


\subsection{Parameter space}
The density and viscosity ratios are fixed to the values for water and air, $\rho_w/\rho_a = 850, \mu_w/\mu_a = 51.15$, and the input slope is fixed at a nominal value of $S=0.55$, leaving the remaining two groups, Re, Bo to be varied. Thus, we investigate the independent effects of variation in surface tension through Bo and viscosity through Re. The fixed value of $S$ is chosen to be sufficiently large to force the wave into a plunging breaker \citep{Deike2015}. We refer the reader to \citep{Deike2016} for an extensive study on the role of the wave slope $S$ at constant Re,Bo. The parameters are shown in Table \ref{tab:config}, and correspond to low (Bo=200), medium (Bo=500) and high (Bo=1000) Bond number, and low (Re=40000) and high (Re=100000) Reynolds numbers. Cases run to test grid-convergence span moderate ($L=10$) and fine ($L=11$) resolutions respectively. Some additional cases at a variety of Reynolds numbers are also run for the energetics comparison in \S\ref{sec:energetics}. We reach a maximum separation of defined scales (wavelength to Hinze scale) of a factor $\sim550$. The grid size for the $L=11$ case reaches 181 million cells, for a maximum runtime (excluding scheduling and queueing times) of 1.4 months and a cost of half a million CPU-hours. {\color{black}These highest resolution cases were run on the Stampede2 cluster at the Texas Advanced Computing Center of the University of Texas, typically on between 192 and 768 cores of the Skylake node system. (Portions of these simulations were also run on the high performance computing resources of the National Computing Center for Higher Education (CINES)). Lower resolution cases ($L=10$) were run on the TigerCPU cluster at Princeton University using typically between 160 and 320 cores.} Note that while these simulations are expensive, they still save several orders of magnitude over a uniform- or fixed-grid approach, which would require a prohibitively large grid size of 8.6 billion cells in the highest-resolution case.

\begin{table}
  \begin{center}
    \begin{tabular}{p{1.5cm}p{0.9cm}p{0.7cm}p{1.2cm}p{1.2cm}p{1.6cm}p{1.2cm}p{2cm}}
      \hline
      Re & Bo & $L$    & $\lambda_0/r_H$ &  $\Delta/r_H$    & $\Delta/l_c$  & $r_H/l_c$ & Cost (CPU-h)   \\  
      \hline
      $40\times 10^3$  & $200$   &  11   & $284$  & $0.139$ & $0.043$ & $0.312$  & $1.75\times10^5$   \\
      $40\times 10^3$  & $200$   &  10   & $143$ & $0.279$ & $0.087$ & $0.311$  & $3.22\times10^4$ \\
      $40\times 10^3$  & $500$   &  11   & $501$ & $0.489$ & $0.069$ & $0.280$ & $2.58\times10^5$ \\
      $40\times 10^3$  & $500$   &  10   & $251$ & $0.489$ & $0.137$ & $0.281$  & $3.83\times10^4$  \\
      $100\times 10^3$ & $500$   &  11   & $501$ & $0.245$ & $0.069$ & $0.280$ & $5.26\times10^5$ \\
      $100\times 10^3$ & $500$   &  10   & $251$ & $0.484$ & $0.137$ & $0.284$ & $6.42\times10^4$ \\
      $100\times 10^3$ & $1000$  &  11   & $767$ & $0.375$ & $0.097$ & $0.259$ & $5.56\times10^5$ \\
      $100\times 10^3$ & $1000$  &  10   & $384$ & $0.738$ & $0.194$ & $0.263$ & $8.76\times10^4$ \\
      \hline
      Total Cost (CPU-h)    &         &       &       &         &     &        & $1.76\times 10^6$ \\
      \hline
    \end{tabular}
  \end{center}
  \caption{Computational matrix of parameter space for 3D breaking waves. The slope for each case is $S=0.63$ modeling a strong plunging breaker. {\color{black} The column labels are as follows: Re - Reynolds number; Bo - Bond number; $L$ - maximum level of grid refinement; $\lambda_0/r_H$, ratio of wavelength to Hinze scale; $\Delta/r_H$ - ratio of smallest grid size to Hinze scale; $\Delta/l_c$ - ratio of smallest grid size to the capillary length, defined as $l_c^2 = 1/(k^2Bo)$ where $k=2\pi/\lambda_0$ is the wavenumber; $r_H/l_c$ - ratio of Hinze scale to capillary length.}}
\label{tab:config}
\end{table}

\subsection{General flow characteristics}\label{sec:qualitative} 

The wave evolves in a manner similar to that seen in previous studies with similar initialization \citep{Deike2015,Deike2016}. Figure \ref{fig:breakersequence} shows a sequence of stills at different stages of the breaking process. The initially planar wave steepens nonlinearly to a point where it locally develops a vertical interface (a,b). The wave then overturns, forming a jet which projects forward into the upstream water surface (c), and impacts onto it (d), breaking the initially planar symmetry. At this moment, a large tube of air is ingested into the liquid bulk which we refer to as the \emph{main cavity}. The wave now also forms a fine-scale 3D structure at the point of impact, while ingesting the tubular cavity. This cavity persists for some time until it breaks along its length into an array of large bubbles (at $t/T=1 - 1.2$ e,f). In the meantime, the continuing breaking process on the surface creates a splash-up jet, as the wave proceeds into the strongly dissipative phase of the active breaking process (f) and develops into a fully-developed 3D flow (g,h) from $t/T=1.4$ onwards. At late times, most of the wave energy has been dissipated in the breaking process, but the turbulent regions persist for some time, during which a very large array of spray and especially bubbles is formed (f,g,h). All the presented cases produce a large quantity of bubbles of various sizes, but spray is abundantly produced, particularly at higher Bond numbers.

These qualitative aspects of the breaking wave dynamics are crucial for a faithful representation of the breaking process. In this respect, the evolution and dynamics of the breaker closely resembles those of laboratory experiments, notwithstanding certain Bond- and Reynolds-number influences and despite the different initializations across studies. The overturning phenomenon is very similar to that seen in \cite{Bonmarin1989,Rapp1990,Drazen2008}; the size and shape of the main ingested cavity matches very closely that seen in a large array of theoretical, numerical and experimental studies \citep{Longuet-Higgins1982,New1983,New1985,Dommermuth1988,Bonmarin1989}; and the subsequent droplet-producing splash sequence closely mirrors that seen in \cite{Erinin2019} (see \S\ref{sec:droplets}). This accurate reproduction of the breaker will be further reflected in various quantitative statistical comparisons with theory and experiment in the remainder of this paper, and moreover builds high confidence in the validity of our new results.


\section{Energetics and  transition to three-dimensional turbulent flow} \label{sec:energetics}
We determine the effect of $\textrm{Re}$ (and $\textrm{Bo}$) on the development of the 3D turbulent flow underneath the breaking wave by direct comparisons of the 3D simulations with 2D counterparts. 

\subsection{Energy dissipation by breaking}
\begin{figure}
  \centering
   \begin{overpic}[width=0.48\linewidth]{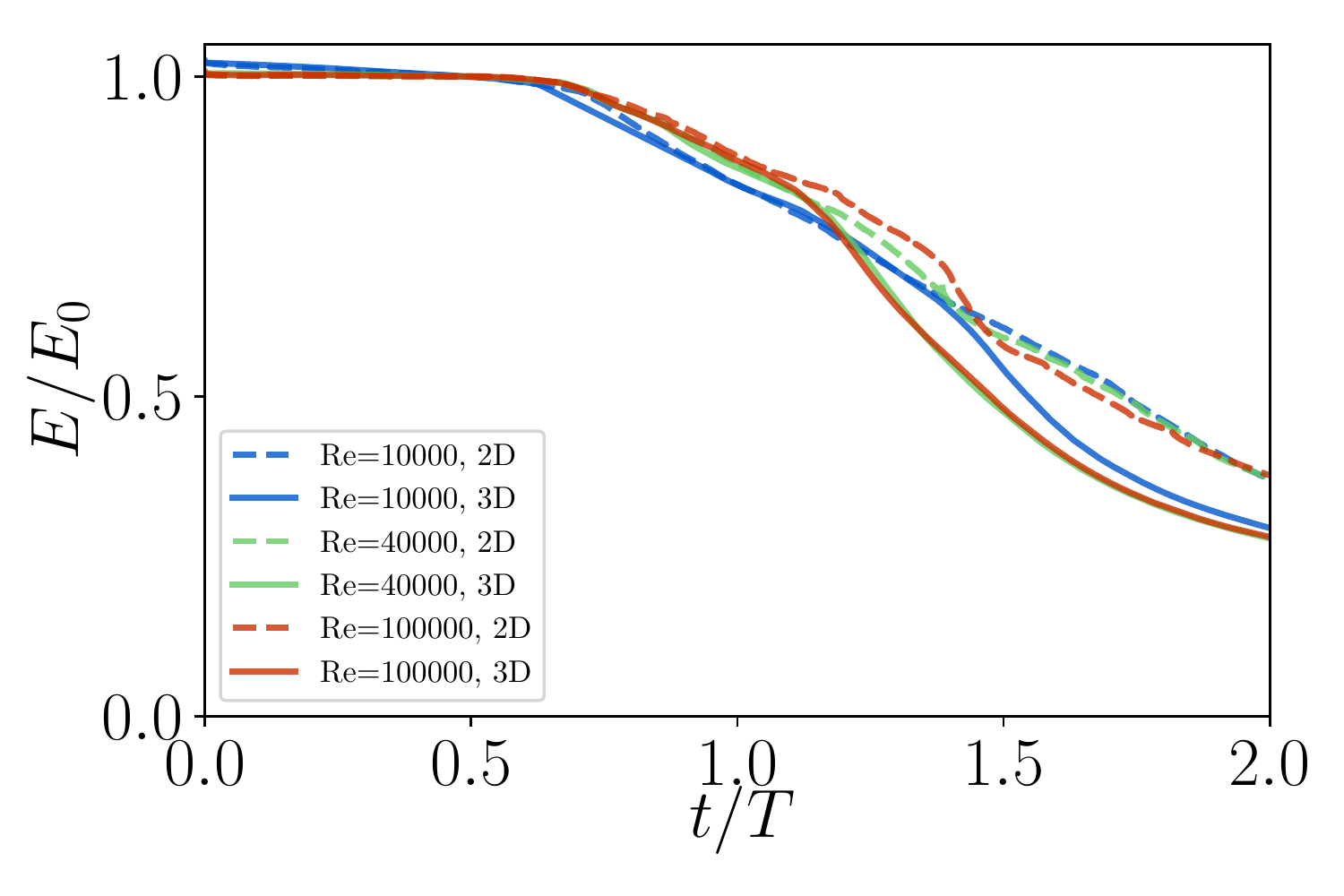}
     \put(0,60){(a)}
   \end{overpic}
   \begin{overpic}[width=0.48\linewidth]{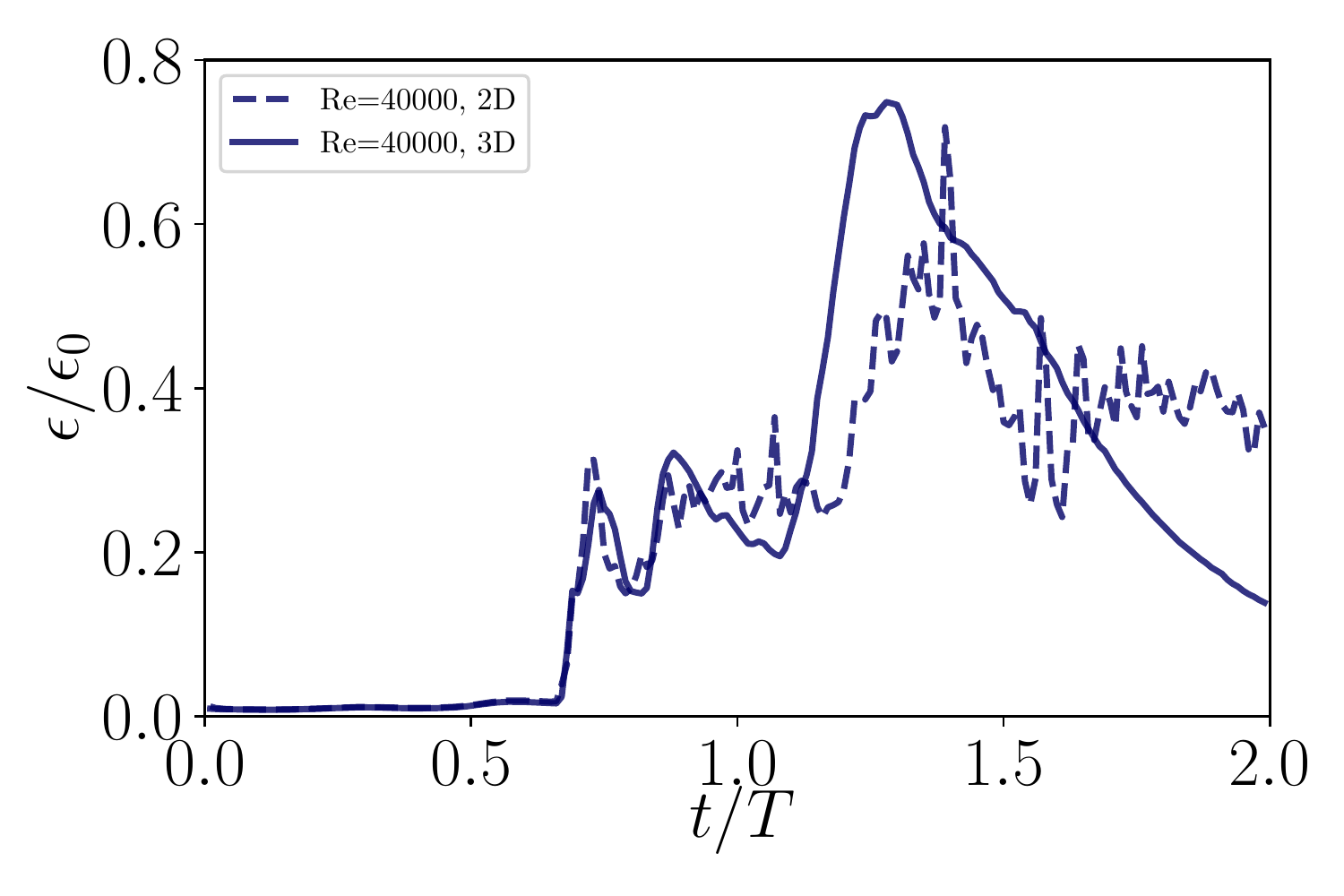}
     \put(0,60){(b)}
   \end{overpic}
  \caption{\color{black}Energy budget for a breaking wave. (a) Energy budgets comparing 2D and 3D for $\textrm{Bo}=500$ and $\textrm{Re}=1, 4, 10$ $\times 10^4$. (b) Corresponding instantaneous resolved dissipation rate comparing 2D and 3D for $\varepsilon$ for $\textrm{Bo}=500, \textrm{Re}=4\times10^4$. All simulations are at Level 11. Grid convergence studies are presented in Supplementary Materials.}
  \label{fig:E}
\end{figure}

The wave mechanical energy is $E=E_P + E_K$, where $E_P = \int_V \rho g (z-z_0) dV$ is the gravitational potential energy with a gauge $z_0$ chosen such that $E_P=0$ for the undisturbed water surface, $E_K = \int_V \rho (\mathbf{u}\cdot\mathbf{u}/2)dV$ is the kinetic energy, and the integrals are taken over the liquid volume $V$ \citep{Deike2015,Deike2016}. The instantaneous dissipation rate in the water is $\varepsilon \equiv \sum_{i,j} \varepsilon_{ij}$ where {\color{black}$\varepsilon_{ij} = (\nu_w/2\mathcal{V}_0) \int_{V} (\partial_i u_j + \partial_j u_i)^2dV$}, with $\partial_i \equiv \partial/\partial x_i$. We decompose $\varepsilon$ into in- and out-of-plane components $\varepsilon_{in} + \varepsilon_{out}$ where $\varepsilon_{in}=\sum_{i,j=x,z}\varepsilon_{ij}$ contains just those contributions of the deformation tensor which lie entirely in the streamwise ($x$) and vertical ($z$) directions, and $\varepsilon_{out}=\varepsilon_{3D} - \varepsilon_{in}$ comprises the remainder (i.e. the sum of terms $\varepsilon_{iy},\varepsilon_{yi}$ for $i=x,y,z$, $y$ being the spanwise direction). A planar flow features only the in-plane contribution $\varepsilon_{3D} = \varepsilon_{in}$, and a 3D flow features an additional contribution $\varepsilon_{out}$ (while in 2D, $\varepsilon_{2D} \equiv \varepsilon_{in}$).

Figure \ref{fig:E}a shows the budget of $E$ over time for increasing Reynolds number ($\textrm{Re}=10^4, 4\times10^4, 10^5$) and constant Bond number ($\textrm{Bo}=500$), with a direct comparison between the 2D and 3D cases. For each case, $E$ remains approximately flat at the earliest times, which correspond to the pre-broken wave where the dissipation is entirely due to the viscous boundary layer at the surface, which is properly resolved here given the high resolution in the boundary layer near the interface and has been verified for low amplitude waves  (see \citet{Deike2015}). Breaking begins as the wave steepens and overturns at $t/T \simeq 0.6$ and extends through $t/T = 2$ and afterwards, corresponding to the impact of the wave, and the active breaking part with air entrainment and generation of turbulence. Only a small amount of energy is dissipated in the air, amounting to approximately $5\%$ or less of the total energy budget. At small $\textrm{Re}$, viscosity is strong and the 2D and 3D budgets are in close agreement throughout the breaking process. For larger $\textrm{Re}$ (smaller viscosity), the 2D and 3D curves begin to strongly diverge at a time $t/T \simeq 1.2$, with the discrepancy becoming more pronounced at larger $\textrm{Re}$. The percentage of energy dissipated for this high slope breaker is about 70\%, close to the amount of energy dissipated in high slope plunging breakers in laboratory experiments \citep{Drazen2008,Rapp1990}.



Numerical convergence of the simulations for the energy budget and instantaneous dissipation rates are fully discussed in supplementary materials. From those results, the budget and dissipation rate at $\textrm{Re}=4 \times 10^4$ are numerically converged in $3D$ between $L=10, 11$, as well as for $\textrm{Re}=10^5$ between $L=10, 11$ in either 2D or 3D. The comparison of dissipation rates is very good for 2D between $L=11, 12$ at all Re. A 3D simulation at $L=12$ at the highest Re is not feasible to run currently given the computational cost. We note that the precise time evolution of the dissipation rate is sensitive to the precise shape at impact.


Numerical resolution of characteristic dissipative scale can also be discussed. Considering Batchelor's estimate for the viscous sublayer under the prebroken wave $\delta \sim \lambda_0/\sqrt{\textrm{Re}}$, our results indicate that at $\textrm{Re}=4\times10^4$ an effective resolution of 5 cells (at $L=10$) in the sublayer suffices for grid convergence. By the same estimation, we attain 6.5 cells in the viscous sublayer for $\textrm{Re}=10^5$ at $L=11$, suggesting grid convergence at this increased resolution. A resolution criteria for traditional single-phase DNS in the literature \citep{Pope2000,Dodd2021} involves the Kolmogorov length scale $\eta = (\nu_w^3/\varepsilon)^{1/4}$, with $k_{max}\eta > 1.5$ is considered sufficiently resolved, where $k_{max}=\pi 2^L/\lambda_0$ is the maximum resolved wavenumber. For the present simulations, for $\textrm{Re}=4\times 10^4, 10^5$, at $L=11$ this corresponds to $k_{max}\eta \simeq 3.4, 1.8$ respectively, which satisfies the criterion; and is similar to resolution used in direct numerical simulations of bubble deformation in turbulence \citep{Farsoiya2021}. For details, see supplementary materials.

Without a parallel (and currently not feasible) investigation of AMR convergence with respect to uniform-grid representation at these high resolution levels, and given these are individual realizations of multiphase turbulent flows, not ensembles, some caution in the interpretation of the present data is required. Nonetheless, using these different estimate of numerical convergence, the convergence characteristics are reasonable, given the complexity of the problem.

\subsection{Transition to three-dimensional turbulent flow}
\begin{figure}
  \centering
    \begin{overpic}[width=0.45\linewidth]{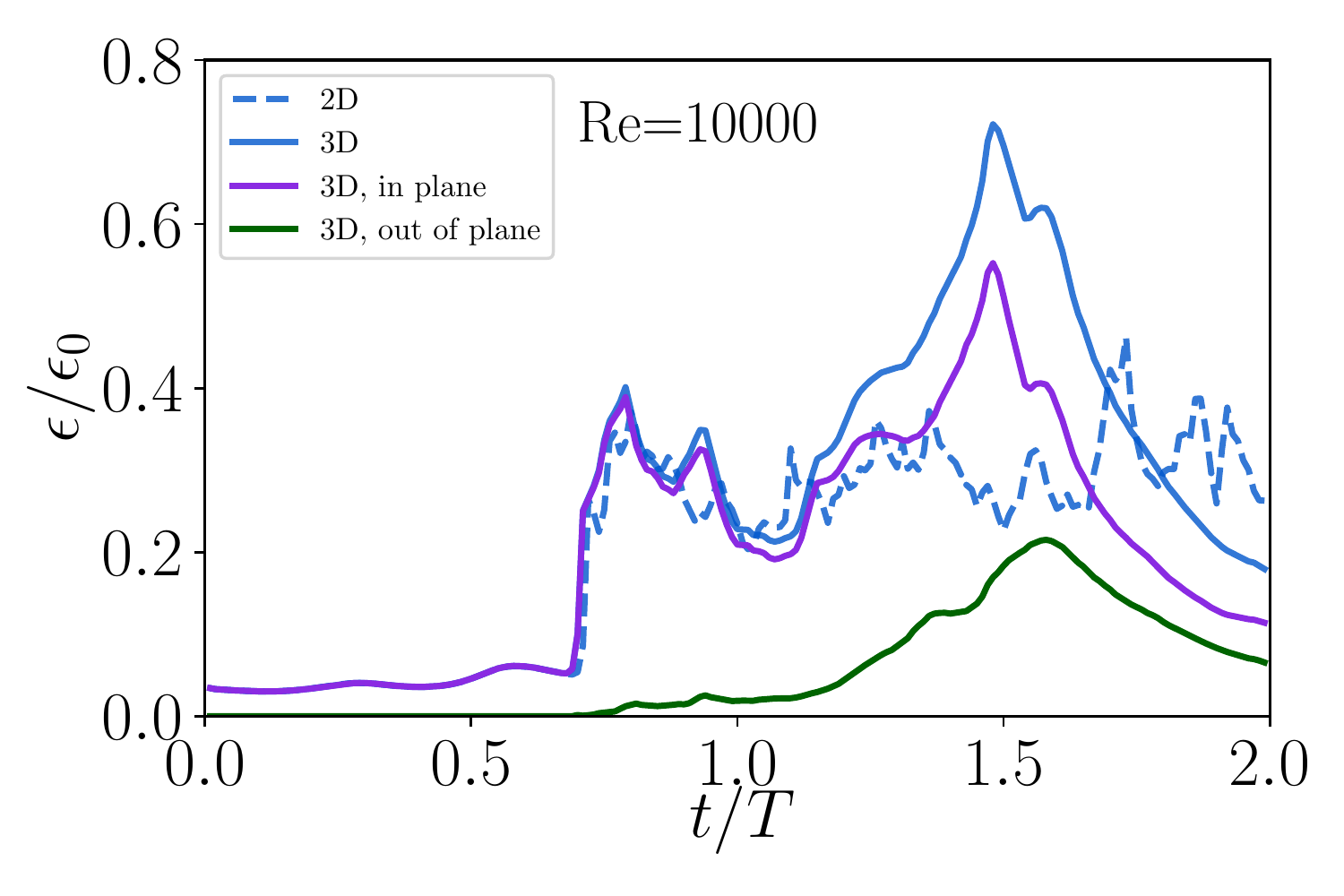}
      \put(0,60){(a)}
    \end{overpic}
    \begin{overpic}[width=0.45\linewidth]{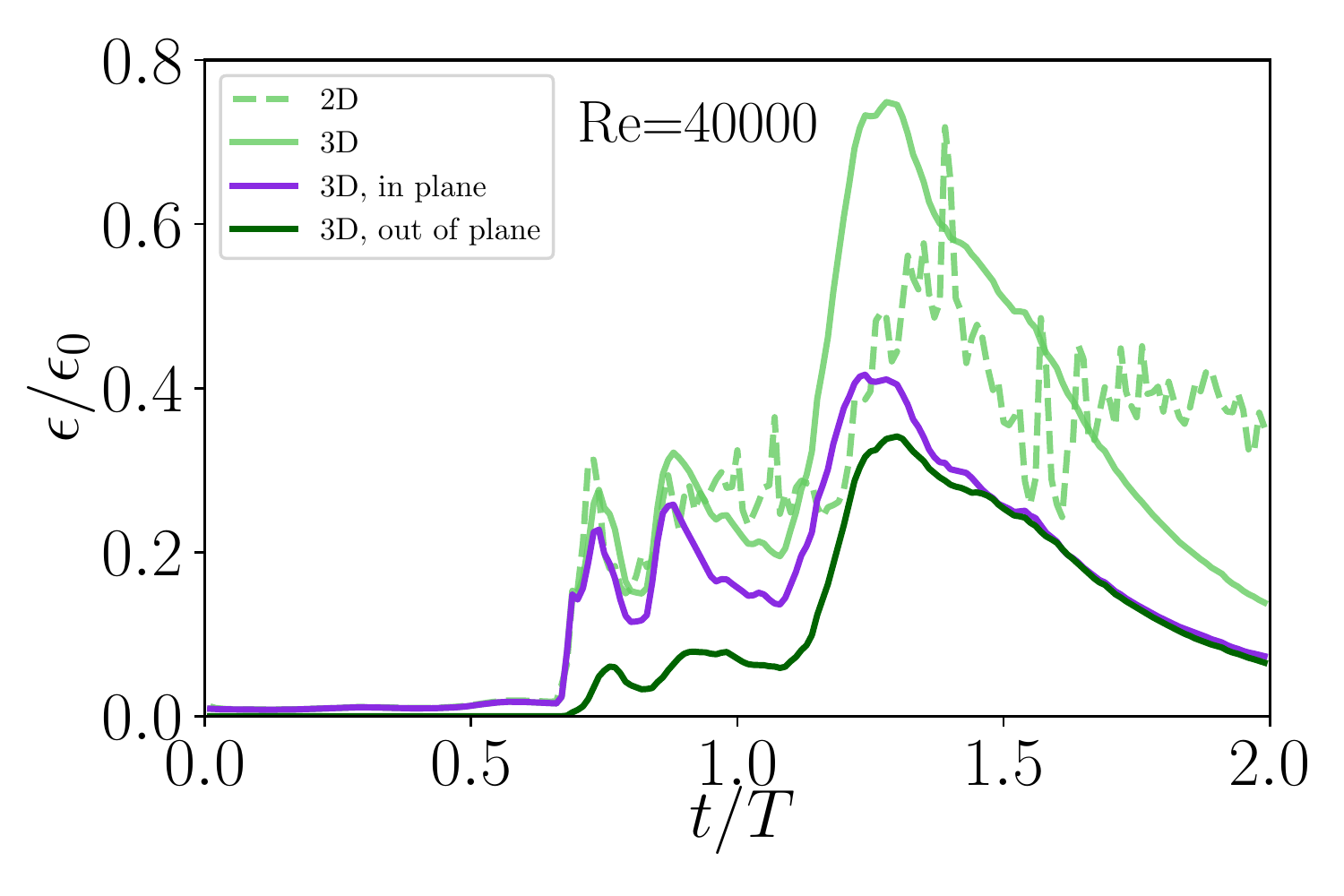}
      \put(0,60){(b)}
    \end{overpic}\\
    \begin{overpic}[width=0.45\linewidth]{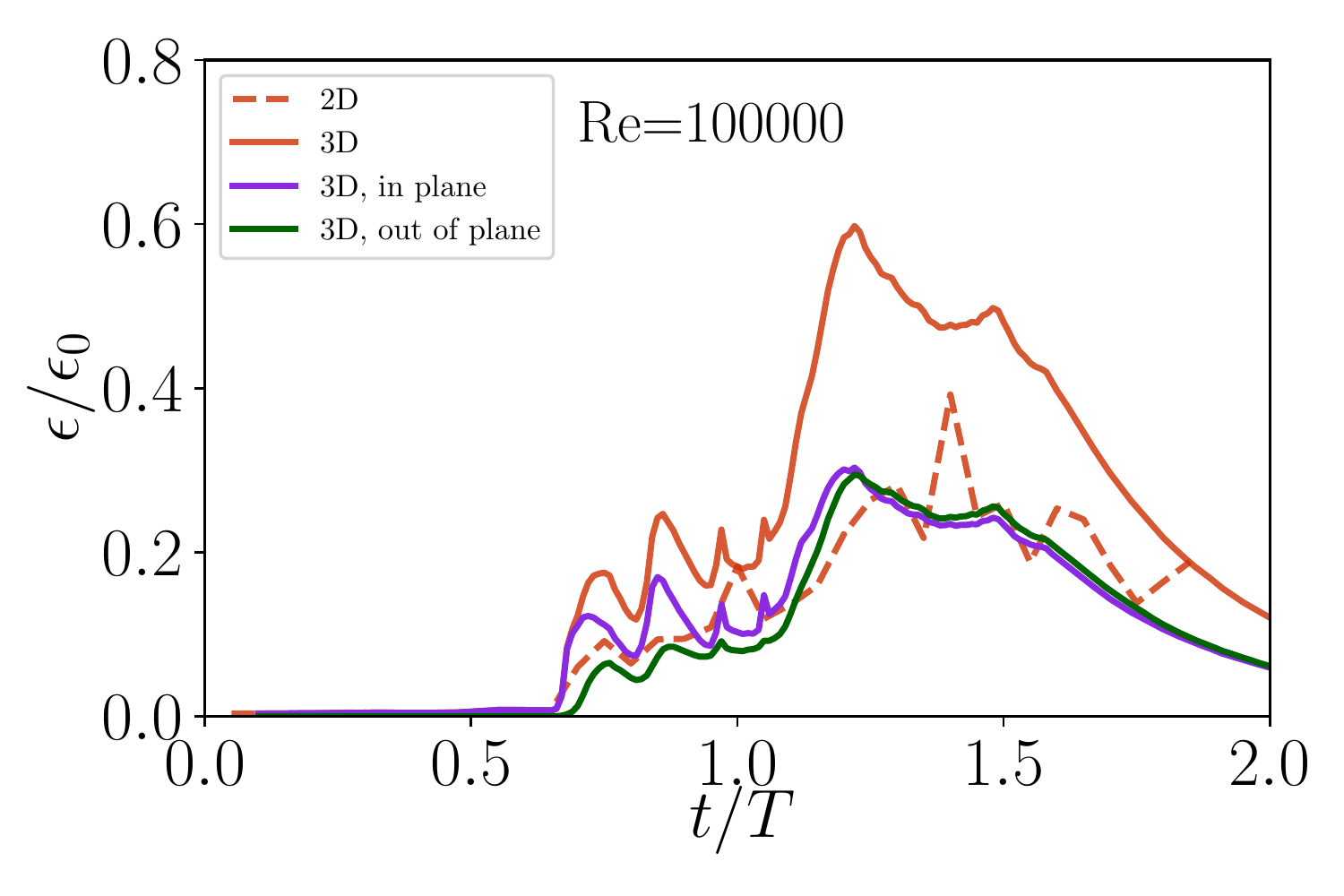}
      \put(0,60){(c)}
    \end{overpic}
    \begin{overpic}[width=0.45\linewidth]{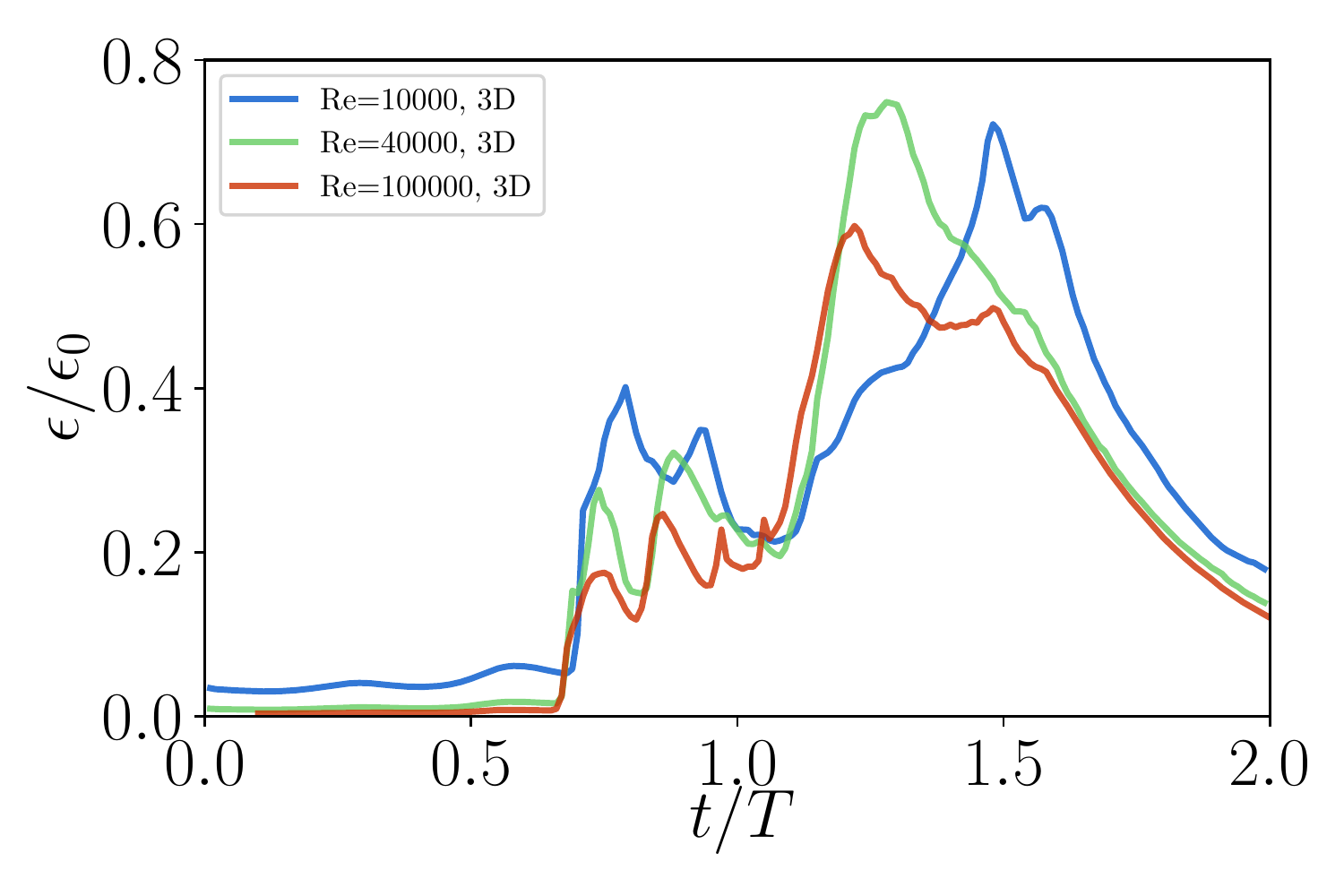}
      \put(0,60){(d)}
    \end{overpic}
  \caption{\color{black}Resolved instantaneous dissipation rates for breaking waves, showing in-plane and out-of-plane contributions to the 3D dissipation rate, along with the corresponding 2D case. $\textrm{Bo}=500$, (a) $\textrm{Re}=1\times10^4$; (b) $\textrm{Re}=4\times10^4$; (c) $\textrm{Re}=1\times 10^5$. The effective resolutions for each case are $1024^3, 2048^3, 2048^3$ respectively. (d) Overlaid total instantaneous dissipation rates for each of the cases from (a),(b),(c), showing similar dissipation rate time-evolution, especially for the two highest $\textrm{Re}$ values. Larger $\textrm{Re}$ corresponds with a more rapid transition from a planar initial flow to fully-developed 3D flow. $\varepsilon(t)$ is normalized by $\varepsilon_0$, the turbulent dissipation rate predicted by the inertial scaling argument (eq. 1.3).}
  \label{fig:Eij}
\end{figure}


{\color{black}Figure \ref{fig:Eij} shows the time evolution of the components of the dissipation rate for increasing $\textrm{Re}$. For each case, prior to breaking, the wave is planar and $\varepsilon_{in}$ is the only (small) contribution to $\varepsilon_{3D}$, but the evolution of $\varepsilon_{in}, \varepsilon_{out}$ on and after jet impact depends on the particular $\textrm{Re}$.  For $\textrm{Re}=10^4$, figure \ref{fig:Eij}a exhibits an almost entirely planar flow, with $\varepsilon_{out}$ becoming significant only late in the breaking process, when $\varepsilon_{in}, \varepsilon_{out}$ both grow rapidly to their respective peak values. Before this time, the total dissipation $\varepsilon_{3D}$ approximately matches $\varepsilon_{2D}$ for much of the time that the flow is planar, but deviates at later times. 

 At higher $\textrm{Re}=4\times10^4$, shown in figure \ref{fig:Eij}b, 3D effects arise earlier and are much more important: $\varepsilon_{out}$ grows gradually from the moment of impact, and at the moment of peak dissipation, $\varepsilon_{in}$ and $\varepsilon_{out}$ are comparable. At late times, they remain similar in magnitude, suggesting that the flow has become fully 3D and turbulent by $t/T=1.3-1.4$. As before, $\varepsilon_{3D}$ diverges from $\varepsilon_{2D}$ at the time of rapid growth of $\varepsilon_{in}, \varepsilon_{out}$, reaching a maximum value almost double that of $\varepsilon_{2D}$. 

Figure \ref{fig:Eij}c, showing $\textrm{Re}=10^5$, is similar to figure \ref{fig:Eij}b, but it does not exhibit any phase of latent planar flow where $\varepsilon_{in} \gg \varepsilon_{out}$, and the transition to a fully 3D flow is much faster after jet impact at $t/T=0.6$. Note that in this case while each of $\varepsilon_{in}, \varepsilon_{out}$ is similar to $\varepsilon_{2D}$, the in- and out-of-plane contributions are not  analogous to 2D processes. 


Finally, figure \ref{fig:Eij}d shows an overlay of each of the total instantaneous dissipation rates from figure \ref{fig:Eij}a-c, suggesting that the total dissipation rate evolution and maximum value is similar between the two highest $\textrm{Re}$ cases. Note however that since these cases are individual realizations of turbulent flow fields, these suggestions should be quantified further by the production and analysis of turbulent ensembles, which are prohibitively expensive to produce at these Reynolds and Bond numbers in the present investigation.}

\begin{figure}
  \centering
   \begin{overpic}[width=0.55\linewidth]{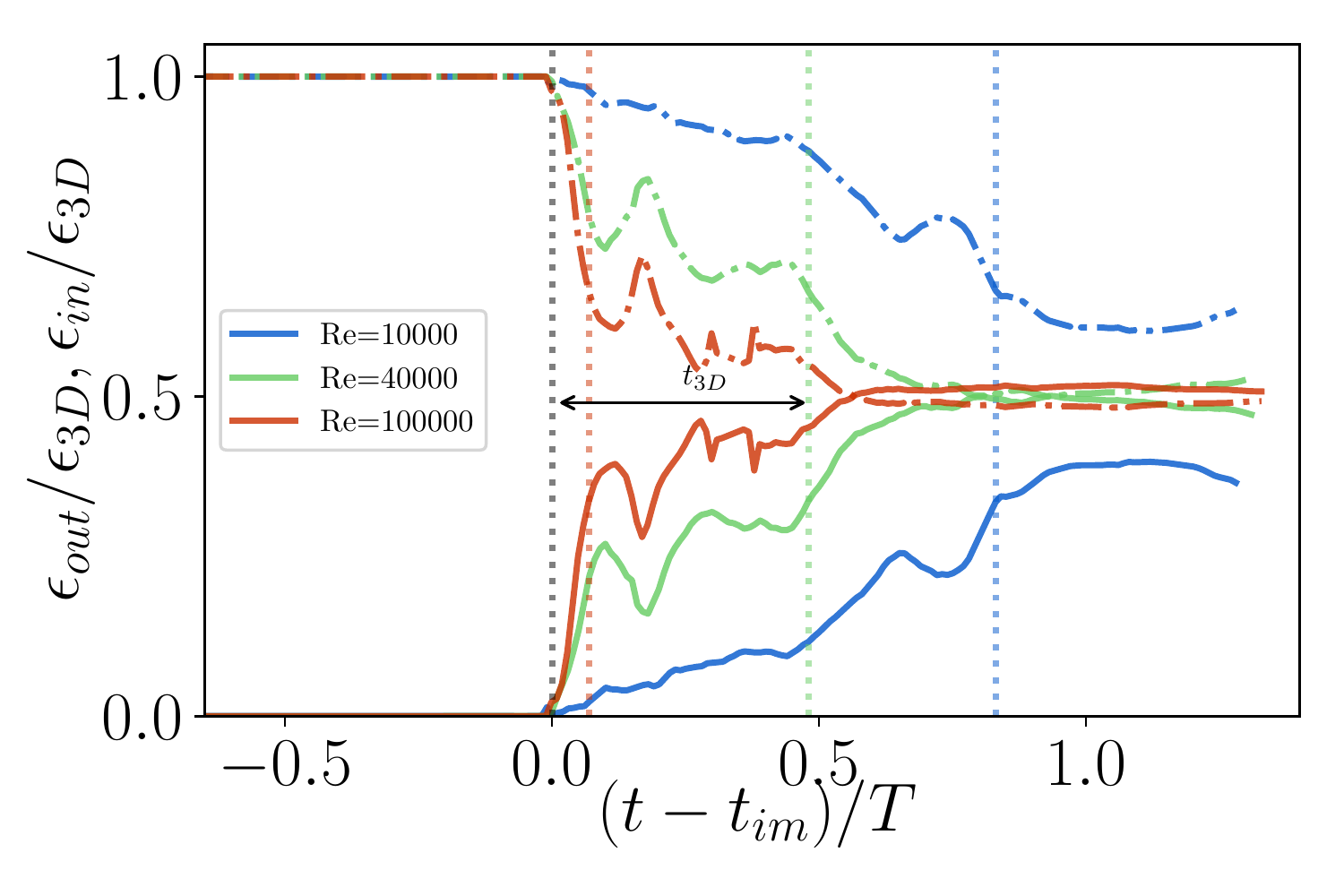}
     \put(-2,60){(a)}
   \end{overpic}
   \begin{overpic}[width=0.55\linewidth]{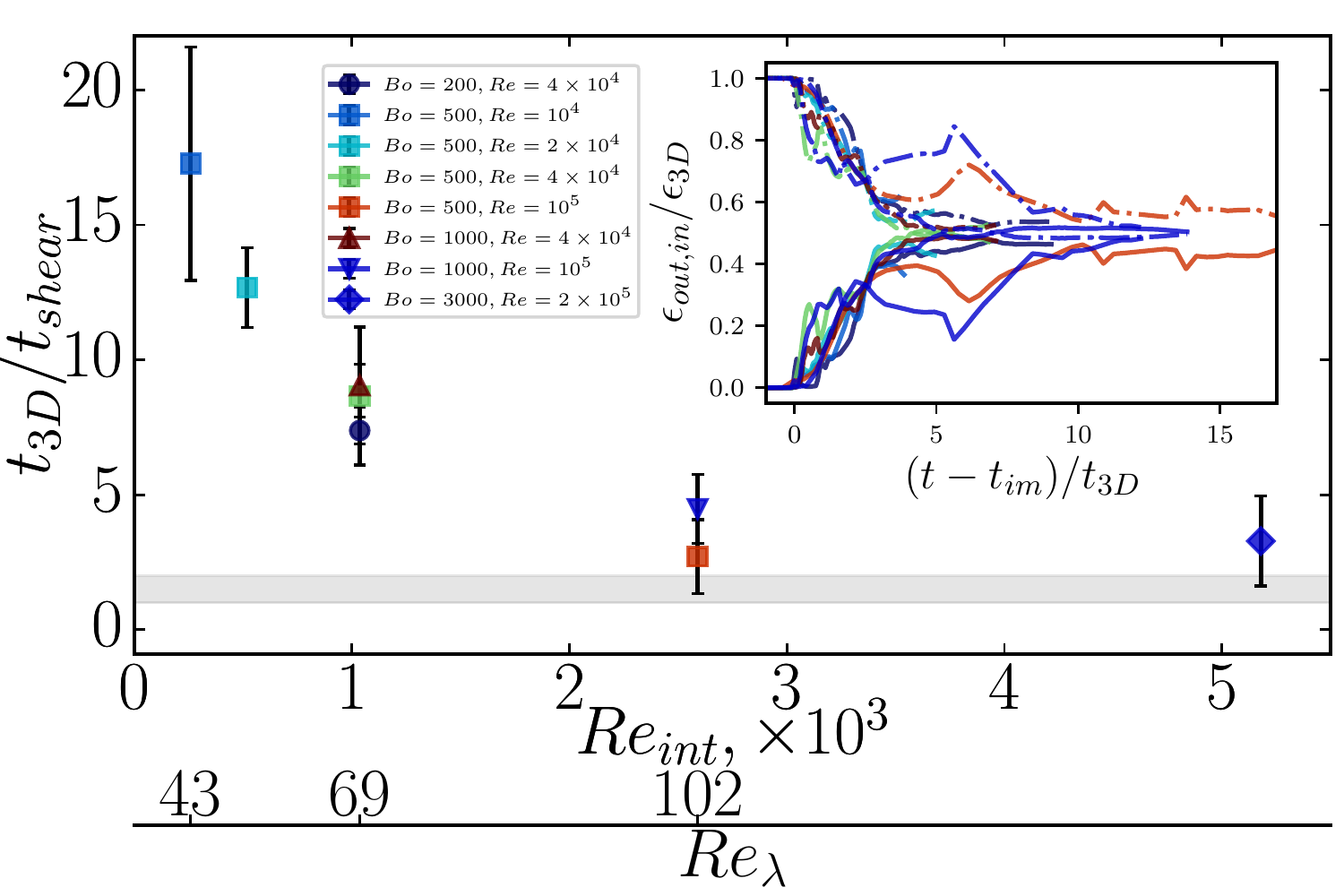}
     \put(-2,60){(b)}
   \end{overpic}
  \caption{\color{black}(a) Transition to fully 3D flow, measured as the relative contribution to dissipation rate $\varepsilon_{out}/\varepsilon_{3D}$ (solid) and $\varepsilon_{in}/\varepsilon_{3D}$ (dash-dotted) as a function of time for various Re number. The transition time is estimated as $\varepsilon_{out}/\varepsilon_{in}=\hat{c}=0.5$ and indicated as vertical dotted lines. Larger $\textrm{Re}$ drives more rapid transition. (b) Transition time $t_{3D}/t_{shear}$ as a function of $\textrm{Re}_{int}$ for all cases: an asymptotic value seems to be reached at high Re, coherent with experimental estimations (grey line). Inset shows the transitions dynamics with time rescaled as $(t-t_{im})/t_{shear}$ with line legend as in (a). Colors in inset are those in main plot.}
  \label{fig:metric}
\end{figure}

The values of $\varepsilon_{3D}$ are similar for the highest $\textrm{Re}$ suggesting the breaking process has achieved an asymptotic behaviour in terms of dissipation rate. The dissipation rate shown in figure 3 and 4 are normalized by that predicted by the scaling argument {\color{black}$\varepsilon_0=\left(\sqrt{gh}\right)^3/h$ (eq. 1.3)}, which describes experimental and numerical data for a wide range of breaking waves \citep{Drazen2008,Romero2012,Deike2016}. As such, our results are compatible with the inertial argument, experimental studies for a wide range of breakers, and previous numerical studies. 

We now investigate the development to 3D flow underneath the breakers. Figure \ref{fig:metric}a shows the relative increase of the out-of-plane contributions, $\varepsilon_{out}/\varepsilon_{3D}$ with time as well as the concomitant decrease of $\varepsilon_{in}/\varepsilon_{3D}$, for increasing Reynolds number. The terminal turbulent state is reached when either curve plateaus; this state occurs earlier for larger $\textrm{Re}$, showing the rapidity of development from planar to 3D flow. This indicates that viscosity mediates the 3D instabilities involved in the transition to turbulence at low $\textrm{Re}$. 

We define a heuristic development time to 3D turbulent flow $t_{3D}$ which is the time from impact until the moment that $\varepsilon_{out}/\varepsilon_{in} = \hat{c}$, where $\hat{c}$ is some representative percentage of the turbulence dissipation rate. For $\hat{c}=0.5$, the value of $t_{3D}$ is indicated in each case of fig. \ref{fig:metric}a. We assess the sensitivity of our time transition definition by varying $\hat{c}$ from $0.4$ to $0.6$. Secondly, small fluctuations in $\varepsilon_{out}/\varepsilon_{3D}$ could affect $t_{3D}$, so we filtered the data with moving averages of window sizes of 3, 5, 7, 9, and 11 points to estimate how $t_{3D}$ responds to gradual smoothing of the curve. The error bars are then estimated as the range of $t_{3D}$ as estimated across both of these methods and are shown in figure \ref{fig:metric}b. Finally, we also studied dependence of $t_{3D}$ on numerical resolution for the cases $\textrm{Bo}=200, \textrm{Re}=4\times10^4$; $\textrm{Bo}=500, \textrm{Re}=2\times 10^4$; and $\textrm{Bo}=500, \textrm{Re}=10^5$; we found that variation of $t_{3D}$ remained within the error bars. For the case $\textrm{Bo}=3000, \textrm{Re}=2\times10^5$, numerical convergence cannot be assessed. 



{\color{black}The transition to 3D turbulence can be analyzed in terms of a turbulence Reynolds number. We plot the transition time in fig. \ref{fig:metric}b for the various initial conditions as a function of two representative turbulence Reynolds number: using the integral length and velocity scales given by the breaking height $h$ and ballistic velocity $\sqrt{gh}$ \citep{Drazen2008}, an integral Reynolds number is $\textrm{Re}_{int} \simeq g^{1/2}h^{3/2}/\nu_w$. The Taylor length-scale characterizing the inertial range is estimated as $\lambda \simeq a\sqrt{10/Re_{int}}$ and fluctuations at this scale as $ v = \lambda\sqrt{\varepsilon/(15\nu_w)}$ \citep{Sreenivasan1984,Dimotakis2005} with $\varepsilon$ a characteristic dissipation rate taken as the peak value of $\varepsilon_{3D}/(\mathcal{V}_0\rho)$. This yields an estimate of the turbulent Reynolds number at the Taylor micro-scale $\textrm{Re}_\lambda = \lambda v/\nu_w \simeq 43, 69, 102$ for the wave Reynolds numbers $1,4,10 \times 10^4$.}

{\color{black} At $\textrm{Re}=4\times10^4$, the $\textrm{Bo}=500,1000$ points are identical, while the case $\textrm{Bo}=200$ shows a slightly lower value of $t_{3D}$, suggesting that, for Bond numbers above $500$, surface tension does not play a significant role in the transition to 3D flow and hence that $\textrm{Re}$ is the main controlling parameter of this process. The inset in fig. \ref{fig:metric}b shows the relative contributions as functions of the rescaled time $(t-t_{im})/t_{3D}$, including the different Bo, showing good collapse between all cases. The 3D transition time can be rationalized in the $\textrm{Re}$-asymptotic limit in terms of a Kelvin-Helmholtz scaling. Considering a uniform density shear layer driven by the breaker speed $\approx c=\sqrt{g/k}$ over the depth of the turbulent cloud $\approx h$, we get $t_{shear} \simeq 1/s$ where $s = k_{KH}U$ with $U \simeq \mathcal{A} c$ and $\mathcal{A}$ is an $O(1)$ constant and $k_{KH} \simeq 2/h$. This shear time $t_{shear}$ is used to normalize the axis in fig. \ref{fig:metric}b, and since the $O(1)$ constant is not precisely known, we indicate $t_{shear}$ with a shaded zone between $1$ and $2$ on figure \ref{fig:metric}b.} 

The transition time $t_{3D}$ seems to plateau at the highest Reynolds number we were able to test, $\textrm{Re}_{\lambda}\approx 50-100$. Further support of the asymptotic regime in $\textrm{Re}$ number is given by considering laboratory experiments of breaking waves  \citep{Loewen1994,Deane2002,Drazen2008,Rapp1990}, with $\lambda_0\sim 1-2$m, leading to $\textrm{Re}\approx 10^6$, and wave slopes $0.4-0.5$ inducing a turbulent flow with $\textrm{Re}_{\lambda}\approx 500$ \citep{Drazen2009}. From optical and acoustic records in these experiments, we estimate the transition time $t_{3D}^{exp}\approx 0.35\pm 0.1$s, consistent with $t_{shear}^{exp}\approx t_{3D}^{exp}$. The transition to $\textrm{Re}$-independent flow suggests a mixing transition Reynolds number $\textrm{Re}_{\lambda}$ \citep{Sreenivasan1984,Dimotakis2005} in the flow underneath the breaking wave, supported by the similarity of $\varepsilon_{3D}$ curves in fig. \ref{fig:Eij}b and \ref{fig:Eij}c and the possibly asymptotic behaviour in fig. \ref{fig:metric}b. This suggests that $\textrm{Re}_{\lambda}\simeq 50 - 100$ in the developed flow corresponds with a transition to $\textrm{Re}$-independent turbulent flow under a breaking wave, which would be consistent with observations of the mixing transition in grid-generated turbulence \citep{Sreenivasan1984} and scalar transport in turbulence \citep{Pullin2000,Dimotakis2005}.

\section{Air entrainment and bubble statistics}
\label{sec:bubbles}

\subsection{Cavity shape at entrainment}
\begin{figure}
\centering
  \begin{overpic}[width=0.8\linewidth]{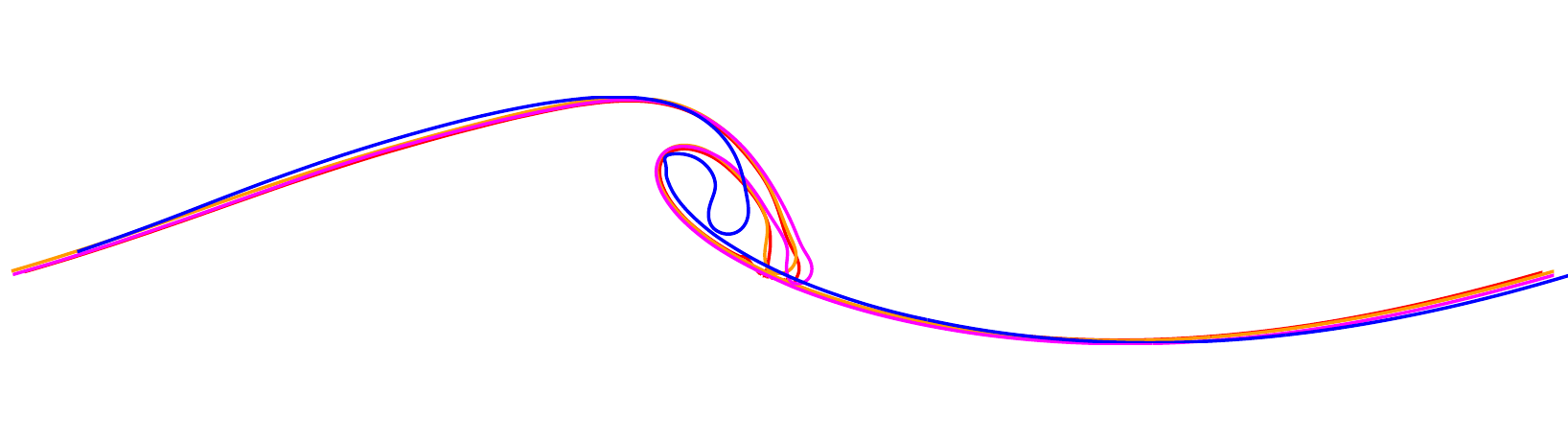}
    \put(0,15){\color{black}(a)}
  \end{overpic}

\begin{overpic}[width=0.8\linewidth]{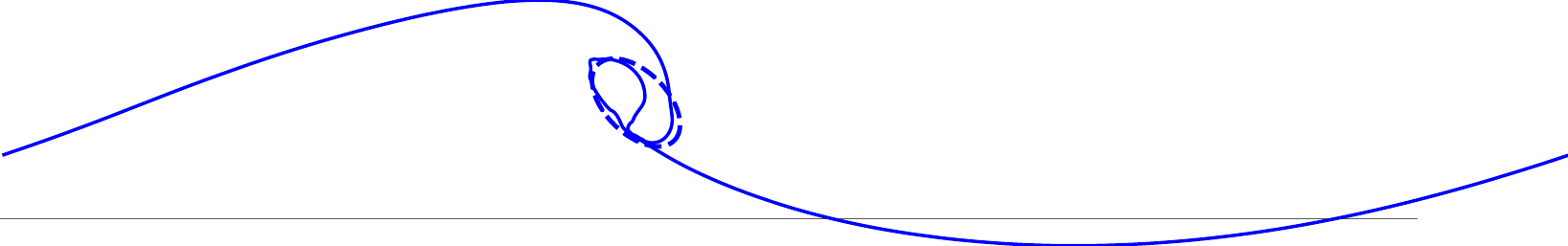}
    \put(0,15){(b)}
  \end{overpic}
\begin{overpic}[width=0.8\linewidth]{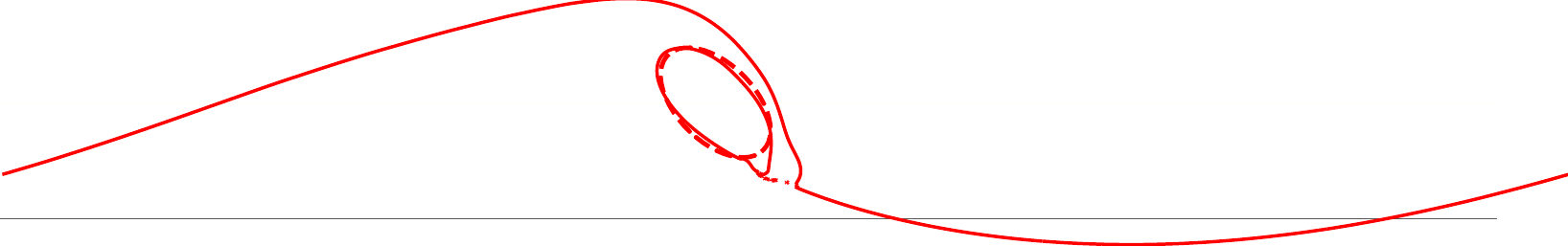}
    \put(0,15){(c)}
  \end{overpic}
\begin{overpic}[width=0.8\linewidth]{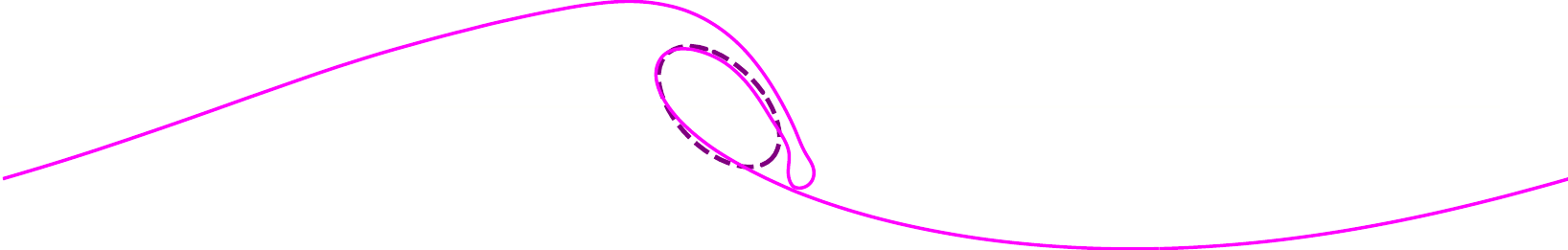}
    \put(0,15){(d)}
  \end{overpic}
\begin{overpic}[width=0.5\linewidth]{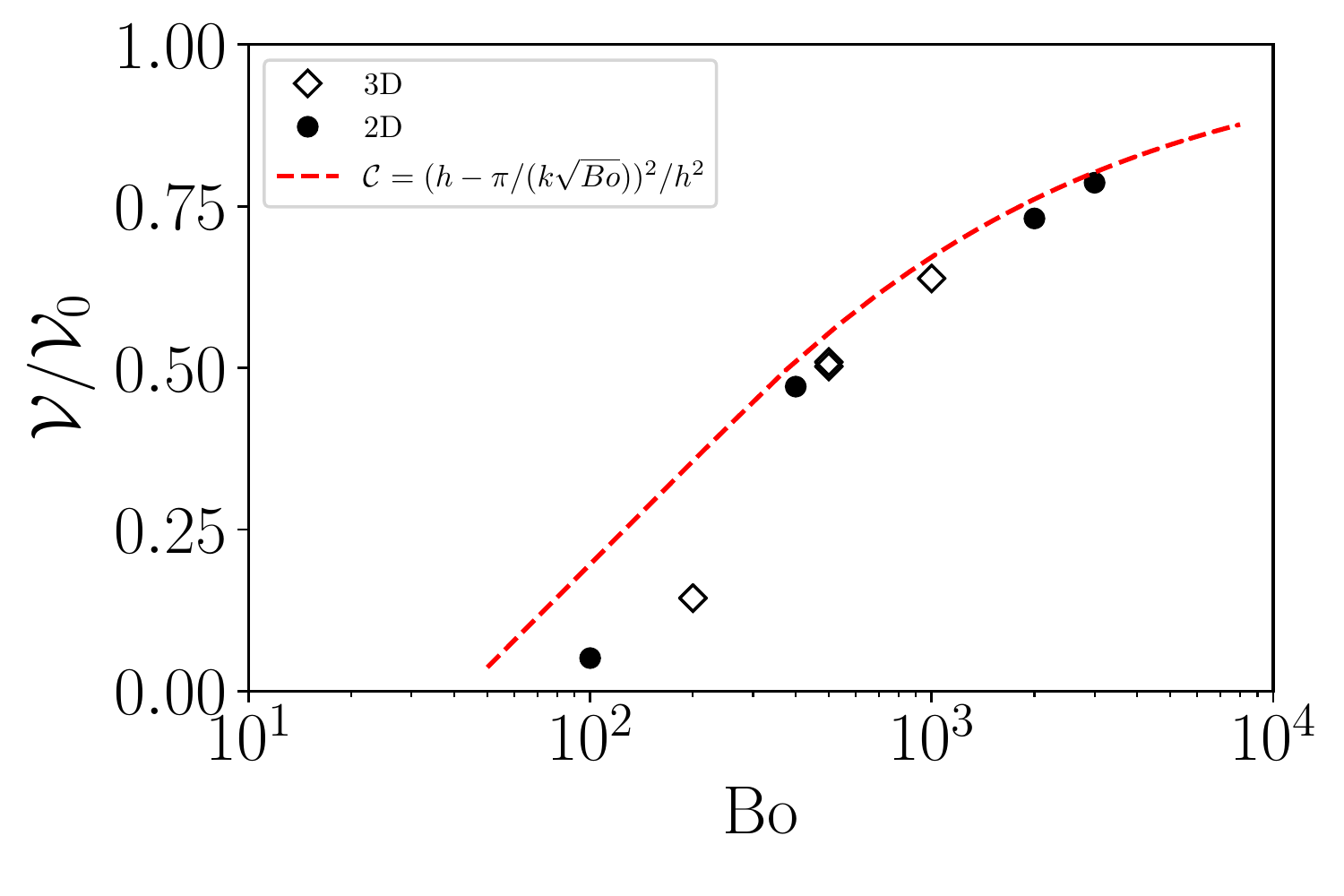}
  \put(0,65){\color{black}(e)}
  \end{overpic}
\caption{(a) Profiles of the volume-of-fluid interface at the moment of wave impact, obtained from 2D simulations. Blue: $Bo=200, Re=10^5$. Orange: $Bo=500, Re=4\times 10^4$. Red: $Bo=500, Re=10^5$. Purple: $Bo=1000, Re=10^5$. $L=11$ for each case. Note the dependence of cavity size on Bond number, but not on Reynolds number (compare red and orange curves).  (b,c,d): Wave profiles at the moment of impact with superimposed fitting ellipses for the cases, (b) $Bo=1000$, (c) $Bo=500$, (d) $Bo=200$, with $Re=10^5$ in all cases. (e) Plot of cavity area over different Bond numbers. Dashed: Corrective scaling from \eqref{cavity}.}
\label{fig:ImpactSnapshots}
\end{figure}

In this section, we describe air entrainment and bubble statistics. We begin by discussing the shape of the cavity at impact, which controls the size of the main cavity and the associated maximum volume of air entrained \citep{Lamarre1991,Deike2016}. Studies using a fully non-linear potential flow formulation, i.e. inviscid conditions and neglecting surface tension effects, have been able to reproduce the shape of the breaking wave at impact to a high level of precision \citep{Dommermuth1988}, with discussion on the elliptical or parametric cubic shape of the cavity \cite[]{New1983, Longuet-Higgins1982}. However, these methods do not resolve the post-impact process. \citet{Lamarre1991,Blenkinsopp2007,Deike2016} discuss that the maximum volume of air entrained is constrained by the length of breaking crest $L_c$ and $A$ is the cross-sectional area of the initially ingested cavity in the breaking process. In particular, $A$ controls the amount of entrained air initially available for subsequent breakup into a bubble size distribution. It has been assumed that the cross section area of entrained air scale as $A \propto \pi h^2/4$ \citep{Duncan1981,Lamarre1991,Blenkinsopp2007,Deike2016}, implicitly arguing that the height of the wave is large compared to the width of the jet.

As  already noted in previous work, when considering a two-phase solver able to resolve post impact, moderate Bo leads to a jet thicker than observed in the laboratory \citep{Chen1999,Song2004}. Such moderate Bond numbers were nevertheless considered in most previous studies when dealing with three-dimensional breaking waves. This is because larger $\textrm{Bo}$ exhibits increased separation between the wave length and Hinze scales and thus incurs a prohibitive numerical expense if all scales are to be resolved \citep{Wang2016,Deike2016}. Here, we use the high numerical efficiency gained through AMR and increased computing power, and are thus able to resolve breakers showing greater separation between length scales. Figure \ref{fig:ImpactSnapshots}a shows again that as $\textrm{Bo}$ increases, the wave jet becomes thinner and projects further forward ahead of the wave. When increasing the Bond number, the jet at impact appears thinner and more closely similar to those observed in laboratory experiments. {\color{black}It is important to remark that, by comparison of the orange and red curves in figure \ref{fig:ImpactSnapshots}a, at Bo=500, the jet thickness is independent of Reynolds number, which confirms that jet thickening is due to capillary effects.}



The thicker jet can be interpreted by comparing the wave height with the capillary length. For breakers in the laboratory, $h\sim 10$cm and {\color{black}$l_c=\sqrt{\gamma/\rho g}\sim 3$mm (the capillary length)} so that $h/l_c\approx 33$; in the DNS for Bo=200, we have $h/l_c \approx 7$ which indicates the importance of capillary effects. By increasing to Bo=1000, we get to $h/l_c\approx 16$ which is closer to laboratory conditions (but still smaller than waves from large scale breakers in the field). 

We therefore propose a correction of entrained area $A$ based on the {\color{black}width of the jet $l_j$}. First, the cavity shape is not truly circular but closely approximates an ellipse (alternatively a parametric cubic function \citep{Longuet-Higgins1982}) with an aspect ratio of $\sqrt{3}$ with its major axis rotated at an angle of approximately $\pi/4$ to the horizontal \cite[]{New1983,New1985}. The cavity area is then $A = \pi p^2/(4\sqrt{3})$ where $p$ is the major axis of the ellipse. If we now assume that, due to the thickness of the jet, the major axis is given by $p \simeq 3^{1/4}(h - K l_c)$, where $K$ is a positive $O(1)$ constant, which we set to $\pi$, we obtain $A = \pi (h-\pi/(k\sqrt{Bo}))^2/4$. This retrieves the usual relation for the cavity volume in the limit $\textrm{Bo} \to \infty$.

Figure \ref{fig:ImpactSnapshots}b-d show wave profiles at the moment of impact with a superimposed '$\sqrt{3}$-ellipse' rotated at $\pi/4$ and with the major axis given by our estimate, $3^{1/4}(h - \pi/(k\sqrt{Bo}))$. For higher Bond numbers ($500, 1000$), the ellipse fits very well, suggesting that our proposed cavity scaling is appropriate at high Bond numbers. Note that for the lowest Bond number ($200$), it approximates the shape of only the very rear of the cavity. This suggests that the $Bo=200$ case is qualitatively distinct from higher $Bo$ cases, in that capillary effects are sufficiently strong to change the morphology of the plunging breaker. Nevertheless, the good fit observed at higher Bond numbers supports the conjecture by \cite{New1983}, further supported by \cite{Dommermuth1988}, that the evolution of the overturning wave is independent of the details of the interior flow. Furthermore, since the $\sqrt{3}$-ellipse has been {\color{black}frequently observed in the above-cited literature}, our result also confirms that this evolution is independent of the details of the initial conditions.

This leads to the cavity correction for the entrained volume, defined as the ratio of the actual entrained cavity $\mathcal{V}$ over its asymptotic value at high Bo number $\mathcal{V}_0$,
\begin{equation}
\mathcal{C} = \frac{\mathcal{V}}{\mathcal{V}_0} = \frac{ (h - \pi/(k\sqrt{Bo}))^2}{h^2}. \label{cavity}
\end{equation}
This new scaling is compared with numerical data in figure \ref{fig:ImpactSnapshots}e and shows good agreement at high Bond numbers, with weaker agreement at lower Bond numbers as expected from figure \ref{fig:ImpactSnapshots}b-d. Note that 2D and 3D simulations are considered in figure \ref{fig:ImpactSnapshots}e and the cavity shape is identical, since the 3D transition of the flow takes place after impact, as discussed in \S\ref{sec:energetics}. The cavity shape is well grid converged as shown in Supplementary Materials.



%

\subsection{Number of bubbles}
\label{sec:number-of-bubbles}
\begin{figure}
\centering
  \begin{overpic}[width=0.45\linewidth]{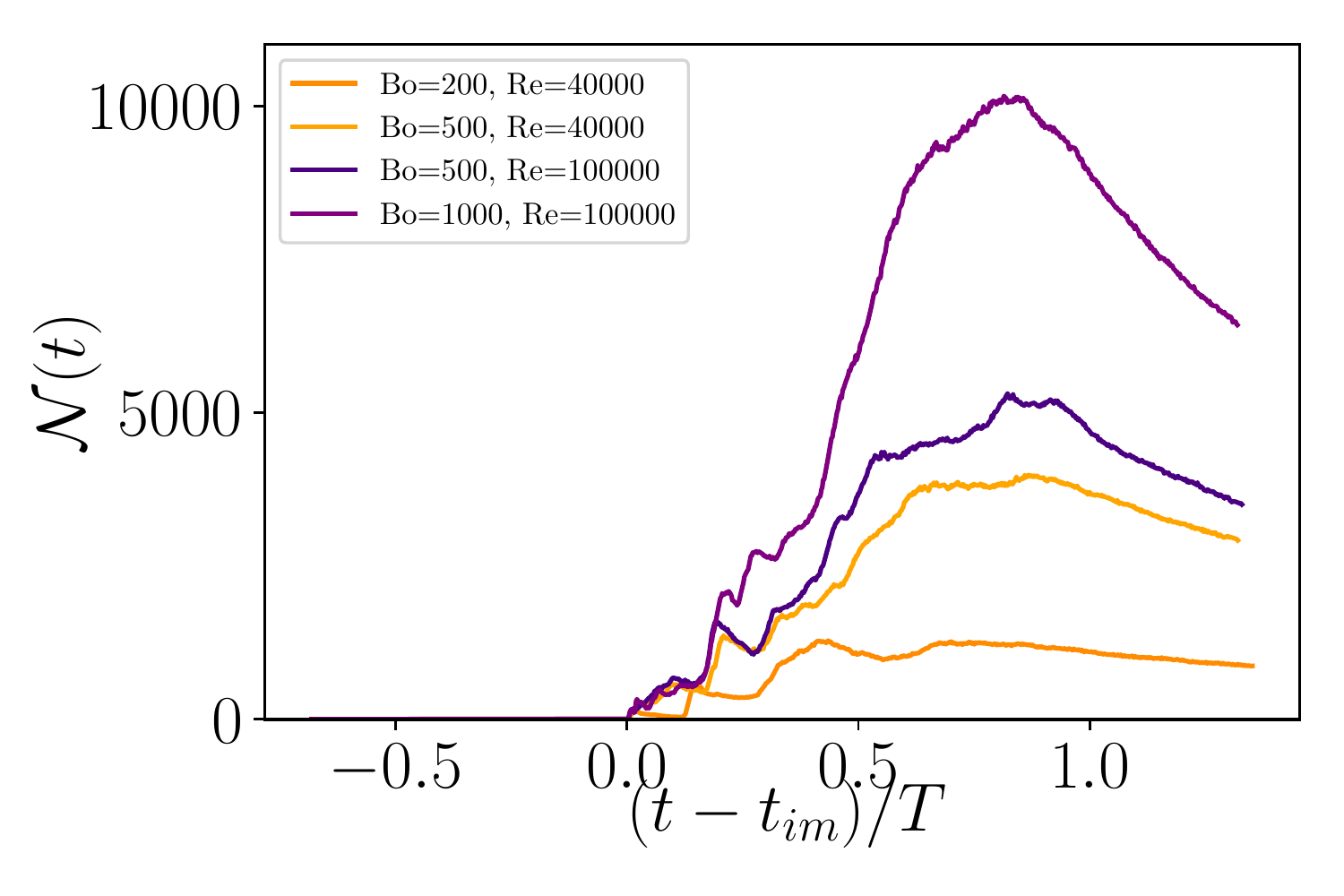}
    \put(0,50){(a)}
  \end{overpic}
  \begin{overpic}[width=0.45\linewidth]{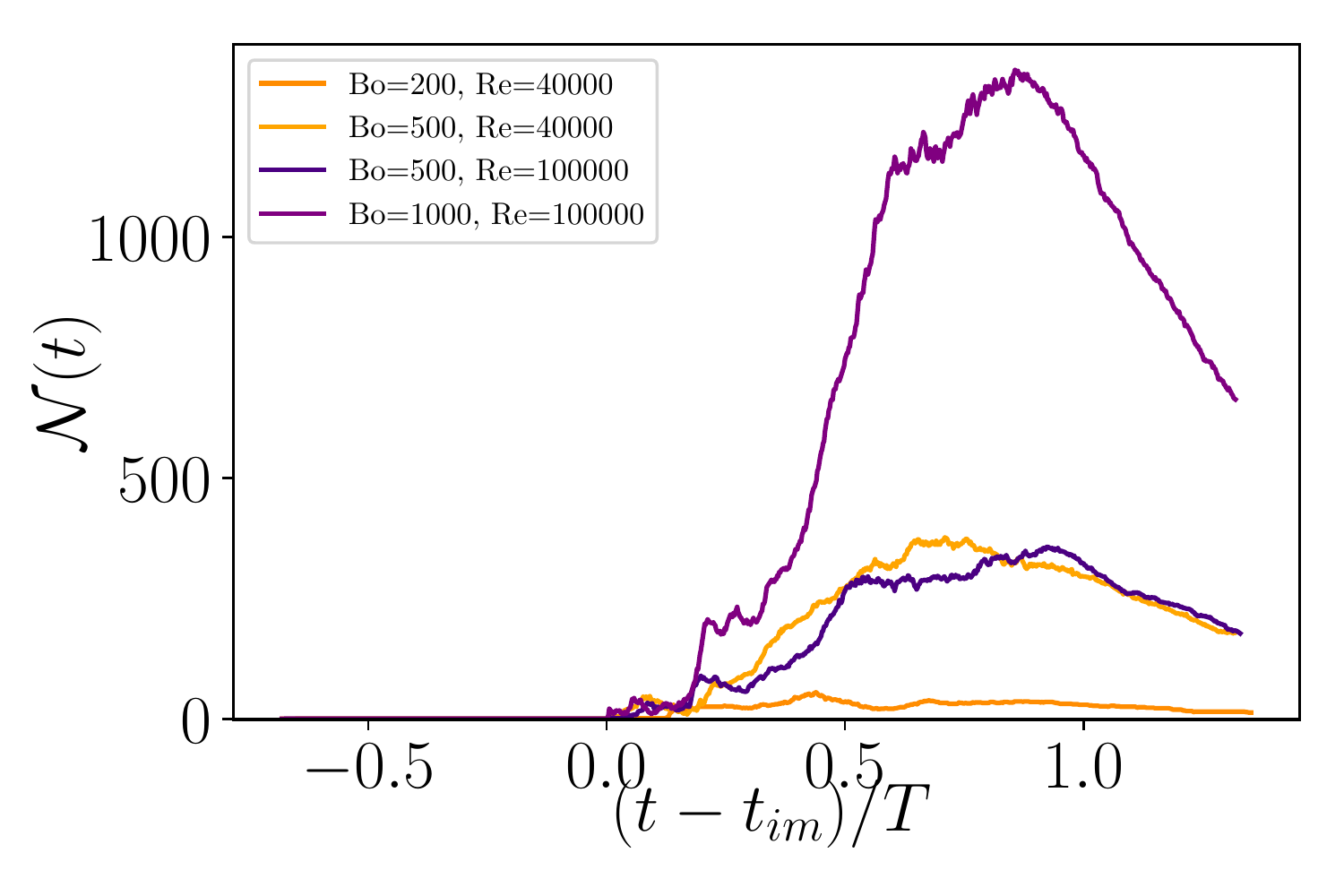}
    \put(0,50){(b)}
  \end{overpic}
  \begin{overpic}[width=0.45\linewidth]{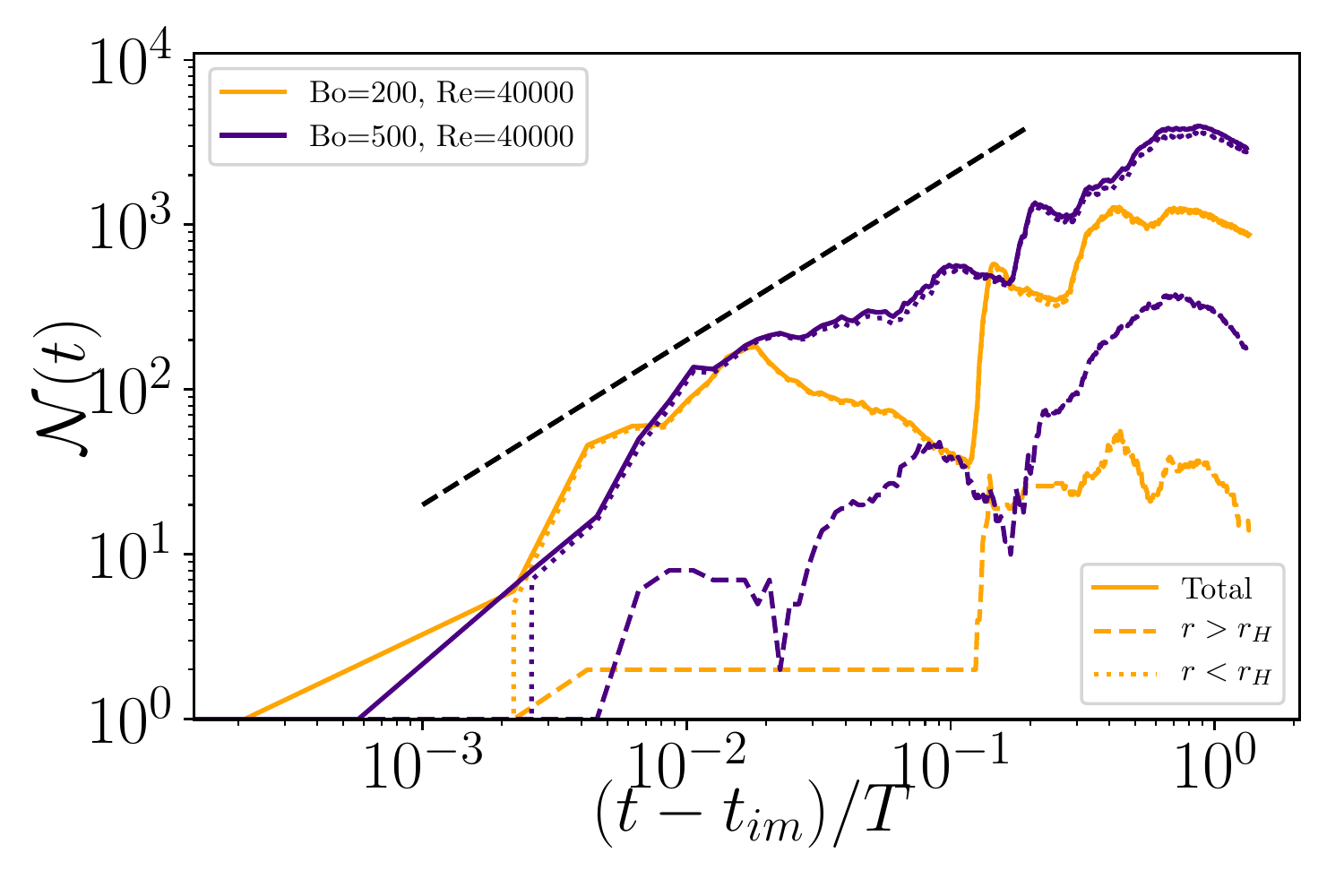}
    \put(0,50){(c)}
  \end{overpic}
  \begin{overpic}[width=0.45\linewidth]{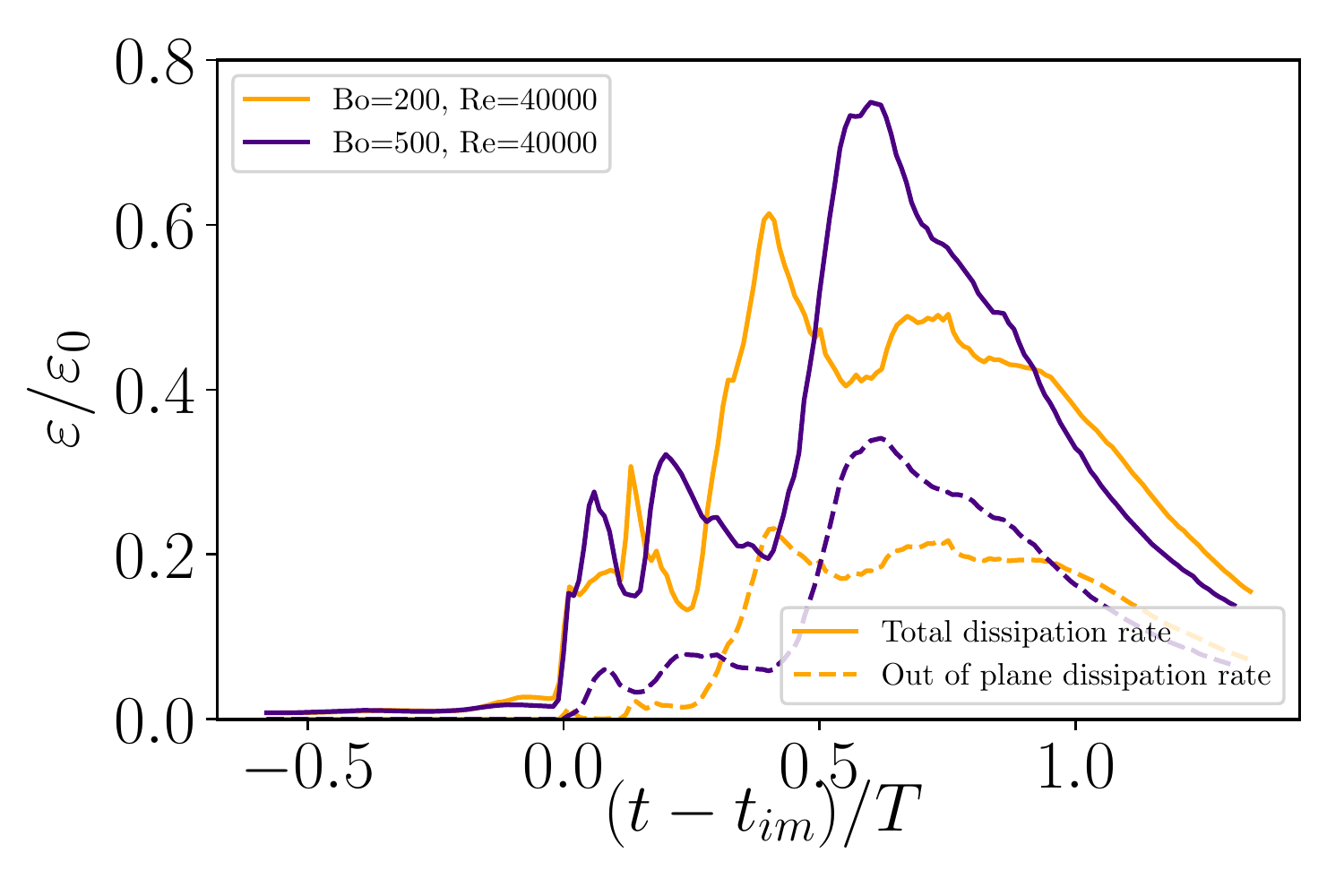}
    \put(0,50){(d)}
  \end{overpic}
  \caption{(a) Total number of bubbles as a function of time $(t-t_{im})/T$. (b) Number of bubbles of size greater than the Hinze scale, $r>r_H$ as a function of time. In both a and b, more bubbles are observed for higher Bo number, corresponding to the larger cavity. The bubble count is similar for the two Re number tested at Bo=500. (c) Detailed count breakdown for two cases, in log-log scales, showing the number of bubbles larger than the Hinze scale, $r>r_H$ and the total count, as a function of time, measured from the moment of breaking, for various cases. A nearly linear increase in number of bubbles is observed. (d) Turbulent dissipation rate as a function of time, showing both the total dissipation rate and the out-of-plane contribution, for the same cases as (c). Maximum $\varepsilon$ is obtained when the cavity has fully collapsed.}
  \label{fig:BTotals}
\end{figure}

\begin{figure}
\centering
\begin{minipage}{0.35\linewidth}
\begin{overpic}[width=\linewidth]{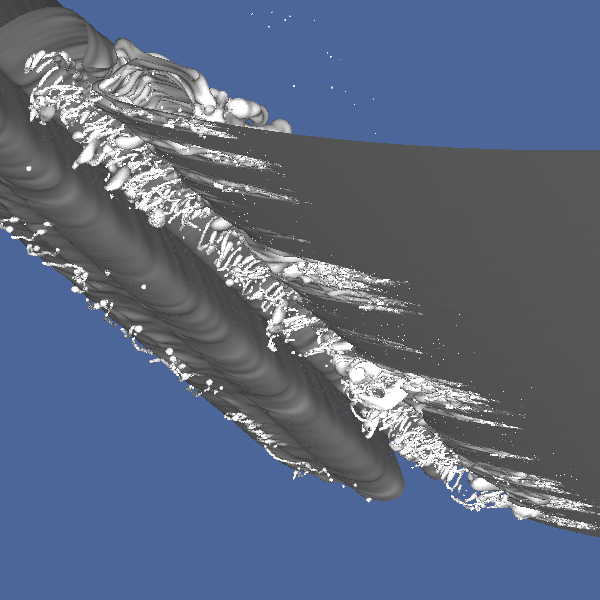}
    \put(0,95){(a)}
  \end{overpic}
\end{minipage}
\begin{minipage}{0.35\linewidth}
  \begin{overpic}[width=\linewidth]{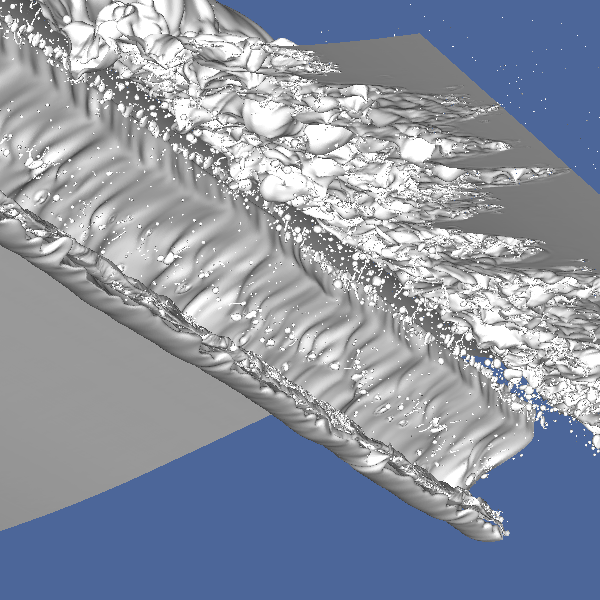}
    \put(0,95){(c)}
  \end{overpic}
\end{minipage}\hfill\\
\begin{minipage}{0.35\linewidth}
  \begin{overpic}[width=\linewidth]{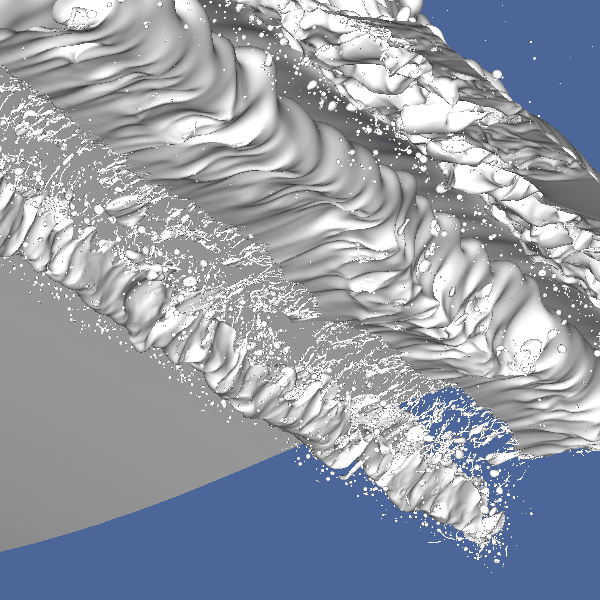}
    \put(0,95){(b)}
  \end{overpic}
\end{minipage}
\begin{minipage}{0.35\linewidth}
  \begin{overpic}[width=\linewidth]{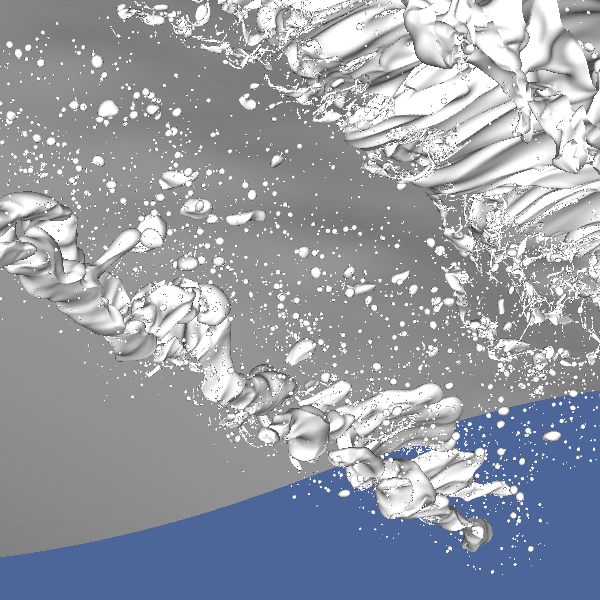}
    \put(0,95){(d)}
  \end{overpic}
\end{minipage}
 \caption{Snapshots of the liquid-gas interface, from below, for the case $\textrm{Bo}=1000, \textrm{Re}=10^5$, at times $(t-t_{im})/T$ of (a) $0.06$, (b) $0.28$, (c) $0.38$, (d) $0.48$. Immediately after breaking, showing the fully-resolved scales of bubble phenomena. Note in particular the very small bubbles visible at and in front of the leading edge of the breaker. At later stages the air cavity collapses and lead to a wide range of bubble sizes.}
  \label{fig:BSnapshots}
\end{figure}

%

We now discuss the formation of bubbles and the time evolution of their number from impact. \textcolor{black}{Numerical convergence is verified for the time evolution and time averaged bubble size distribution in the Supplementary Materials.} 

Figure \ref{fig:BTotals}a shows the total number of bubbles $\mathcal{N}$ as a function of time $(t-t_{im})/T$. More bubbles are produced with increasing Bo, showing an order of magnitude variation in peak bubbles produced, while the production is less sensitive to Reynolds number. The total number of bubbles begins increasing at the moment of impact and peaks at the end of the active breaking stage, between 0.75T and T after impact. Particularly at higher Bond numbers, there is an increase in production rate at $0.4T$ which persists until $\sim 0.75T$. Similar observations are made when considering only the super-Hinze scale bubbles $r>r_H$, as shown in panel (b). The number of super-Hinze scale bubbles is much smaller than the total count, between 20 at Bo=200 to 750 at Bo=1000.

The increase of bubble production rate at $(t-t_{im})/T = 0.4$ correlates with the breakup of the main cavity. Figure \ref{fig:BSnapshots} shows a view of the surface from below from $(t-t_{im})/T = 0.06$ to $0.48$. In (a) and (b), the main cavity is mostly intact, with some minor shedding of bubbles appearing off a limb of the cavity in (b). Due to the turbulence around the cavity, it deforms and ruptures dramatically in (c), creating a large number of bubbles of many sizes. The remaining parts of the cavity then destabilize further in (d), and eventually break up entirely by $0.7T$ after impact. Note that a significant number of bubbles is produced before this time: figure \ref{fig:BSnapshots}a shows a snapshot of the breaker from below, where many small bubbles have been entrained at the leading edge of the breaker but well before the main cavity (visible to the rear of the wave) has begun to disintegrate. Some chains of larger bubbles are also visible near the main cavity and under the primary splash-up.

Returning to figure \ref{fig:BTotals}, (c) shows the breakdown between sub- and super-Hinze scale bubbles for two particular cases Bo$=200$ and Bo$=500$, both at Re$=40000$. The total count is dominated by sub-Hinze scale bubbles. The number of bubbles increases rapidly and at a roughly constant rate from the moment of impact until $0.7T$ or $0.8T$ after impact, when it begins to decay. The increase in production between $0.4T$ and $0.7T$ is subtle (on the log-log scale); before then, the bubble production rate appears to follow a broadly linear trend (indicated by the dashed black line). 

Finally, figure \ref{fig:BTotals}d shows the energy dissipation rate during the breaking process for the same two cases as (c), similarly to figure \ref{fig:Eij}. Note again that the energy dissipation rate increases rapidly from $0.4T$ to $~0.6T$ after impact, along with the out-of-plane contribution. The turbulence dissipation rate (as well as its out of plane contribution) is maximum when the cavity is fully broken. 



This discussion suggests two effects controlling the bubble production and resulting size distribution: i) the initial air entrainment and impact, which will control initial sub-Hinze scale production, and ii) the fragmentation process of the cavity, which depends on the cavity size and the turbulence being produced during impact.



\begin{figure}
\begin{minipage}{0.45\linewidth}
  \begin{overpic}[width=\linewidth]{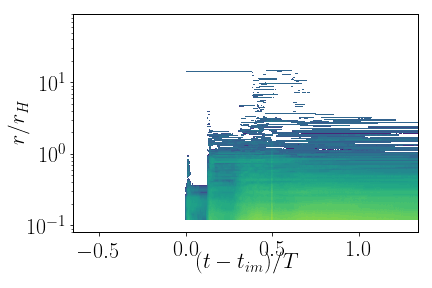}
    \put(0,58){(a) $\qquad \textrm{Bo}=200, \textrm{Re}=40000$}
  \end{overpic}
\end{minipage}
\begin{minipage}{0.45\linewidth}
\begin{overpic}[width=\linewidth]{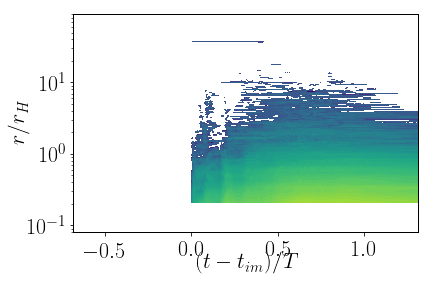}
    \put(0,58){(b) $\qquad \textrm{Bo}=500, \textrm{Re}=40000$}
  \end{overpic}
\end{minipage}\hfill\\
\begin{minipage}{0.45\linewidth}
 \begin{overpic}[width=\linewidth]{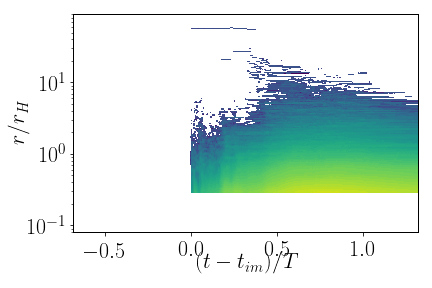}
    \put(0,58){(c) $\qquad \textrm{Bo}=1000, \textrm{Re}=100000$}
  \end{overpic}
\end{minipage}
\begin{minipage}{0.45\linewidth}
  \begin{overpic}[width=\linewidth]{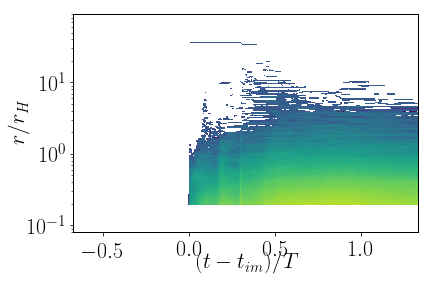}
    \put(0,58){(d) $\qquad \textrm{Bo}=500, \textrm{Re}=100000$}
  \end{overpic}
\end{minipage}\hfill\\
\begin{minipage}{\linewidth}
  \centering
  \includegraphics[height=0.07\linewidth]{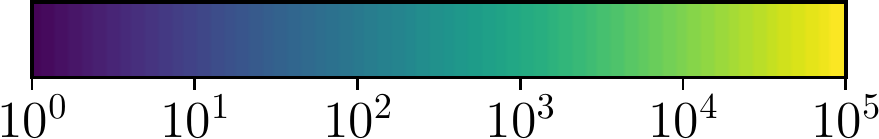}
\end{minipage}
\caption{Contours of bubble size distribution over time, (a) $Bo=200, Re=40\times10^3$; (b) $Bo=500, Re=40\times10^3$, (c) $Bo=1000, Re=100\times10^3$, (d) $Bo=500, Re=100\times10^3$. $L=11$ for each case. With increasing Bond number, the main cavity size increases compared to the Hinze scale. The bubble statistics is similar for the two Re numbers at Bo=500. Small sub-Hinze scale bubbles are produced at impact, while a broad bubble cascade occurs once the cavity collapses.}
\label{fig:BContours}
\end{figure}

We more closely discuss the relative roles of the initial sub-Hinze production and the later multiscale fragmentation processes of the main cavity, and examine the statistics of the bubble populations. For each case, the number $N$ and sizes of bubbles are sampled at various times $t$ and binned by equivalent bubble radius $r$ into bins of size $\Delta r$, resulting in a time-dependent size distribution $N(r/r_H,t/T)$, where $r_H$ is the Hinze scale given by \eqref{Hinze} and $T$ is the wave period, and which has been normalized by bin size such that $\int N(r/r_H, t/T)dr \simeq \sum N(r/r_H, t/T)\Delta r = \mathcal{N}(t/T)$, where $\mathcal{N}(t/T)$ is the total number of bubbles at time $t$ and summation is done across all radius bins. 

Figure \ref{fig:BContours} shows the contours resulting from plotting $N(r/r_H, t/T)$ over time and radius, for the cases (a) $Bo=200$, $Re=4\times10^4$; (b) $Bo=500$, $Re=4\times10^4$; (C) $Bo=500$, $Re=10^5$; (d) $Bo=1000$, $Re=10^5$. In each case for $(t-t_{im})/T < 0$ there are no bubbles because the wave has not broken. The moment of impact corresponds with the generation of an array of sub-Hinze scale bubbles along with a single large ``bubble'', visible as an isolated line on the plot, which is the main cavity (see \S \ref{sec:qualitative}). (Individual or small numbers of similarly-sized bubbles are visible as isolated lines on the plot.) This persists until $(t-t_{im})/T \simeq 0.4$; it is illustrated by figure \ref{fig:BSnapshots}a. At $(t-t_{im})/T \simeq 0.4$, the cavity destabilizes and breaks into an array of large bubbles (see figure \ref{fig:BSnapshots}c,d), which themselves break up and further populate the size distribution, so that at $t/T=0.6-0.7$ there is a broad array of large and small bubbles, with the distribution weighted towards the small bubbles. At late times, ($(t-t_{im})/T=1$ onwards), the number of large bubbles reduces as they break up or reach the surface and burst. The small bubbles remain mostly entrained in the liquid for the remainder of the simulation. For a sufficiently long simulation time, all the small bubbles would eventually rise to the surface and burst, however the resolution of these bursting events would require even higher resolution (on the individual bubble) \citep{Berny2020} and are not considered here. \textcolor{black}{We note that the dynamics of entrainment of the small bubbles at impact will present similarities with the physics of air entrainment by falling jets, as discussed by \citet{Kiger2012}.}

Significant qualitative differences in the distributions between the different cases are apparent only with respect to Bond number; larger $\textrm{Bo}$ corresponds to a smaller Hinze scale $r_H$, so that the distributions are generally larger relative to $r_H$. A clear indicator is the size in $r/r_H$ of the main cavity. $\textrm{Re}$ does not affect the shape of the bubble size distribution (compare the two $Bo=500$ cases, at $Re=4\times10^4, 10^5$, since the mean turbulent dissipation rate $\epsilon_l$ (which informs $r_H$) is not sensitive to $\textrm{Re}$ for sufficiently large $\textrm{Re}$ - see \S\ref{sec:energetics}.

\subsection{Bubble size distribution over the active breaking time: scalings}

Having discussed qualitatively the bubble production and size distribution as a function of time, we now turn to quantitative evaluations of the size distribution and its scale dependence. We focus on time averaged distributions over the active break-up time as statistical convergence of the data in the time evolution remains challenging and would require ensemble averages (requiring substantive computing time). We aim to scale the number of bubbles in the system. Figure \ref{fig:ImpactSnapshots} shows that the cavity shape changes at small Bo due to capillary effects, resulting in a smaller cavity, and a smaller volume of air entrained. This is confirmed by the bubble count in figure \ref{fig:BTotals}.

The time evolution of the bubble size distribution can be described as an extension of the model proposed by \citet{Deike2016}, for the super-Hinze bubble size distribution, based on a turbulence-buoyancy balance
\begin{equation}
  N(r, t) = B \frac{AL_c}{2\pi} \frac{\varepsilon(t - \Delta \tau)}{W g} r^{-10/3} r_m^{-2/3}, \label{DeikeN}
\end{equation}
where $A$ is the cross-sectional area of the initially ingested cavity in the breaking process; $L_c$ is the length of breaking crest; $\varepsilon(t - \Delta \tau)$ is the energy dissipation rate and $\Delta \tau$ is the time between breaker impact and peak energy dissipation rate, which corresponds to the cavity collapse time; $W\approx h/\tau$ is a dissipation-weighted vertical mean velocity of the bubble plume over the active breaking period; with $\tau$ the active breaking period, $B$ is a dimensionless constant. 

Following \cite{Deike2016} the timescale of the cavity collapse is evaluated as $\Delta \tau \sim r_{m}^{2/3}\varepsilon^{-1/3}$, where $r_m$ is the cavity size, evaluated using the scaling of the cavity length scale, $r_m = h-l_j$, i.e. at high Bo number it will be independent of the Bond number (that is, the cavity of large scale breakers does not depend on surface tension), while at moderate to low Bond number, surface tension effects become important. The cross section area $A$ controls the amount of entrained air initially available for subsequent breakup into a bubble size distribution, which we estimate from the cavity shape, so that {\color{black}$A L_c \equiv \mathcal{V} \propto r_m^2 L_c$}. This leads to the geometric scaling $N(r)\propto r_m^{4/3}$, which indeed indicates that the number of bubbles will increase with the size of the cavity.


Introducing the Hinze scale as characteristic length scale, eq. \ref{DeikeN} can be written in a non-dimensional form as,
\begin{equation}
  N(r/r_H, t/T) = \frac{B}{2\pi} \frac{\varepsilon(t - \Delta \tau)}{W g} \left(\frac{r}{r_H}\right)^{-10/3} \frac{\mathcal{V}}{r_H^3} \left(\frac{r_H}{r_m}\right)^{2/3}. \label{DeikeNTime}
\end{equation}


As described in \citet{Deike2016}, the factor $\varepsilon(t - \Delta \tau)/(W g)$ describes the time evolution while the number of bubbles is determined by the strength of the break-up process and the scale separation between the initial cavity size and the Hinze scale, $ (\mathcal{V}/r_H^3)(r_H/r_m)^{2/3} \propto \left(r_m/r_H\right)^{4/3}(L_c/r_H)$.

{\color{black} 
We note that the controlling parameter in bubble break-up is the Weber number, which defines the ratio between the inertial turbulent stresses and the surface tension. When analyzing the cavity collapse, a Weber number can be defined, based on the cavity radius of $r_m$, which depends on $\textrm{Bo}$ but, as we have suggested above, approaches a constant value $h/2$ for sufficiently large $\textrm{Bo}$. The cavity's Weber number is then,
    $\textrm{We}_m = \frac{\mathcal{C}_1 \rho \varepsilon^{-2/3} h^{5/3}}{\sigma}$, 
  where $\mathcal{C}_1$ is a constant, and $\varepsilon$ is the energy dissipation rate. Since the dissipation rate $\varepsilon$ scales with the wave height $h$, we obtain $\textrm{We}_m = \mathcal{C}_1 \frac{\rho g S^2}{\sigma k^2} = \mathcal{C}_1 \textrm{Bo}S^2$; and we further note that the scale separation $r_m/r_H$ is linked to the Weber number by $\frac{r_m}{r_H} \propto \left(We_m\right)^{3/5}$. This links the driving Weber number of the bubble statistics and breakup processes with the Bond number and slope of the wave.}

\textcolor{black}{Separate studies of bubbles and droplets break-up in turbulence have demonstrated that one can observe the $N(r)\propto r^{-10/3}$ scaling in other contexts from breaking waves \citep{Mukherjee2019,Soligo2019,Riviere2021}, suggesting a universal character of the break-up cascade, provided the injection size is much larger than the Hinze scale, $r_m\gg r_H$. Numerical and experimental results have shown that the number of child bubbles formed by the break-up of a large super-Hinze bubble in turbulence follows a simple power-law scaling, expressed in terms of the bubble Weber number, $\mathcal{N} \propto (r_{m}/r_H)^{\alpha}$, with $\alpha$ between 1 and 2 \citep{Vejravzka2018,Riviere2021}, which appears compatible with our results; since from eq. 4.3, $\frac{\mathcal{V}}{r_H^3}(\frac{r_H}{r_m})^{2/3} \sim \frac{L_c r_m^{4/3}}{r_H^{7/3}} \sim (\frac{r_m}{r_H})^{4/3}(\frac{L_c}{r_H})$.  Note also that the DNS from \citet{Riviere2021} observe a nearly linear increase of the number of bubbles during the fragmentation process at high Weber number, analogous to the behaviour observed for the cavity collapse.}

\begin{figure}
  \centering
   \includegraphics[width=0.6\linewidth]{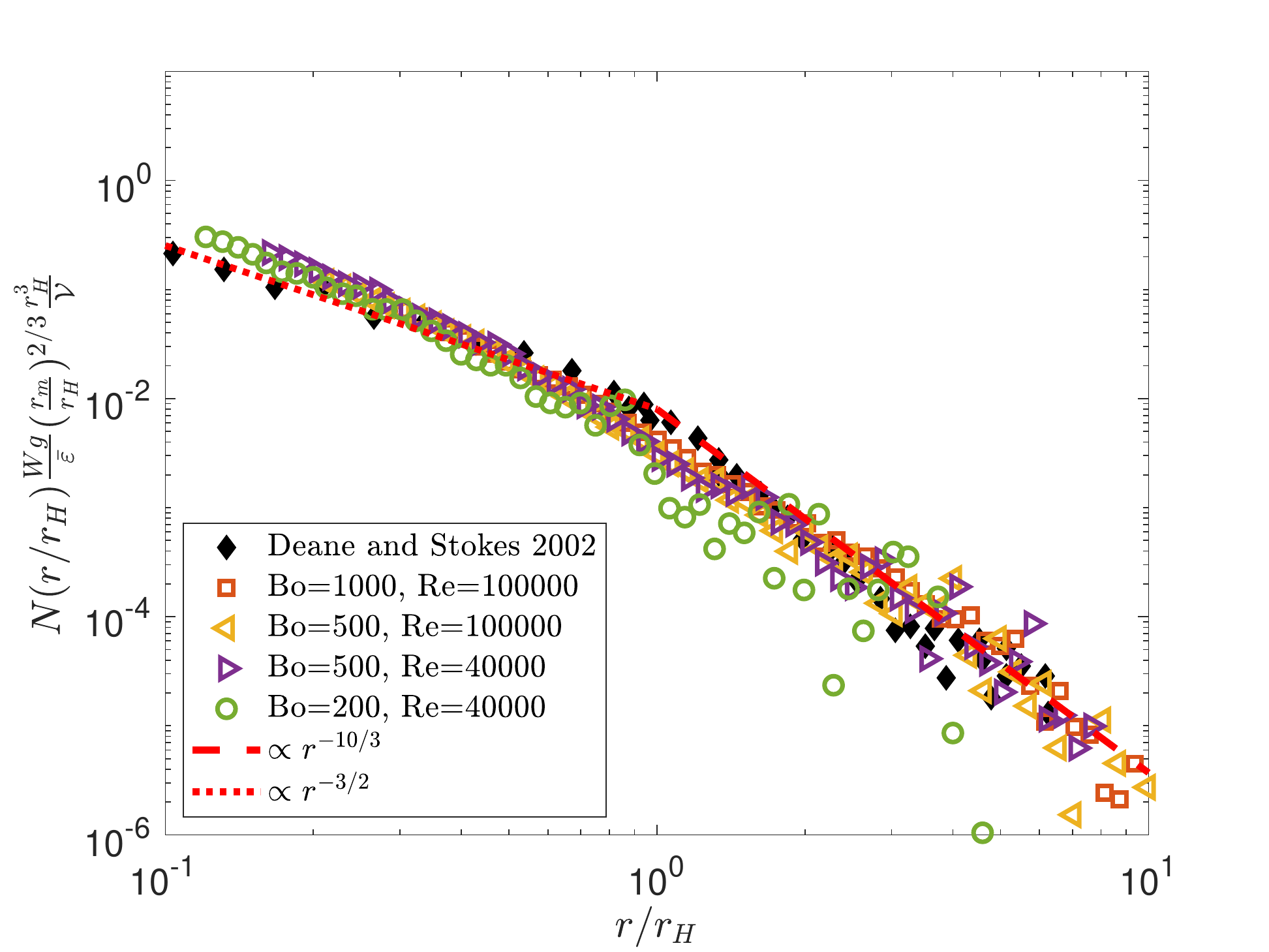}
   \caption{Time-averaged bubble size distributions over the full time window, $N(r/r_H)$, together with the experimental data from \citet{Deane2002}. Experimental data is plotted over $r/r_H$ on the abscissa, and scaled according to eq. \ref{DeikeNtAvg-super}. All data collapse reasonably onto a single curve given the complexity of the problem. The sub-Hinze volume is about 6\% of the total entrained volume. }
\label{fig:TAvg}
\end{figure}


We consider the time-averaged version of eq. \ref{DeikeNTime}, analogous to the equation proposed by \citet{Deike2016}, to rescale the data onto a universal scaling,
\begin{equation}
   N_{super}(r/r_H) = \frac{B}{2\pi}\frac{\bar\varepsilon}{W g}  \left(\frac{r}{r_H}\right)^{-10/3} \frac{\mathcal{V}}{r_H^3} \left(\frac{r_m}{r_H}\right)^{-2/3}. \label{DeikeNtAvg-super}
\end{equation}
for super-Hinze scale bubbles. The sub-Hinze scale follows a $r^{-3/2}$ scaling. Since the super- and sub-Hinze distributions must be continuous at the Hinze scale, we obtain
\begin{equation}
  N_{sub}(r/r_H) = \frac{B}{2\pi} \frac{\bar\varepsilon}{W g}  \left(\frac{r}{r_H}\right)^{-3/2} \frac{\mathcal{V}}{r_H^3} \left(\frac{r_m}{r_H}\right)^{-2/3}. \label{DeikeNtAvg-sub}
\end{equation}

\subsection{Bubble size distribution over the active breaking time: comparison with laboratory experiments}

We rescale the experimental distribution by the estimated cavity volume, as \citet{Deane2002} report a bubble size distribution $n(r)$ in units of number of bubbles per bin size, per unit volume. For the present comparison, we consider that all bubbles are initially contained in the cavity volume $\mathcal{V}_0$. 

The time-averaged bubble size distribution, for all Re and Bo cases, over the active breaking time $t/T \in [0:1.2]$ are shown in figure  \ref{fig:TAvg}, and compare with the laboratory experiments from \citet{Deane2002}. For all Bo numbers, the bubble size distribution follows the direct cascade scaling for super-Hinze bubbles, $N(r/r_H)\propto (r/r_H)^{-10/3}$. We resolve up to one order-of-magnitude below the Hinze scale at $L=11$, in the Bo=200 case. For all cases, within this range, the size distributions have developed a shape which is clearly less steep than the super-Hinze results, close to the $r^{-3/2}$ scaling, but the transition between the two regimes is not as sharp as observed in the experimental data. Note that our simulations stop at the end of the active breaking period, and as such do not describe the late-time plume evolution and steepening of the bubble size distribution, which evolves due to both degassing and further break-up, as discussed by \citet{Deane2002,Deike2016,gaylo2021effects}. For Bo=200, where the numerical resolution is sufficient to allow for a discussion of the sub-Hinze scale bubbles, we observe a scaling compatible with the experimental data set from \citet{Deane2002}, $N(r/r_H)\propto (r/r_H)^{-3/2}$. The size distribution is normalized such that $\int N(r/r_H) d(r/r_H)=\mathcal{N}$, the total number of bubbles. The partitioning in volume of air entrained is about 94\% within the super-Hinze range of scale and about 6\% of the air within the sub-Hinze bubbles, similar to the discussion of \citet{Deane2002}. Figure \ref{fig:TAvg} shows that the distribution in the super-Hinze regime between the Bo=1000 and the experimental \citet{Deane2002} data agree reasonably well in the super-Hinze region and suggests that the asymptotic regime in Bo observed for the cavity volume in figure \ref{fig:TAvg} has been reached. All data in figure \ref{fig:TAvg} are reasonably well collapsed onto a single curve including the experimental data of \citet{Deane2002}, given the uncertainties in the measurements and estimations of the various terms in the scaling model.

\section{Droplet statistics}
\label{sec:droplets}
\subsection{Stages of droplet production}

We now discuss droplet production. Although all breaking waves in this study produce some droplets, large numbers of droplets only appear at larger $\textrm{Bo}$. Figure \ref{fig:DStages} shows qualitatively some of the different production mechanisms observed in these cases. Some droplets are produced immediately on impact (figure \ref{fig:DStages}a); from a secondary splash-up (\ref{fig:DStages}b); a sustained surface splashing in the developed breaker (\ref{fig:DStages}c); and some jet droplets, which are partially resolved in these simulations (\ref{fig:DStages}d). \textcolor{black}{Numerical convergence of droplet statistics is notoriously difficult to achieve because of the near-singular nature of droplet breakup and pinch-off physics \citep{Herrmann2013,Pairetti2020} while fundamental two-dimensional studies can provide detailed dynamics of the rupturing sheet and filaments \citep{jian2020air}. Numerical convergence of our data is discussed in details in Supplementary Materials.} 


\begin{figure}
\centering
\begin{minipage}{0.35\linewidth}
  \begin{overpic}[width=\linewidth]{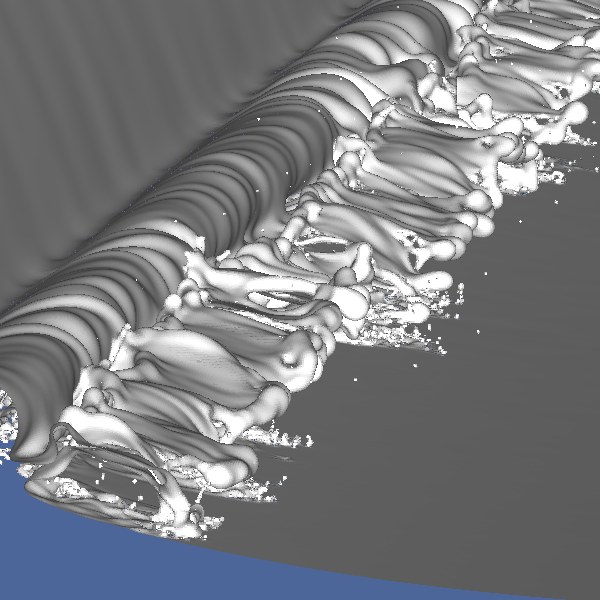}
    \put(0,95){(a)}
  \end{overpic}
\end{minipage}
\begin{minipage}{0.35\linewidth}
  \begin{overpic}[width=\linewidth]{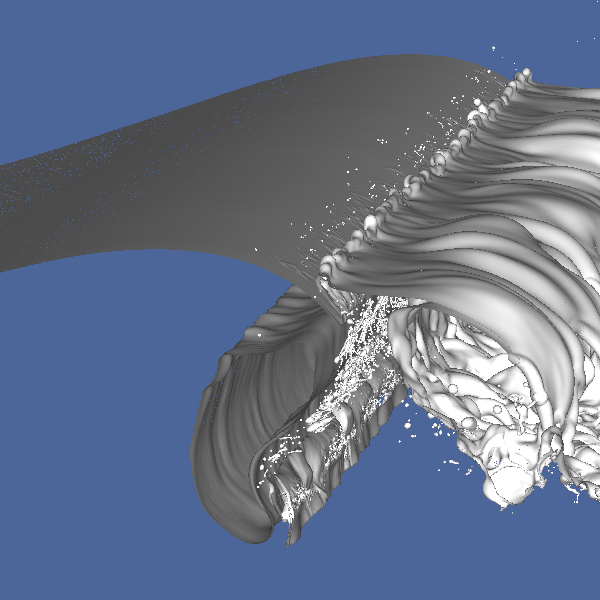}
    \put(0,95){(b)}
  \end{overpic}
\end{minipage}\hfill\\
\begin{minipage}{0.35\linewidth}
  \begin{overpic}[width=\linewidth]{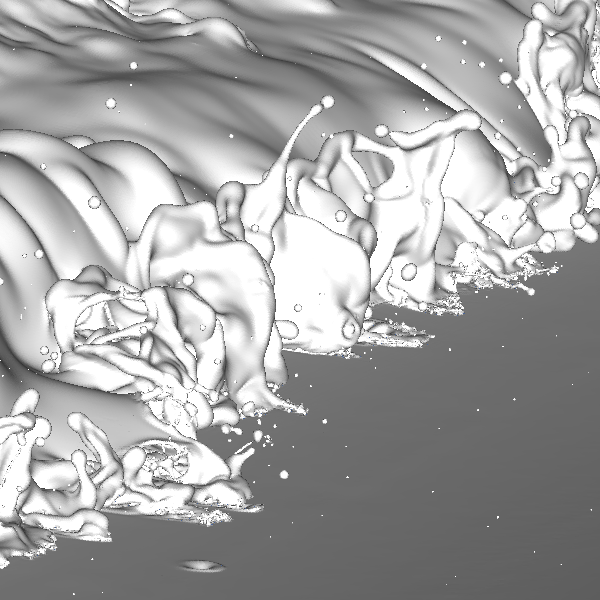}
    \put(0,95){(c)}
  \end{overpic}
\end{minipage}
\begin{minipage}{0.35\linewidth}
  \begin{overpic}[width=\linewidth]{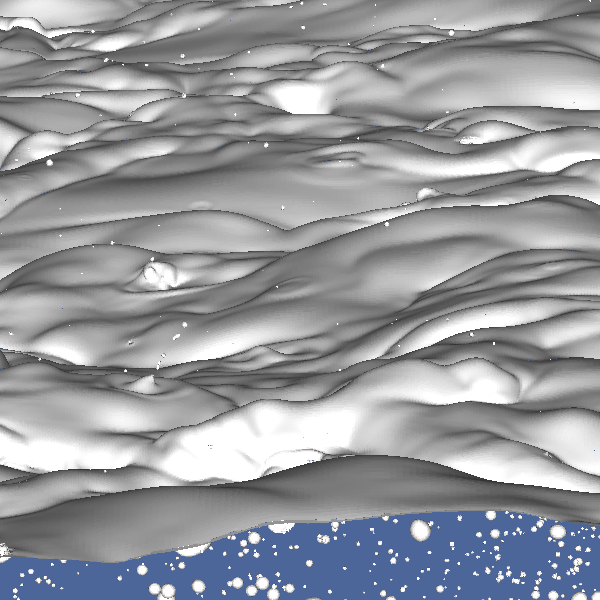}
    \put(0,95){(d)}
  \end{overpic}
\end{minipage}
  \caption{Snapshots of the liquid-gas interface at different magnifications and different times, showing different stages of droplet production, for the case $\textrm{Bo}=1000, \textrm{Re}=10^5$. (a) $(t-t_{im})/T= 0.06$, splashing produced by the initial impact at the front of the breaker; (b) $(t-t_{im})/T=0.2$, secondary splash-up shortly after impact producing a peak in droplet count, (c) $(t-t_{im})/T=0.52$, sustained droplet production later in the active breaking phase, (d) $(t-t_{im})/T=1.1$, jet droplet production at late times.}
  \label{fig:DStages}
\end{figure}

Figure \ref{fig:DRes}a shows the sizes of droplets produced by the secondary splash relative to the mesh size, suggesting that many of these droplets in particular have radii of approximately the smallest mesh size, hence the have to be considered with caution. Figure \ref{fig:DRes}b shows a fragmenting jet produced later in the breaking process, with only the largest droplets exhibiting a radius of more than double the mesh size. The largest droplets appear during the sustained splashing phase (corresponding to figure \ref{fig:DStages}c) and statistics for such droplets are numerically converged.

\begin{figure}
\centering
\begin{minipage}{0.35\linewidth}
  \begin{overpic}[width=\linewidth]{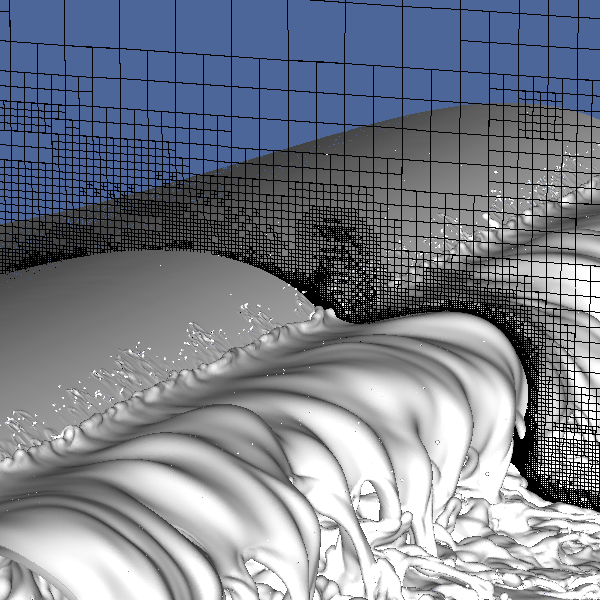}
    \put(0,95){(a)}
  \end{overpic}
\end{minipage}
\begin{minipage}{0.35\linewidth}
  \begin{overpic}[width=\linewidth]{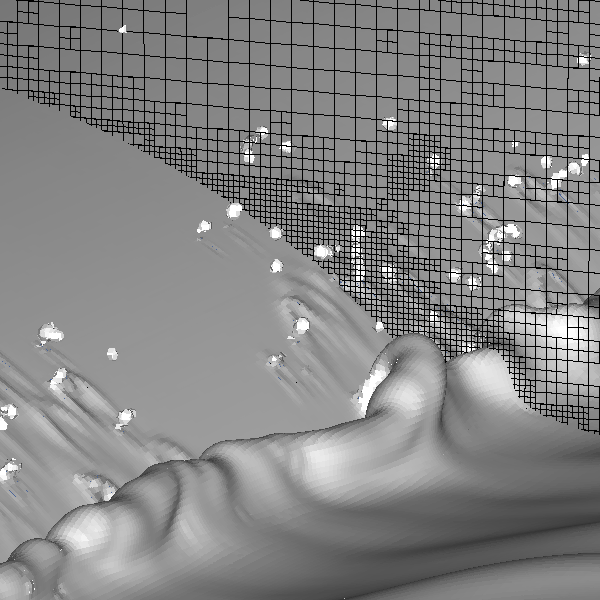}
    \put(0,95){(b)}
  \end{overpic}
\end{minipage}\hfill\\
\begin{minipage}{0.35\linewidth}
  \begin{overpic}[width=\linewidth]{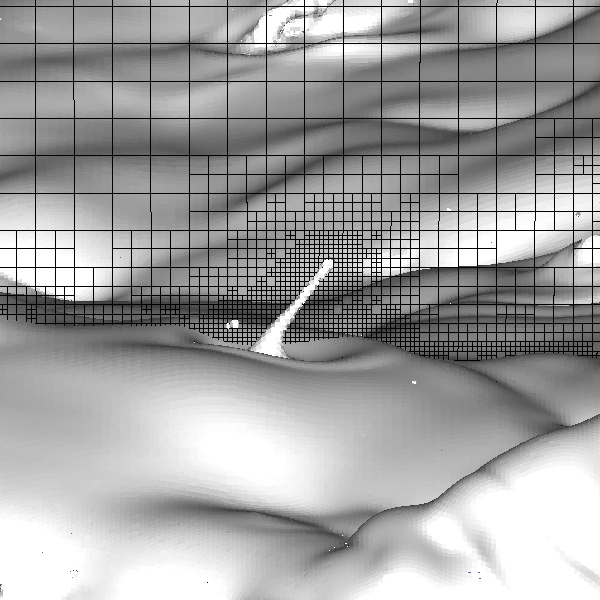}
    \put(0,95){(c)}
  \end{overpic}
\end{minipage}
\begin{minipage}{0.35\linewidth}
  \begin{overpic}[width=\linewidth]{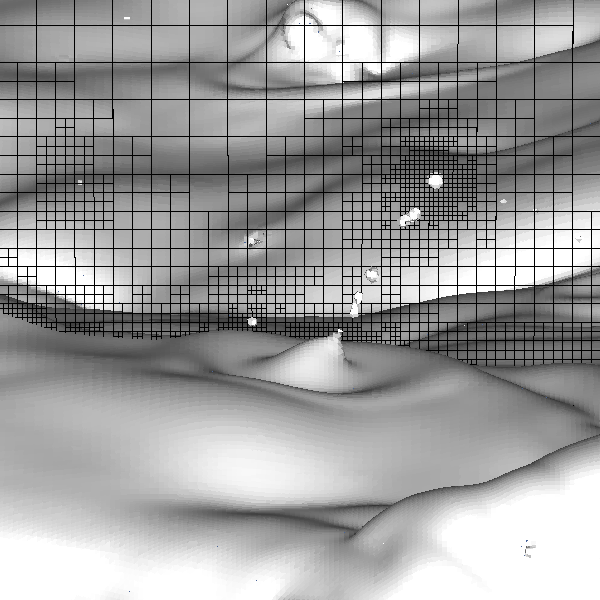}
    \put(0,95){(d)}
  \end{overpic}
\end{minipage}
  \caption{Snapshots of the liquid-gas interface for two droplet production stages, showing overlaid section of the numerical mesh, for the case $\textrm{Bo}=1000, \textrm{Re}=10^5$. (a,b) $(t-t_{im})/T= 0.2$, production of fine droplets by the secondary splash-up. Many of these droplets are resolved to less than four mesh cells per droplet diameter, for which numerical convergence is difficult to achieve. (c,d) $(t-t_{im})/T=1.2$, jet and droplet production after bubble bursting. Note that the largest droplets exceed four mesh cells per diameter.}
  \label{fig:DRes}
\end{figure}

\textcolor{black}{The total droplet production over time is shown in Figure \ref{fig:DTotals}. Fewer droplets are produced for all cases compared to the bubble count (fig. \ref{fig:BTotals}), and for $\textrm{Bo}=200$, fewer than 100 droplets are produced over time, which prevents any statistical convergence of the distribution. The number of droplets produced increases with Bo number, and for $\textrm{Bo}=500$ both $\textrm{Re}$ show a similar time evolution in the number of drops, with about 200 drops at most. The $\textrm{Bo}=1000$ case shows the largest droplet counts, with many droplets produced at early times after impact and up to 800 drops.} 


For the higher Bond number cases, figure \ref{fig:DTotals}a shows two prominent peaks in the droplet production. The first, sharp peak occurs at approximately the same time for both $Bo=500, 1000$, at $(t-t_{im})/T \simeq 0.2$. Figure \ref{fig:DStages}b shows qualitatively the flow around this time for the $Bo=1000$ case: shortly after the initial impact, which produces a small amount of droplets, there is a secondary impact between the splash-up and the bulk of the wave; this causes a second splash-up which projects directly upwards from the surface and produces many droplets. For $Bo=200$, while this same process occurs, surface tension is too strong to allow this secondary splash-up to generate droplets. This corresponds with the first peak in figure \ref{fig:DTotals}a, and explains why it only appears for large Bond numbers. The peak is sharp because the droplets are produced in a single well-defined process and they are quickly destroyed as they fall back to the surface. The second peak is broader and occurs for all cases at around $(t-t_{im})/T = 0.5$ to $0.6$. The state around this time is shown qualitatively for $Bo=1000$ in figure \ref{fig:DStages}c. It occurs as the wave proceeds through its active breaking phase and is made up of many small-scale splashing events and the bursting of large bubbles that were ingested earlier in the process. Since this process is longer and not as well-defined in space or time, the peak in \ref{fig:DTotals}b is accordingly broader.

\begin{figure}
  \centering
  \begin{minipage}{0.5\linewidth}
    \begin{overpic}[width=\linewidth]{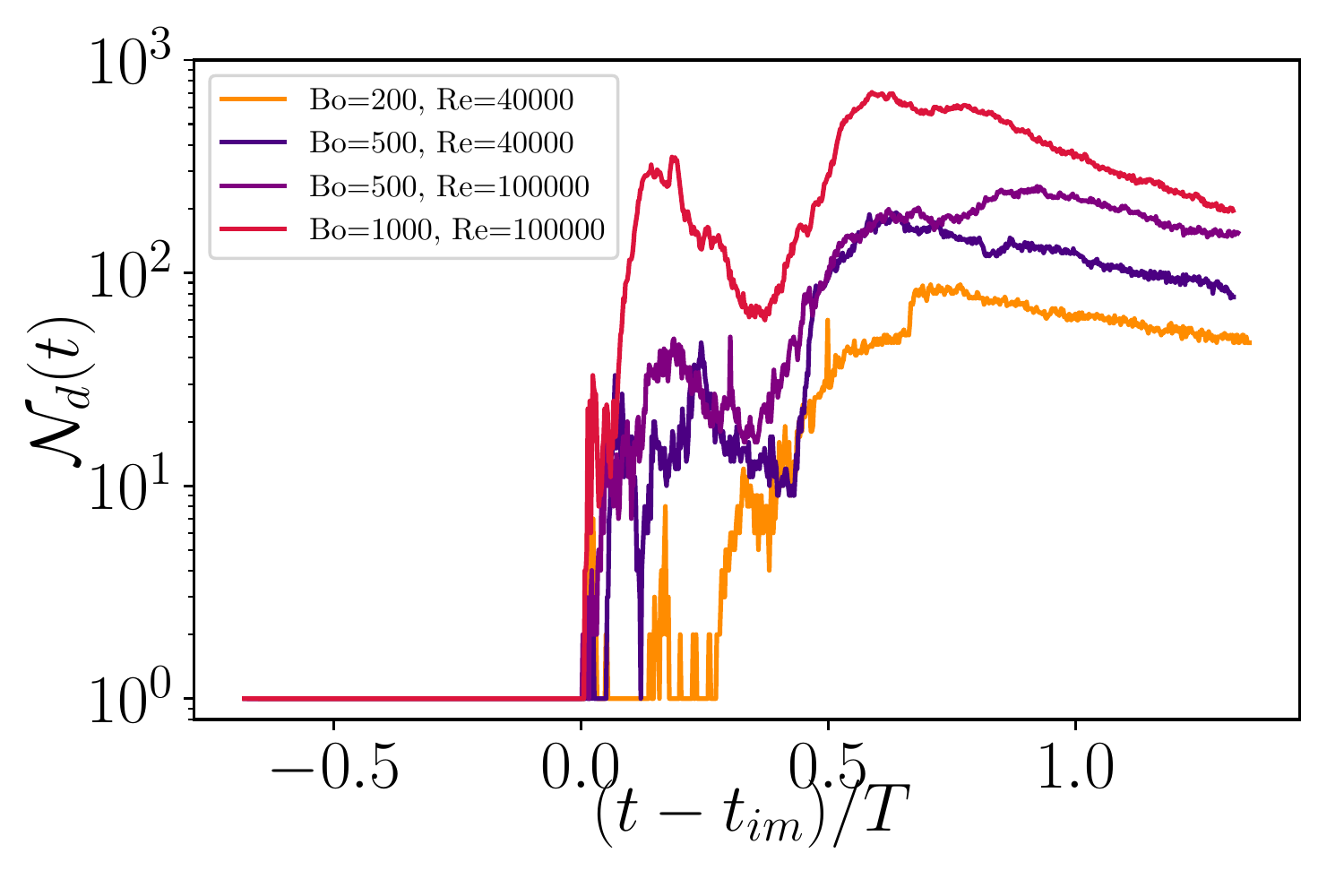}
      \put(0,50){(a)}
    \end{overpic}
  \end{minipage}\hfill\\
  \begin{minipage}{0.5\linewidth}
    \begin{overpic}[width=\linewidth]{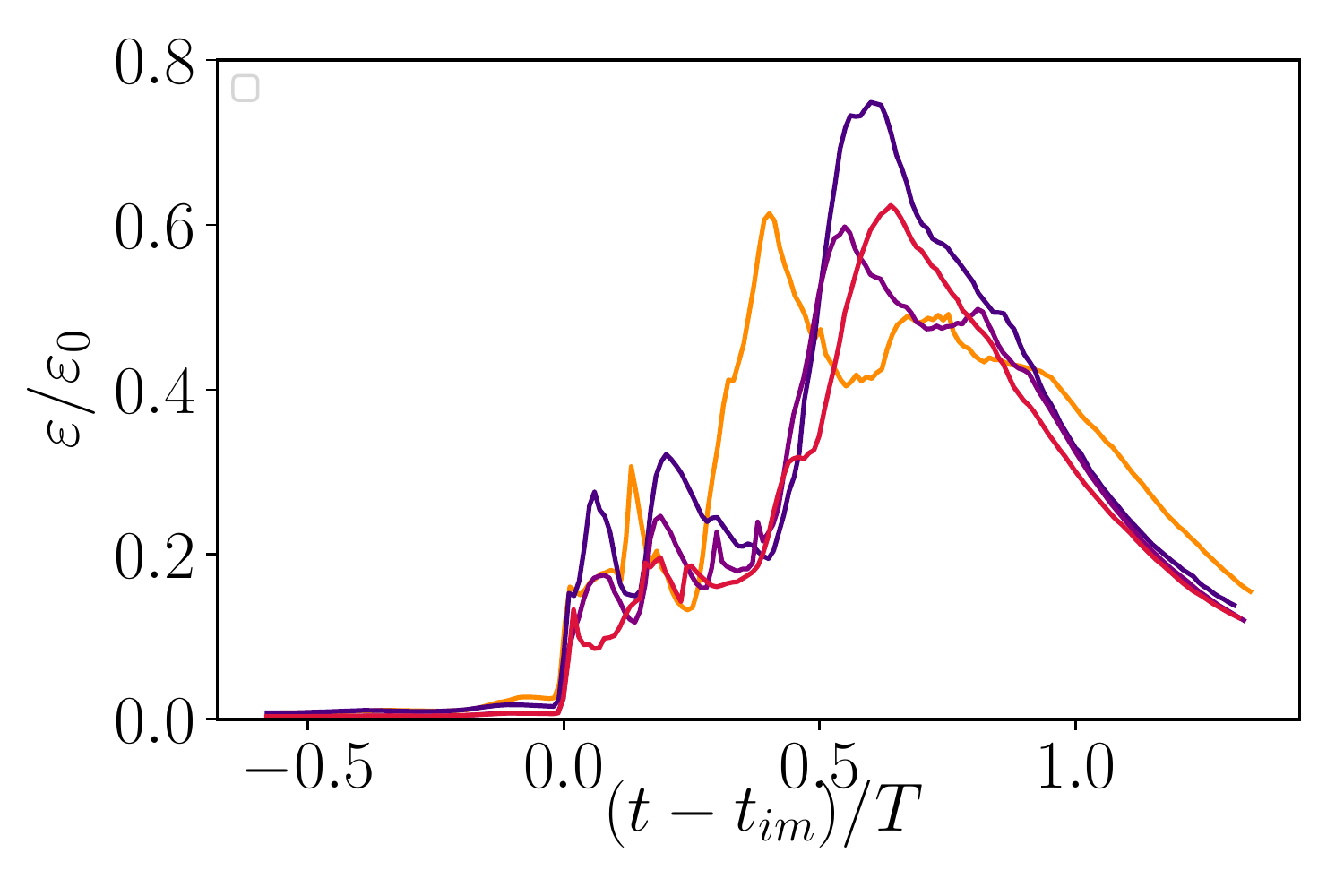}
      \put(0,50){(b)}
    \end{overpic}
\end{minipage}
  \caption{(a) Total number of droplets plotted over time for various cases, measured from moment of impact. (b) Energy dissipation rates for the same cases, showing the total contribution.}
  \label{fig:DTotals}
\end{figure}


Figure \ref{fig:DTotals}b shows the energy dissipation rates for the same cases as in (a). In contrast to the close connection between the bubble statistics and energy dissipation rate, there is no clear correlation between the dissipation rate and the droplet production. 

We now discuss the droplet statistics. The droplets are gathered and binned similarly to the bubbles, into distributions $N_d(r_d/l_c, t/T)$, where $l_c$ is the capillary length. The droplet populations are strongly influenced by the strength of the breaker \citep{Erinin2019}, and by the impact of the (ballistic) jet, particularly at early times, suggesting to use the gravity-capillary length as the relevant length scale. Note also that the lack of clear dependence on Reynolds number in the drop production suggests that viscosity does not play a role in the drop formation process. Figure \ref{fig:DContours} shows the contour maps for the droplet size distributions for the cases $\textrm{Bo}=200, \textrm{Re}=4\times10^4$; $\textrm{Bo}=500, \textrm{Re}=4\times 10^4$; $\textrm{Bo}=500, \textrm{Re}=10^5$; $\textrm{Bo}=1000, \textrm{Re}=10^5$. These corroborate the picture drawn from figure \ref{fig:DTotals}; there are two main peaks of droplet production, which produce short-lived drops; the first peak is sharp and the second is broader. We also observe that these peaks, and especially the second peak, are the source of large droplets. There is a slight Bond number dependency seen in the sizes of the droplets produced in the first peak; that is, increased Bond number produces more droplets (as in figure \ref{fig:DTotals}a) as well as larger ones.


\begin{figure}
\begin{minipage}{0.45\linewidth}
  \begin{overpic}[width=\linewidth]{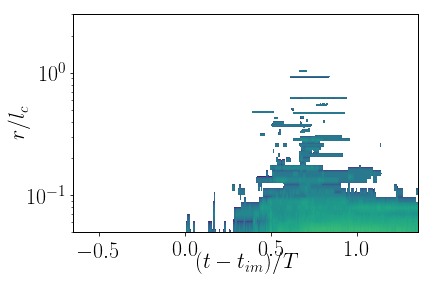}
    \put(0,58){(a) $\qquad \textrm{Bo}=200, \textrm{Re}=40000$}
  \end{overpic}
\end{minipage}
\begin{minipage}{0.45\linewidth}
  \begin{overpic}[width=\linewidth]{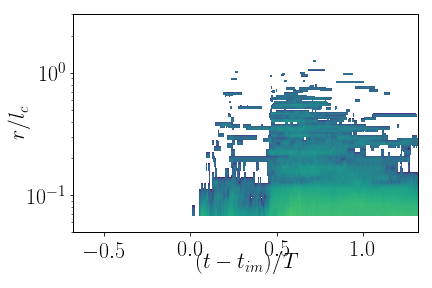}
    \put(0,58){(b) $\qquad \textrm{Bo}=500, \textrm{Re}=40000$}
  \end{overpic}
\end{minipage}\hfill\\
\begin{minipage}{0.45\linewidth}
  \begin{overpic}[width=\linewidth]{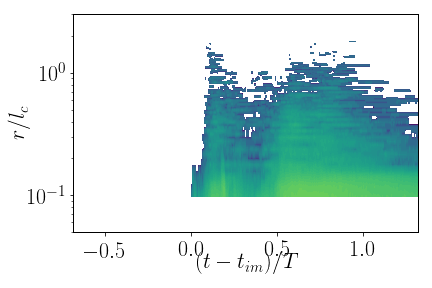}
    \put(0,58){(c) $\qquad \textrm{Bo}=1000, \textrm{Re}=100000$}
  \end{overpic}
\end{minipage}
\begin{minipage}{0.45\linewidth}
  \begin{overpic}[width=\linewidth]{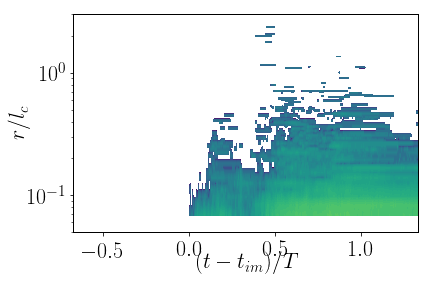}
    \put(0,58){(d) $\qquad \textrm{Bo}=500, \textrm{Re}=100000$}
  \end{overpic}
\end{minipage}\hfill\\
\begin{minipage}{\linewidth}
  \centering
  \includegraphics[height=0.07\linewidth]{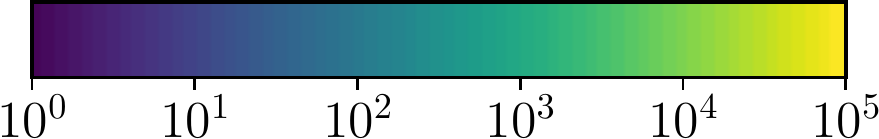}
\end{minipage}
\caption{Contours of droplet size distribution over time. (a) $Bo=200, Re=40 \times 10^3$; (b) $Bo=500, Re=40\times10^3$; (c) $Bo=500, Re=100\times10^3$; (d) $Bo=1000, Re=100\times10^3$. $L=11$ for all cases.}
\label{fig:DContours}
\end{figure}

\subsection{Time-averaged distribution and comparison with Erinin et al 2019}
\label{sec:timeaverageddroplets}
We now seek to compare the present numerical data with experiment. For this purpose, we consider the droplet size distributions time-averaged over $(t-t_{im})/T \in [0.2, 1]$. The experimental data presented in \citet{Erinin2019} are reported as a droplet count per bin size, per unit length of breaking crest. In order to compare to the numerical data, we multiply \citet{Erinin2019} by the wave tank width (1.15m), which yields an absolute number of drop distribution, per unit bin-size. We consider only the Part I data from Erinin \emph{et al.} which corresponds to the earlier splashing stage which is best resolved in our data and we do not consider the later drop production stage which corresponds to jet drop production. \textcolor{black}{We observe a reasonable agreement between our numerical data and those of \citet{Erinin2019} in the range of drop size $0.08r_d/l_c$ to $r_d/l_c$, in terms of total number of ejected drops and scaling with radius. This observation is extremely encouraging, as we note that the breaker from Erinin is at a slightly smaller wave slope than our breakers, and a slope (or wave height, or falling jet speed) dependency is expected. Note that again, the effect of Reynolds number is small, since the two $\textrm{Bo}=500$ cases at $\textrm{Re}=4\times 10^4, 10^5$ collapse well. }


\begin{figure}
  \centering
   \includegraphics[width=0.5\linewidth]{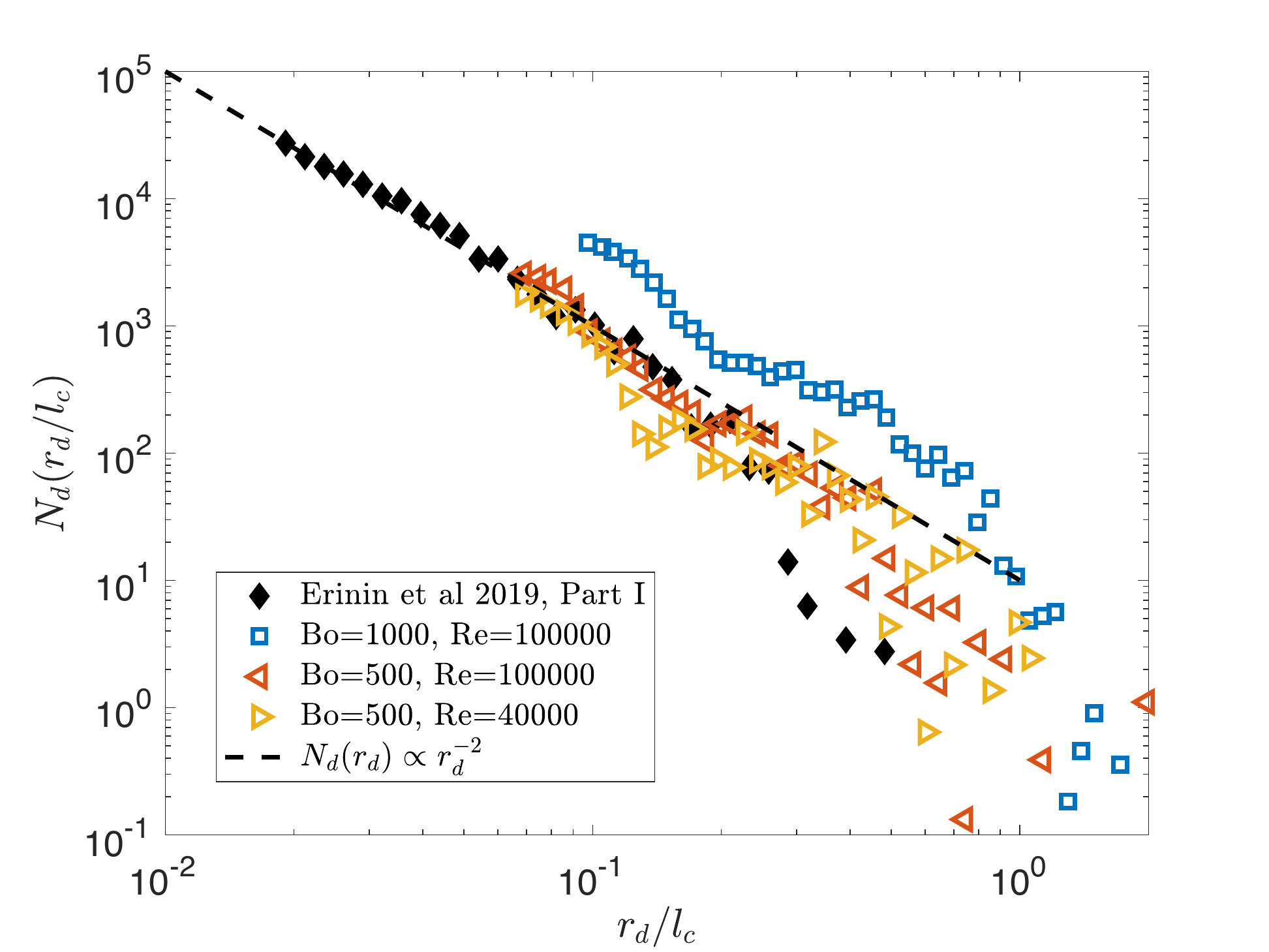}
  \caption{Time-average of droplet size distributions over the time window $t/T \in [0.2, 1.0]$, and experimental data from \cite{Erinin2019}. Experimental and numerical data is scaled consistently.}
  \label{fig:DComparison}
\end{figure}

To compare the scale of drops being produced, we normalized the drop size by the capillary length $l_c$, and therefore present $N_d(r_d/l_c)$ as a function of $r_d/l_c$, shown in figure \ref{fig:DComparison}. We observe a remarkable general agreement in shape in overall number of drops, as well as range of drops produced, while the data of Erinin \emph{et al.} extend to smaller droplets that we are not yet able to resolve. Some Bo-number dependency is evident in the numerical data which can be attributed to the enhanced surface tension effects that reduce the fragmentation process at low Bo-number. As with the bubble size distributions, there probably exists a high-Bo regime independent of surface tension, but this critical Bo number has not yet been identified. The expected dependency in slope also complicates the analysis. Understanding these effects requires both experimental and numerical data at various slopes. However, these open questions do not reduce the importance of having achieved direct numerical simulations of drop production by a splashing process, which are well-resolved numerically and agree reasonably well with the experimental data in terms of the range of drop size produced and their total number, \emph{despite significant differences in the details of the initialization between them}. {\color{black} We do remark that the presence of wind, not accounted for in this study, would likely generate spume, which would affect the droplet size distributions, but would require the resolution of the turbulent boundary layer forcing the wave.}





\subsection{Droplet velocity statistics}
\label{sec:droplet_velocities}
Lastly, we consider the statistics of droplet velocities. Figure \ref{fig:DVel}a shows a contour plot of the droplet velocities (normalized by the deep water phase speed $c_{ph}=\sqrt{g/k}$) over time. It shows that smaller velocities on the order of $v \sim c_{ph}$ are prevalent throughout the breaking process, with larger droplet velocities $\sim 3c_{ph} - 4 c_{ph}$ appearing during the secondary splash (see figure \ref{fig:DStages}b) and the sustained splashing later in the breaking period (see figure \ref{fig:DStages}c). Comparison with figure \ref{fig:DContours}d shows these larger velocities are attained at the same time that large droplets appear. Indeed, the joint distribution of velocities and droplet radii, during the time of the sustained splashing ($(t-t_{im})/T \simeq 0.6$) shown in figure \ref{fig:DVel}c suggests that the highest speeds are attained by the largest droplets, though there are not many such droplets. Large droplets may also be very slow. Most droplets are small (as confirmed by the marginal size distribution, shown in figure \ref{fig:DVel}b and matching earlier figures) but these vary broadly in speed. Finally, the marginal velocity distribution is shown in figure \ref{fig:DVel}d, showing a peak in droplets that have low speeds $\sim c_{ph}$ with a drop-off at very small speeds and a skew toward high speeds. The distribution is not governed by a power law, unlike the size distributions, but appears to be best described by a Gamma distribution (solid line) or by a log-normal distribution (dotted), both of which have been observed in many fragmentation processes \citep{Ling2017,Villermaux2020}.

\begin{figure}
\centering
  \begin{overpic}[width=0.45\linewidth]{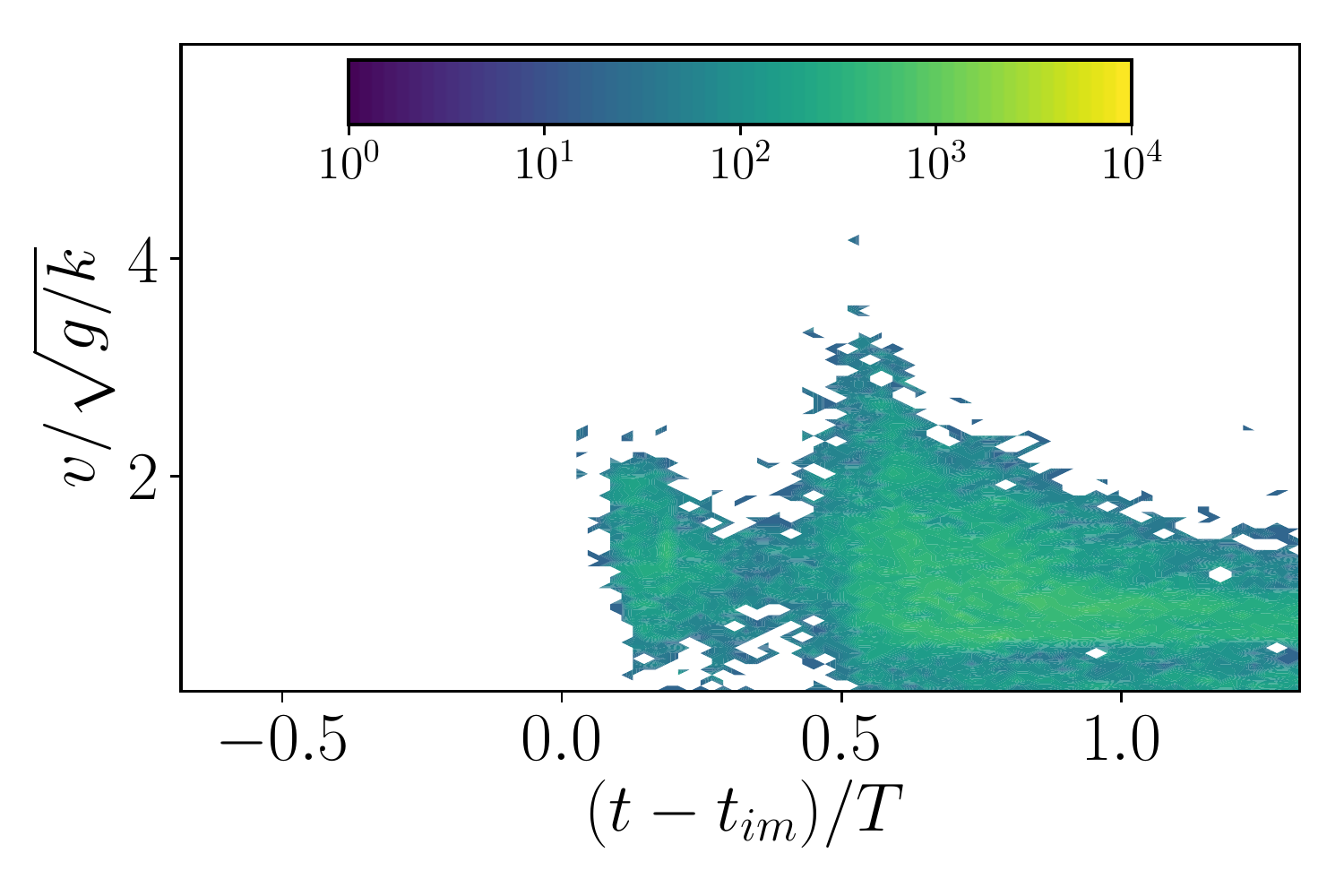}
    \put(0,60){(a)}
  \end{overpic}
  \begin{overpic}[width=0.45\linewidth]{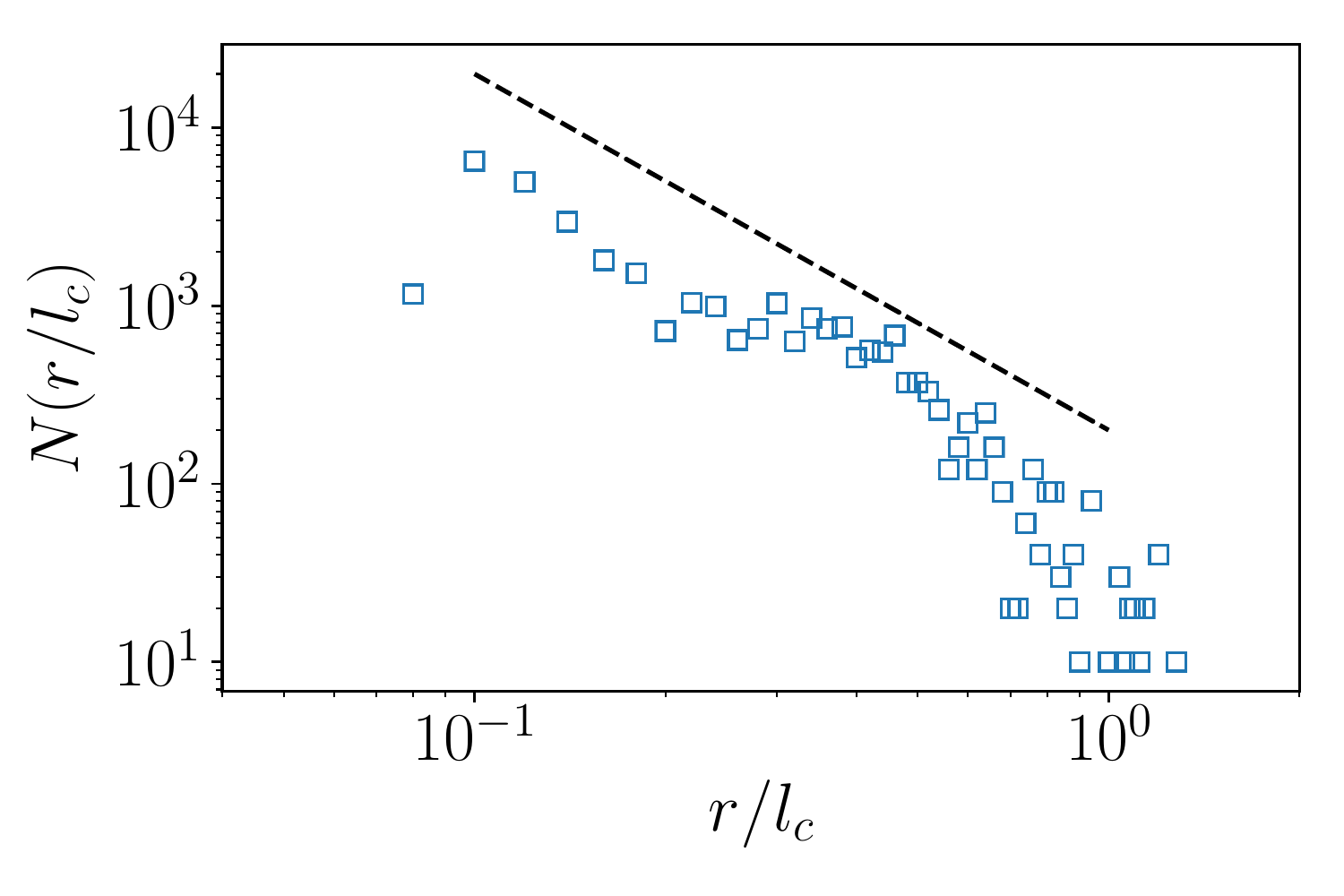}
    \put(0,60){(b)}
  \end{overpic}
  \begin{overpic}[width=0.45\linewidth]{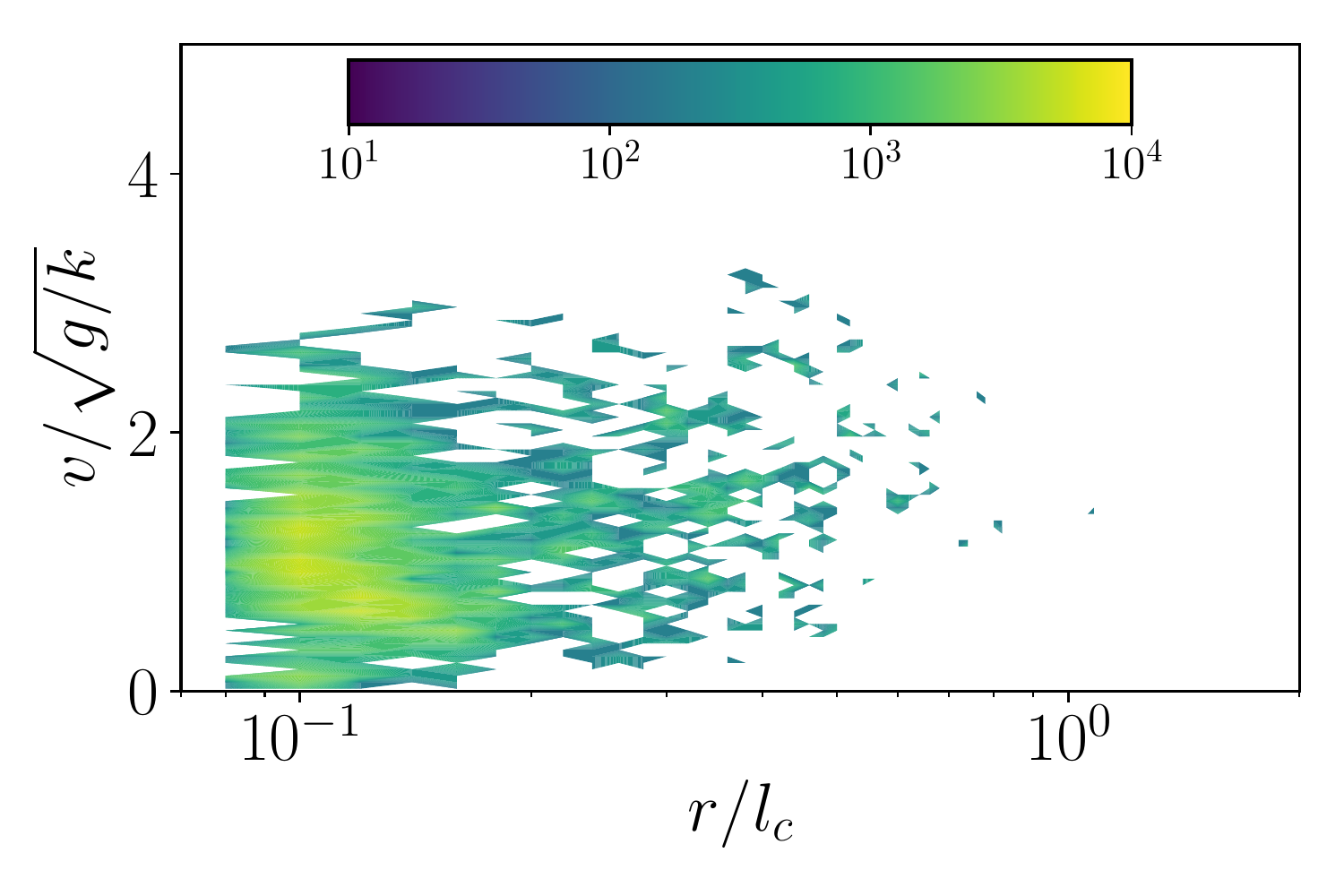}
    \put(0,60){(c)}
  \end{overpic}
  \begin{overpic}[width=0.45\linewidth]{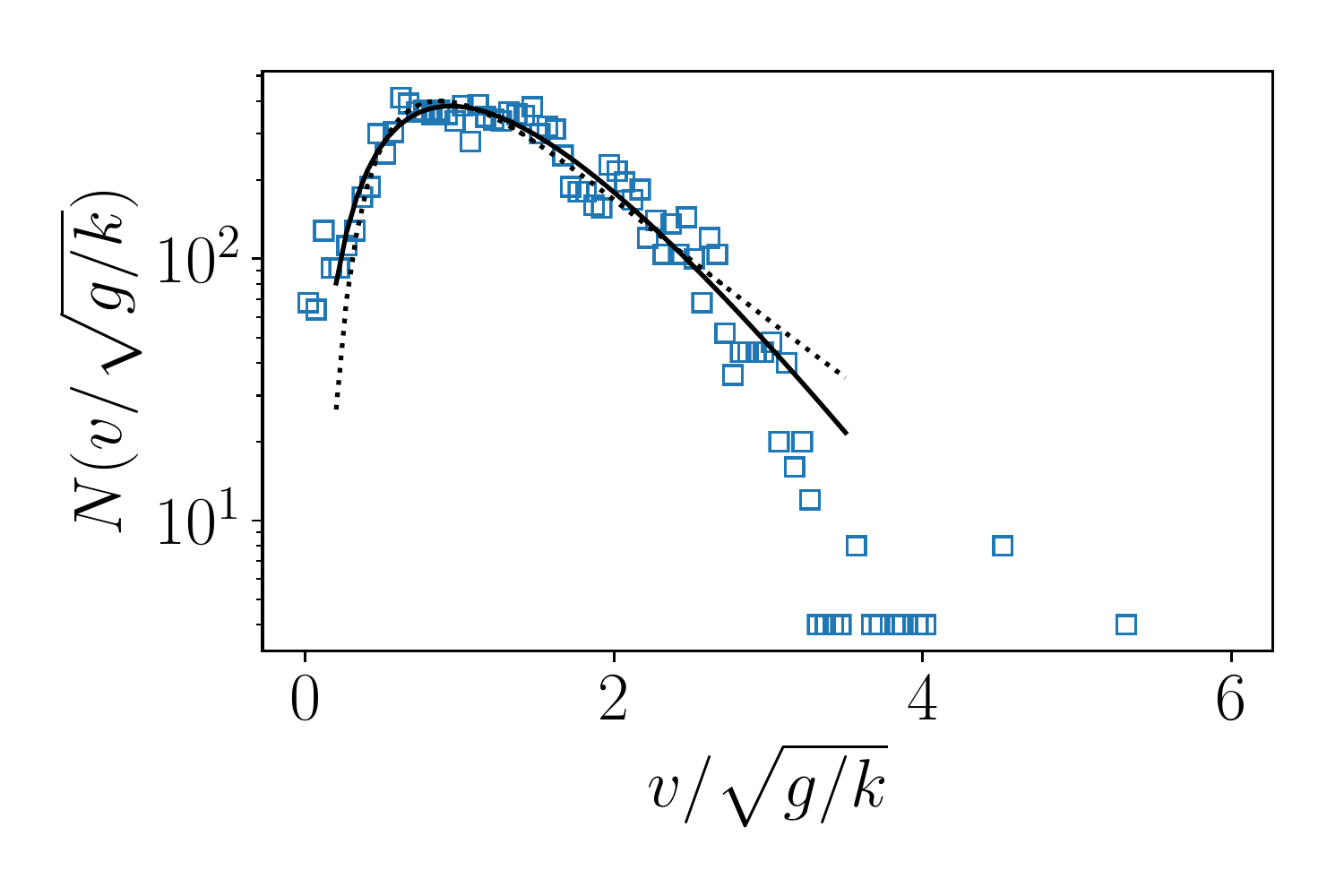}
    \put(0,60){(d)}
  \end{overpic}
  \caption{(a) Contour of droplet velocities for the case $Bo=1000, Re=10^5$, plotted over time on the horizontal axis and velocity normalized by the wave phase speed on the vertical. (b) The droplet size distribution at the time $(t-t_{im})/T=0.6$ averaged over a time width of $\Delta t/T = 0.1$. Dashed line: power law with exponent $-2$, as in Region I of \cite{Erinin2019}. (c) The velocity distribution for the same time as (b). Lines: Fits for gamma (solid) and log-normal (dotted) distributions. (d) The joint size--velocity distribution for the same time as (b). }
  \label{fig:DVel}
\end{figure}

It should be noted that the velocities presented in figure \ref{fig:DVel} are those of all droplets in the gas phase, and therefore represent droplets at all points in their ballistic trajectories. The data therefore does not in general represent only ejection speeds per se. Nevertheless, it can be assumed that the largest droplet velocities observed in figures \ref{fig:DVel}a,c are those of ejecting droplets, since no larger velocities are ever observed. Thus the fastest ejection speeds in the data are of the order of $3c_{ph} - 4c_{ph}$, and they mostly occur for droplets larger than approximately $0.15 l_c$ and up to $0.4l_c - 0.5l_c$. Complete statistical separation of the just-ejected droplets from the rest of the droplet population remains to be conducted in a future study. 


\section{Concluding remarks}
\label{sec:conclusion} 
We have presented high-resolution simulations of breaking waves using DNS of the two-phase Navier-Stokes equations with surface tension exhibiting transition in a multiphase environment from laminar to turbulent flow, for a wide range of Reynolds numbers. By varying Bond and Reynolds numbers at high numerical resolution, we discuss the energetics of the breaker as well as statistics for bubble and droplet populations. For the energy, we have analysed the transition to 3D flow in terms of the volume-integrated dissipation rate in the water phase, and showed a Reynolds-number dependency for values of the wave Reynolds number less than $10^5$, which corresponds to a mixing transition at a turbulence Reynolds number of $Re_\lambda \simeq 100$, analogous to results in a variety of canonical single-phase turbulent flows. We characterize the transition time scale, which is associated with a shear mechanism, the horizontal breaker speed and the vertical breaker height. The result thus appears generic for highly-energetic breaking waves at high slope. The shear-layer instability mechanism driving the transition is local and is expected to be independent of the type of breaker (spilling or plunging). Other features of the energetics such as a large peak in dissipation rate during the active breaking phase can be explained in terms of the breakup of the main cavity entrained by the plunging breaker. This critically contextualizes prior observations in the literature that the energetics of numerical 2D breakers approximate those of 3D breakers \citep{Song2004,Iafrati2009,Iafrati2011,Deike2015}. 

Regarding the bubble statistics, we resolve across multiple scales extending from the main cavity to below the Hinze scale, particularly at low wave Bond numbers, and find reasonable agreement with experiments \citep{Deane2002} across the full range of resolved bubble sizes. We describe capillary effects on the plunging jet and ingested cavity, and characterize an asymptotic Bond number. We extend the bubble size distribution model from \cite{Deike2016} to account for variation due to capillary effects in the size of the main cavity ingested by the breaker, and in the subsequent fragmentation and breakup cascade of the cavity. Incorporated in the scaling, and as noted by \cite{Deike2016}, is the close connection between the bubble statistics and the energy dissipation rate in the bulk liquid. The scaling shows good collapse of the data and, again, good agreement with experiments. 

We also present statistics on the droplet populations produced by the breakers. We find good agreement in the shape of the droplet size distributions with the recent experiments of \cite{Erinin2019}, although some slope and Bond number effects are present and remain to be precisely quantified. Statistics on the droplet velocities are discussed, and it is found that the fastest-ejecting droplets travel at up to four times the phase speed of the wave, and are also some of the largest droplets; these are produced during the most intense splashing periods of the breaker. 

The bubble and droplet size distribution seem to be both independent of the Reynolds number, once above the critical Reynolds number identified in studying the 3D turbulence transition. Consistent results in simulations and experiments for the bubble and droplet size distributions, when scaled by the characteristic length scale of the problem, reinforces the discussion in the literature \cite{Deike2015,Deike2016} that the details of these breakers are essentially \emph{local} in the sense that whatever the initial conditions of the breaker, the dissipative, bubble and droplet properties depend only parameters of the wave at the point of breaking and not on the pre-breaking history of the wave. 

We note that the results discussed here are grid converged, thanks to the use of adaptive mesh refinement techniques, which allow an effective grid size of $2048^3$ grid points. These results show the ability to resolve the mixing transition in the turbulent flow in multiphase DNS of 3D breaking waves, and paves the way for realistic direct simulations of turbulent two-phase flows.

\section*{Acknowledgements}
\label{sec:acknowledgements}
This work was supported by the National Science Foundation (Physical Oceanography) under Grant No. 1849762 to L.D. and the Cooperative Institute for Earth System modeling between Princeton and the Geophysical Fluid Dynamics Laboratory (GFDL) NOAA. Computations were partially performed using allocation TG-OCE180010 to W.M. from the Extreme Science and Engineering Discovery Environment (XSEDE), which is supported by NSF Grant No. ACI-1053575; and the HPC resources of CINES and TGCC under the allocations 2019- A0072B07760; 2020-A0092B07760 granted by GENCI, and from the Jean Zay Grand Challenge allocation from IDRIS. {\color{black}Computations were also performed on resources managed and supported by Princeton Research Computing, a consortium of groups led including the Princeton Institute for Computational Science and Engineering and the Office of Information Technology's High Performance Computing Center and Visualization Laboratory at Princeton University. We are grateful to the anonymous reviewers whose comments have helped improving the quality of the manuscript.}

We declare no conflict of interest.

\bibliographystyle{jfm}
\bibliography{oceanBib}

\begin{thebibliography}{110}
\expandafter\ifx\csname natexlab\endcsname\relax\def\natexlab#1{#1}\fi
\def\au#1{#1} \def\ed#1{#1} \def\yr#1{#1}\def\at#1{#1}\def\jt#1{\textit{#1}}
  \def\bt#1{#1}\def\bvol#1{\textbf{#1}} \def\vol#1{#1} \def\pg#1{#1}
  \def\publ#1{#1}\def\arxiv#1{#1}\def\org#1{#1}\def\st#1{\textit{#1}}

\bibitem[Banner {\em et~al.\/}(2014)Banner, Barthelemy, Fedele, Allis,
  Benetazzo, Dias \& Peirson]{banner2014linking}
{\sc \au{Banner, ML}, \au{Barthelemy, Xavier}, \au{Fedele, Francesco},
  \au{Allis, Michael}, \au{Benetazzo, Alvise}, \au{Dias, Frederic} \&
  \au{Peirson, WL}} \yr{2014}  \at{Linking reduced breaking crest speeds to
  unsteady nonlinear water wave group behavior}.  \jt{Physical review letters}
  \bvol{112}~(11),  \pg{114502}.

\bibitem[Banner \& Peirson(2007)]{Banner2007}
{\sc \au{Banner, Michael~L} \& \au{Peirson, William~L}} \yr{2007}  \at{Wave
  breaking onset and strength for two-dimensional deep-water wave groups}.
  \jt{Journal of Fluid Mechanics}  \bvol{585},  \pg{93}.

\bibitem[Bell {\em et~al.\/}(1989)Bell, Colella \& Glaz]{Bell1989}
{\sc \au{Bell, John~B}, \au{Colella, Phillip} \& \au{Glaz, Harland~M}}
  \yr{1989}  \at{A second-order projection method for the incompressible
  navier-stokes equations}.  \jt{Journal of Computational Physics}
  \bvol{85}~(2),  \pg{257 -- 283}.

\bibitem[Berny {\em et~al.\/}(2020)Berny, Deike, S{\'e}on \&
  Popinet]{Berny2020}
{\sc \au{Berny, Alexis}, \au{Deike, Luc}, \au{S{\'e}on, Thomas} \& \au{Popinet,
  St{\'e}phane}} \yr{2020}  \at{Role of all jet drops in mass transfer from
  bursting bubbles}.  \jt{Physical Review Fluids}  \bvol{5}~(3),  \pg{033605}.

\bibitem[Blenkinsopp \& Chaplin(2007)]{Blenkinsopp2007}
{\sc \au{Blenkinsopp, CE} \& \au{Chaplin, JR}} \yr{2007}  \at{Void fraction
  measurements in breaking waves}.  \jt{Proc. R. Soc. A}  \bvol{463}~(2088),
  \pg{3151--3170}.

\bibitem[Blenkinsopp \& Chaplin(2010)]{Blenkinsopp2010}
{\sc \au{Blenkinsopp, C.~E.} \& \au{Chaplin, J.~R.}} \yr{2010}  \at{Bubble size
  measurements in breaking waves using optical fiber phase detection probes.}
  \jt{Oceanic Engineering, IEEE Journal of}  \bvol{35}~(2),  \pg{388--401}.

\bibitem[Bonmarin(1989)]{Bonmarin1989}
{\sc \au{Bonmarin, P}} \yr{1989}  \at{Geometric properties of deep-water
  breaking waves}.  \jt{Journal of fluid mechanics}  \bvol{209},
  \pg{405--433}.

\bibitem[Chan {\em et~al.\/}(2020{\natexlab{{\em a\/}}})Chan, Johnson \&
  Moin]{Chan2020}
{\sc \au{Chan, W.H.R.}, \au{Johnson, P.} \& \au{Moin, P.}}
  \yr{2020{\natexlab{{\em a\/}}}}  \at{The turbulent bubble break-up cascade.
  part 1. theoretical developments}.  \jt{arXiv preprint arXiv:2008.12883} .

\bibitem[Chan {\em et~al.\/}(2020{\natexlab{{\em b\/}}})Chan, Johnson \&
  Moin]{Chan2020a}
{\sc \au{Chan, W.H.R.}, \au{Johnson, P.} \& \au{Moin, P.}}
  \yr{2020{\natexlab{{\em b\/}}}}  \at{The turbulent bubble break-up cascade.
  part 1. theoretical developments}.  \jt{arXiv preprint arXiv:2008.12883} .

\bibitem[Chen {\em et~al.\/}(1999)Chen, Kharif, Zaleski \& Li]{Chen1999}
{\sc \au{Chen, Gang}, \au{Kharif, Christian}, \au{Zaleski, St{\'e}phane} \&
  \au{Li, Jie}} \yr{1999}  \at{Two-dimensional navier--stokes simulation of
  breaking waves}.  \jt{Physics of fluids}  \bvol{11}~(1),  \pg{121--133}.

\bibitem[De~Vita {\em et~al.\/}(2018)De~Vita, Verzicco \& Iafrati]{Vita2018}
{\sc \au{De~Vita, Francesco}, \au{Verzicco, Roberto} \& \au{Iafrati,
  Alessandro}} \yr{2018}  \at{Breaking of modulated wave groups: kinematics and
  energy dissipation processes}.  \jt{Journal of fluid mechanics}  \bvol{855},
  \pg{267--298}.

\bibitem[Deane \& Stokes(2002)]{Deane2002}
{\sc \au{Deane, Grant~B} \& \au{Stokes, M~Dale}} \yr{2002}  \at{Scale
  dependence of bubble creation mechanisms in breaking waves}.  \jt{Nature}
  \bvol{418}~(6900),  \pg{839--844}.

\bibitem[Deike {\em et~al.\/}(2018)Deike, Ghabache, Liger-Belair, Das, Zaleski,
  Popinet \& Seon]{Deike2018b}
{\sc \au{Deike, L.}, \au{Ghabache, E.}, \au{Liger-Belair, G.}, \au{Das, A.~K.},
  \au{Zaleski, S.}, \au{Popinet, S.} \& \au{Seon, T.}} \yr{2018}  \at{The
  dynamics of jets produced by bursting bubbles}.  \jt{Phys. Rev. Fluids}
  \bvol{3}.

\bibitem[Deike \& Melville(2018)]{Deike2018}
{\sc \au{Deike, L.} \& \au{Melville, W.~K.}} \yr{2018}  \at{Gas transfer by
  breaking waves}.  \jt{Geophysical Research Letters}  \bvol{45}~(19),
  \pg{10482--10492}.

\bibitem[Deike {\em et~al.\/}(2016)Deike, Melville \& Popinet]{Deike2016}
{\sc \au{Deike, L.}, \au{Melville, W.~K.} \& \au{Popinet, S.}} \yr{2016}
  \at{Air entrainment and bubble statistics in breaking waves}.  \jt{J. Fluid
  Mech.}  \bvol{801},  \pg{91–129}.

\bibitem[Deike {\em et~al.\/}(2017)Deike, Pizzo \& Melville]{Deike2017}
{\sc \au{Deike, L.}, \au{Pizzo, N.} \& \au{Melville, W.~K.}} \yr{2017}
  \at{Lagrangian transport by breaking surface waves}.  \jt{J. Fluid Mech.}
  \bvol{829},  \pg{364–391}.

\bibitem[Deike {\em et~al.\/}(2015)Deike, Popinet \& Melville]{Deike2015}
{\sc \au{Deike, L.}, \au{Popinet, S.} \& \au{Melville, W. K.}} \yr{2015}
  \at{Capillary effects on wave breaking}.  \jt{J. Fluid Mech.}  \bvol{769},
  \pg{541–569}.

\bibitem[Derakhti \& Kirby(2014)]{Derakhti2014}
{\sc \au{Derakhti, Morteza} \& \au{Kirby, James~T}} \yr{2014}  \at{Bubble
  entrainment and liquid--bubble interaction under unsteady breaking waves}.
  \jt{Journal of fluid mechanics}  \bvol{761},  \pg{464--506}.

\bibitem[Derakhti \& Kirby(2016)]{Derakhti2016}
{\sc \au{Derakhti, M.} \& \au{Kirby, J.~T}} \yr{2016}  \at{Breaking-onset,
  energy and momentum flux in unsteady focused wave packets}.  \jt{J. Fluid
  Mech.}  \bvol{790},  \pg{553--581}.

\bibitem[Derakhti {\em et~al.\/}(2020)Derakhti, Kirby, Banner, Grilli \&
  Thomson]{Derakhti2020}
{\sc \au{Derakhti, Morteza}, \au{Kirby, James~T}, \au{Banner, Michael~L},
  \au{Grilli, Stephan~T} \& \au{Thomson, Jim}} \yr{2020}  \at{A unified
  breaking onset criterion for surface gravity water waves in arbitrary depth}.
   \jt{Journal of Geophysical Research: Oceans}  \pg{p. e2019JC015886}.

\bibitem[Dimotakis(2005)]{Dimotakis2005}
{\sc \au{Dimotakis, Paul~E}} \yr{2005}  \at{Turbulent mixing}.  \jt{Annu. Rev.
  Fluid Mech.}  \bvol{37},  \pg{329--356}.

\bibitem[Dodd {\em et~al.\/}(2021)Dodd, Mohaddes, Ferrante \& Ihme]{Dodd2021}
{\sc \au{Dodd, Michael~S}, \au{Mohaddes, Danyal}, \au{Ferrante, Antonino} \&
  \au{Ihme, Matthias}} \yr{2021}  \at{Analysis of droplet evaporation in
  isotropic turbulence through droplet-resolved dns}.  \jt{International
  Journal of Heat and Mass Transfer}  \bvol{172},  \pg{121157}.

\bibitem[Dommermuth {\em et~al.\/}(1988)Dommermuth, Yue, Lin, Rapp, Chan \&
  Melville]{Dommermuth1988}
{\sc \au{Dommermuth, Douglas~G}, \au{Yue, Dick~KP}, \au{Lin, WM}, \au{Rapp,
  RJ}, \au{Chan, ES} \& \au{Melville, WK}} \yr{1988}  \at{Deep-water plunging
  breakers: a comparison between potential theory and experiments}.
  \jt{Journal of Fluid Mechanics}  \bvol{189},  \pg{423--442}.

\bibitem[Drazen \& Melville(2009)]{Drazen2009}
{\sc \au{Drazen, D.A.} \& \au{Melville, W.~K.}} \yr{2009}  \at{Turbulence and
  mixing in unsteady breaking surface waves}.  \jt{Journal of Fluid Mechanics}
  \bvol{628},  \pg{85--119}.

\bibitem[Drazen {\em et~al.\/}(2008)Drazen, Melville \& Lenain]{Drazen2008}
{\sc \au{Drazen, D.~A.}, \au{Melville, W.~K.} \& \au{Lenain, L.}} \yr{2008}
  \at{Inertial scaling of dissipation in unsteady breaking waves}.  \jt{J.
  Fluid Mech.}  \bvol{611},  \pg{307–332}.

\bibitem[Druzhinin {\em et~al.\/}(2017)Druzhinin, Troitskaya \&
  Zilitinkevich]{Druzhinin2017}
{\sc \au{Druzhinin, Oleg~A}, \au{Troitskaya, Yu~I} \& \au{Zilitinkevich,
  Sergej~S}} \yr{2017}  \at{The study of droplet-laden turbulent airflow over
  waved water surface by direct numerical simulation}.  \jt{Journal of
  Geophysical Research: Oceans}  \bvol{122}~(3),  \pg{1789--1807}.

\bibitem[Duncan(1981)]{Duncan1981}
{\sc \au{Duncan, J.~H}} \yr{1981}  \at{An experimental investigation of
  breaking waves produced by a towed hydrofoil}.  \jt{Proc. R. Soc. Lon.
  Ser.-A.}  \bvol{377}~(1770),  \pg{331--348}.

\bibitem[Duncan {\em et~al.\/}(1999)Duncan, Qiao \& Philomin]{Duncan1999}
{\sc \au{Duncan, J.~H.}, \au{Qiao, H.} \& \au{Philomin, V.}} \yr{1999}
  \at{Gentle spilling breakers : crest profile evolution}.  \jt{J.Fluid Mech}
  \bvol{379},  \pg{191--222}.

\bibitem[Erinin {\em et~al.\/}(2019)Erinin, Wang, Liu, Towle, Liu \&
  Duncan]{Erinin2019}
{\sc \au{Erinin, Martin~A}, \au{Wang, Sophie~D}, \au{Liu, Ren}, \au{Towle,
  David}, \au{Liu, Xinan} \& \au{Duncan, James~H}} \yr{2019}  \at{Spray
  generation by a plunging breaker}.  \jt{Geophysical Research Letters}
  \bvol{46}~(14),  \pg{8244--8251}.

\bibitem[Farsoiya {\em et~al.\/}(2021)Farsoiya, Popinet \& Deike]{Farsoiya2021}
{\sc \au{Farsoiya, Palas~Kumar}, \au{Popinet, St{\'e}phane} \& \au{Deike, Luc}}
  \yr{2021}  \at{Bubble-mediated transfer of dilute gas in turbulence}.
  \jt{Journal of Fluid Mechanics}  \bvol{920}.

\bibitem[Fedele {\em et~al.\/}(2020)Fedele, Banner \& Barthelemy]{Fedele2020}
{\sc \au{Fedele, Francesco}, \au{Banner, Michael~L} \& \au{Barthelemy, Xavier}}
  \yr{2020}  \at{Crest speeds of unsteady surface water waves}.  \jt{Journal of
  Fluid Mechanics}  \bvol{899}.

\bibitem[Fuster {\em et~al.\/}(2009)Fuster, Agbaglah, Josserand, Popinet \&
  Zaleski]{Fuster2009}
{\sc \au{Fuster, Daniel}, \au{Agbaglah, Gilou}, \au{Josserand, Christophe},
  \au{Popinet, Stéphane} \& \au{Zaleski, Stéphane}} \yr{2009}  \at{Numerical
  simulation of droplets, bubbles and waves: state of the art}.  \jt{Fluid
  Dynamics Research}  \bvol{41}~(6),  \pg{065001}.

\bibitem[Fuster \& Popinet(2018)]{Fuster2018}
{\sc \au{Fuster, Daniel} \& \au{Popinet, St{\'e}phane}} \yr{2018}  \at{An
  all-mach method for the simulation of bubble dynamics problems in the
  presence of surface tension}.  \jt{Journal of Computational Physics}
  \bvol{374},  \pg{752--768}.

\bibitem[Garrett {\em et~al.\/}(2000)Garrett, Li \& Farmer]{Garrett2000}
{\sc \au{Garrett, C.}, \au{Li, M.} \& \au{Farmer, D.}} \yr{2000}  \at{The
  connection between bubble size spectra and energy dissipation rates in the
  upper ocean}.  \jt{J. Phys. Oceanogr.}  \bvol{30}~(9),  \pg{2163--2171}.

\bibitem[Gayen \& Sarkar(2010)]{Gayen2010}
{\sc \au{Gayen, Bishakhdatta} \& \au{Sarkar, Sutanu}} \yr{2010}  \at{Turbulence
  during the generation of internal tide on a critical slope}.  \jt{Physical
  review letters}  \bvol{104}~(21),  \pg{218502}.

\bibitem[Gaylo {\em et~al.\/}(2021)Gaylo, Hendrickson \& Yue]{gaylo2021effects}
{\sc \au{Gaylo, Declan~B}, \au{Hendrickson, Kelli} \& \au{Yue, Dick~KP}}
  \yr{2021}  \at{Effects of power-law entrainment on bubble fragmentation
  cascades}.  \jt{Journal of Fluid Mechanics}  \bvol{917}.

\bibitem[Grare {\em et~al.\/}(2013)Grare, Peirson, Branger, Walker,
  Giovanangeli \& Makin]{Grare2013}
{\sc \au{Grare, Laurent}, \au{Peirson, William~L}, \au{Branger, Hubert},
  \au{Walker, James~W}, \au{Giovanangeli, Jean-Paul} \& \au{Makin, Vladimir}}
  \yr{2013}  \at{Growth and dissipation of wind-forced, deep-water waves}.
  \jt{Journal of Fluid Mechanics}  \bvol{722},  \pg{5--50}.

\bibitem[Hao \& Shen(2019)]{Hao2019}
{\sc \au{Hao, Xuanting} \& \au{Shen, Lian}} \yr{2019}  \at{Wind--wave coupling
  study using les of wind and phase-resolved simulation of nonlinear waves}.
  \jt{Journal of Fluid Mechanics}  \bvol{874},  \pg{391--425}.

\bibitem[Hendrickson \& Yue(2006)]{Hendrickson2006}
{\sc \au{Hendrickson, K} \& \au{Yue, DKP}} \yr{2006} Navier-stokes simulations
  of unsteady small-scale breaking waves at a coupled air-water interface.
  \bt{In {\em 26th Symposium on Naval Hydrodynamics\/}}.

\bibitem[Herrmann(2013)]{Herrmann2013}
{\sc \au{Herrmann, M.}} \yr{2013}  \at{On simulating primary atomization}.
  \jt{Atomization Spray.}  \bvol{23}~(11).

\bibitem[Hinze(1955)]{Hinze}
{\sc \au{Hinze, J.~O.}} \yr{1955}  \at{Fundamentals of the hydrodynamic
  mechanism of splitting in dispersion processes}.  \jt{AIChE Journal}
  \bvol{1}~(3),  \pg{289--295}.

\bibitem[Iafrati(2009)]{Iafrati2009}
{\sc \au{Iafrati, A}} \yr{2009}  \at{Numerical study of the effects of the
  breaking intensity on wave breaking flows}.  \jt{Journal of Fluid Mechanics}
  \bvol{622},  \pg{371--411}.

\bibitem[Iafrati(2011)]{Iafrati2011}
{\sc \au{Iafrati, A}} \yr{2011}  \at{Energy dissipation mechanisms in wave
  breaking processes: spilling and highly aerated plunging breaking events}.
  \jt{Journal of Geophysical Research: Oceans}  \bvol{116}~(C7).

\bibitem[Jian {\em et~al.\/}(2020)Jian, Deng \& Thoraval]{jian2020air}
{\sc \au{Jian, Zhen}, \au{Deng, Peng} \& \au{Thoraval, Marie-Jean}} \yr{2020}
  \at{Air sheet contraction}.  \jt{Journal of Fluid Mechanics}  \bvol{899}.

\bibitem[Kiger \& Duncan(2012)]{Kiger2012}
{\sc \au{Kiger, Kenneth~T} \& \au{Duncan, James~H}} \yr{2012}
  \at{Air-entrainment mechanisms in plunging jets and breaking waves}.
  \jt{Annual Review of Fluid Mechanics}  \bvol{44},  \pg{563--596}.

\bibitem[Lamarre \& Melville(1991)]{Lamarre1991}
{\sc \au{Lamarre, E.} \& \au{Melville, W.K.}} \yr{1991}  \at{Air entrainment
  and dissipation in breaking waves}.  \jt{Nature}  \bvol{351},  \pg{469--472}.

\bibitem[de~Leeuw {\em et~al.\/}(2011)de~Leeuw, Andreas, Anguelova, Fairall,
  Lewis, O'Dowd, Schulz \& Schwartz]{deLeeuw2011}
{\sc \au{de~Leeuw, Gerrit}, \au{Andreas, Edgar~L}, \au{Anguelova,
  Magdalena~D.}, \au{Fairall, C.~W.}, \au{Lewis, Ernie~R.}, \au{O'Dowd, Colin},
  \au{Schulz, Michael} \& \au{Schwartz, Stephen~E.}} \yr{2011}  \at{Production
  flux of sea spray aerosol}.  \jt{Reviews of Geophysics}  \bvol{49}~(2),
  \arxiv{arXiv:
  https://agupubs.onlinelibrary.wiley.com/doi/pdf/10.1029/2010RG000349}.

\bibitem[Leifer \& de~Leeuw(2006)]{Leifer2006}
{\sc \au{Leifer, I.} \& \au{de~Leeuw, G.}} \yr{2006}  \at{Bubbles generated
  from wind-steepened breaking waves: 1. bubble plume bubbles}.  \jt{Journal of
  Geophysical Research: Oceans}  \bvol{111}~(C6), c06020.

\bibitem[Lhuissier \& Villermaux(2012)]{Lhuissier2012}
{\sc \au{Lhuissier, H.} \& \au{Villermaux, E.}} \yr{2012}  \at{Bursting bubble
  aerosols}.  \jt{J. Fluid Mech.}  \bvol{696},  \pg{5--44}.

\bibitem[Liang {\em et~al.\/}(2011)Liang, McWilliams, Sullivan \&
  Baschek]{Liang2011}
{\sc \au{Liang, J.‐H.}, \au{McWilliams, J.~C.}, \au{Sullivan, P.~P.} \&
  \au{Baschek, B.}} \yr{2011}  \at{Modeling bubbles and dissolved gases in the
  ocean}.  \jt{J. Geophys. Res.}  \bvol{116}~(C3).

\bibitem[Liang {\em et~al.\/}(2012)Liang, McWilliams, Sullivan \&
  Baschek]{Liang2012}
{\sc \au{Liang, J-H.}, \au{McWilliams, J.C.}, \au{Sullivan, P.P.} \&
  \au{Baschek, B.}} \yr{2012}  \at{Large eddy simulation of the bubbly ocean:
  New insights on subsurface bubble distribution and bubble-mediated gas
  transfer}.  \jt{J. Geophys. Res.}  \bvol{117}~(C4).

\bibitem[Ling {\em et~al.\/}(2017)Ling, Fuster, Zaleski \&
  Tryggvason]{Ling2017}
{\sc \au{Ling, Y.}, \au{Fuster, D.}, \au{Zaleski, S.} \& \au{Tryggvason, G.}}
  \yr{2017}  \at{Spray formation in a quasiplanar gas-liquid mixing layer at
  moderate density ratios: a numerical closeup}.  \jt{Physical Review Fluids}
  \bvol{2}~(1),  \pg{014005}.

\bibitem[Loewen \& Melville(1994)]{Loewen1994}
{\sc \au{Loewen, MR} \& \au{Melville, WK}} \yr{1994}  \at{An experimental
  investigation of the collective oscillations of bubble plumes entrained by
  breaking waves}.  \jt{The Journal of the Acoustical Society of America}
  \bvol{95}~(3),  \pg{1329--1343}.

\bibitem[Loewen {\em et~al.\/}(1996)Loewen, O'Dor \& Skafel]{Loewen1996}
{\sc \au{Loewen, M.~R.}, \au{O'Dor, M.~A.} \& \au{Skafel, M.~G.}} \yr{1996}
  \at{Bubbles entrained by mechanically generated breaking waves.}  \jt{J.
  Geophys. Res.}  \bvol{101}~(C9),  \pg{20759--20769}.

\bibitem[Longuet-Higgins(1982)]{Longuet-Higgins1982}
{\sc \au{Longuet-Higgins, MS}} \yr{1982}  \at{Parametric solutions for breaking
  waves}.  \jt{Journal of Fluid Mechanics}  \bvol{121},  \pg{403--424}.

\bibitem[Longuet-Higgins \& Cokelet(1976)]{Longuet-Higgins1976}
{\sc \au{Longuet-Higgins, Michael~Selwyn} \& \au{Cokelet, ED}} \yr{1976}
  \at{The deformation of steep surface waves on water-i. a numerical method of
  computation}.  \jt{Proceedings of the Royal Society of London. A.
  Mathematical and Physical Sciences}  \bvol{350}~(1660),  \pg{1--26}.

\bibitem[Lubin \& Glockner(2015)]{Lubin2015}
{\sc \au{Lubin, P.} \& \au{Glockner, S.}} \yr{2015}  \at{Numerical simulations
  of three-dimensional plunging breaking waves: generation and evolution of
  aerated vortex filaments}.  \jt{Journal of Fluid Mechanics}  \bvol{767},
  \pg{364–393}.

\bibitem[Lubin {\em et~al.\/}(2006)Lubin, Vincent, Abadie \&
  Caltagirone]{Lubin2006}
{\sc \au{Lubin, Pierre}, \au{Vincent, St{\'e}phane}, \au{Abadie, St{\'e}phane}
  \& \au{Caltagirone, Jean-Paul}} \yr{2006}  \at{Three-dimensional large eddy
  simulation of air entrainment under plunging breaking waves}.  \jt{Coastal
  engineering}  \bvol{53}~(8),  \pg{631--655}.

\bibitem[Martinez-Bazan {\em et~al.\/}(1999)Martinez-Bazan, Montanes \&
  Lasheras]{Martinez-Bazan1999}
{\sc \au{Martinez-Bazan, C.}, \au{Montanes, J.L.} \& \au{Lasheras, J.C.}}
  \yr{1999}  \at{On the breakup of an air bubble injected into a fully
  developed turbulent flow. part 1. breakup frequency}.  \jt{J. Fluid Mech.}
  \bvol{401},  \pg{157--182}.

\bibitem[McWilliams(2016)]{Mcwilliams2016}
{\sc \au{McWilliams, James~C}} \yr{2016}  \at{Submesoscale currents in the
  ocean}.  \jt{Proceedings of the Royal Society A: Mathematical, Physical and
  Engineering Sciences}  \bvol{472}~(2189),  \pg{20160117}.

\bibitem[Melville(1982)]{Melville1982}
{\sc \au{Melville, WK}} \yr{1982}  \at{The instability and breaking of
  deep-water waves}.  \jt{Journal of Fluid Mechanics}  \bvol{115},
  \pg{165--185}.

\bibitem[Melville(1994)]{Melville1994}
{\sc \au{Melville, W.K}} \yr{1994}  \at{Energy dissipation by breaking waves}.
  \jt{Journal of Physical Oceanography}  \bvol{24}~(10),  \pg{2041--2049}.

\bibitem[Melville(1996)]{Melville1996}
{\sc \au{Melville, W~Kendall}} \yr{1996}  \at{The role of surface-wave breaking
  in air-sea interaction}.  \jt{Annual review of fluid mechanics}
  \bvol{28}~(1),  \pg{279--321}.

\bibitem[Melville {\em et~al.\/}(2002)Melville, Veron \& White]{Melville2002}
{\sc \au{Melville, W.~K.}, \au{Veron, F.} \& \au{White, C.~J.}} \yr{2002}
  \at{The velocity field under breaking waves: coherent structure and
  turbulence}.  \jt{J. Fluid. Mech.}  \bvol{454}.

\bibitem[Mostert \& Deike(2020)]{Mostert2020}
{\sc \au{Mostert, W} \& \au{Deike, L}} \yr{2020}  \at{Inertial energy
  dissipation in shallow-water breaking waves}.  \jt{Journal of Fluid
  Mechanics}  \bvol{890}.

\bibitem[Mukherjee {\em et~al.\/}(2019)Mukherjee, Safdari, Shardt,
  Kenjere{\v{s}} \& Van~den Akker]{Mukherjee2019}
{\sc \au{Mukherjee, Siddhartha}, \au{Safdari, Arman}, \au{Shardt, Orest},
  \au{Kenjere{\v{s}}, Sa{\v{s}}a} \& \au{Van~den Akker, Harry~EA}} \yr{2019}
  \at{Droplet--turbulence interactions and quasi-equilibrium dynamics in
  turbulent emulsions}.  \jt{Journal of Fluid Mechanics}  \bvol{878},
  \pg{221--276}.

\bibitem[New(1983)]{New1983}
{\sc \au{New, AL}} \yr{1983}  \at{A class of elliptical free-surface flows}.
  \jt{J. Fluid Mech.}  \bvol{130},  \pg{219--239}.

\bibitem[New {\em et~al.\/}(1985)New, McIver \& Peregrine]{New1985}
{\sc \au{New, AL}, \au{McIver, P} \& \au{Peregrine, DH}} \yr{1985}
  \at{Computations of overturning waves}.  \jt{J. Fluid Mech.}  \bvol{150},
  \pg{233--251}.

\bibitem[Ortiz-Suslow {\em et~al.\/}(2016)Ortiz-Suslow, Haus, Mehta \&
  Laxague]{ortiz2016sea}
{\sc \au{Ortiz-Suslow, David~G}, \au{Haus, Brian~K}, \au{Mehta, Sanchit} \&
  \au{Laxague, Nathan~JM}} \yr{2016}  \at{Sea spray generation in very high
  winds}.  \jt{Journal of the Atmospheric Sciences}  \bvol{73}~(10),
  \pg{3975--3995}.

\bibitem[Pairetti {\em et~al.\/}(2020)Pairetti, Dami{\'a}n, Nigro, Popinet \&
  Zaleski]{Pairetti2020}
{\sc \au{Pairetti, C.I.}, \au{Dami{\'a}n, S.~M.}, \au{Nigro, N.M.},
  \au{Popinet, S.} \& \au{Zaleski, S.}} \yr{2020}  \at{Mesh resolution effects
  on primary atomization simulations}.  \jt{Atomization Spray.}
  \bvol{30}~(12).

\bibitem[Perlin {\em et~al.\/}(2013)Perlin, Choi \& Tian]{Perlin2013}
{\sc \au{Perlin, M.}, \au{Choi, W.} \& \au{Tian, Z.}} \yr{2013}  \at{Breaking
  waves in deep and intermediate waters}.  \jt{Annu. Rev. Fluid Mech.}
  \bvol{45},  \pg{115--145}.

\bibitem[Perrard {\em et~al.\/}(2021)Perrard, Rivi{\`e}re, Mostert \&
  Deike]{Perrard2020}
{\sc \au{Perrard, St{\'e}phane}, \au{Rivi{\`e}re, Ali{\'e}nor}, \au{Mostert,
  Wouter} \& \au{Deike, Luc}} \yr{2021}  \at{Bubble deformation by a turbulent
  flow}.  \jt{Journal of Fluid Mechanics}  \bvol{920}.

\bibitem[Phillips(1985)]{Phillips1985}
{\sc \au{Phillips, O.~M.}} \yr{1985}  \at{Spectral and statistical properties
  of the equilibrium range in wind-generated gravity waves}.  \jt{Journal of
  Fluid Mechanics}  \bvol{156},  \pg{505–531}.

\bibitem[Pizzo(2020)]{Pizzo2020}
{\sc \au{Pizzo, Nick}} \yr{2020}  \at{Theory of deep-water surface gravity
  waves derived from a lagrangian}.  \jt{Journal of Fluid Mechanics}
  \bvol{896}.

\bibitem[Pizzo \& Melville(2019)]{pizzo2019focusing}
{\sc \au{Pizzo, N.} \& \au{Melville, W.K.}} \yr{2019}  \at{Focusing deep-water
  surface gravity wave packets: wave breaking criterion in a simplified{\=a}
  model}.  \jt{Journal of Fluid Mechanics}  \bvol{873}.

\bibitem[Pizzo {\em et~al.\/}(2019)Pizzo, Melville \& Deike]{Pizzo2019}
{\sc \au{Pizzo, N}, \au{Melville, WK} \& \au{Deike, L}} \yr{2019}
  \at{Lagrangian transport by nonbreaking and breaking deep-water waves at the
  ocean surface}.  \jt{Journal of Physical Oceanography}  \bvol{49}~(4),
  \pg{983--992}.

\bibitem[Pope(2000)]{Pope2000}
{\sc \au{Pope, S.B.}} \yr{2000} {\em Turbulent flows\/}.  \publ{Cambridge
  University Press}.

\bibitem[Popinet(2003)]{Popinet2003}
{\sc \au{Popinet, S.}} \yr{2003}  \at{Gerris: a tree-based adaptive solver for
  the incompressible euler equations in complex geometries}.  \jt{J. Comput.
  Phys.}  \bvol{190}~(2),  \pg{572 -- 600}.

\bibitem[Popinet(2009)]{Popinet2009}
{\sc \au{Popinet, S.}} \yr{2009}  \at{An accurate adaptive solver for
  surface-tension-driven interfacial flows}.  \jt{J. Comput. Phys.}
  \bvol{228}~(16),  \pg{5838 -- 5866}.

\bibitem[Popinet(2018)]{Popinet2018}
{\sc \au{Popinet, St{\'e}phane}} \yr{2018}  \at{Numerical models of surface
  tension}.  \jt{Annual Review of Fluid Mechanics}  \bvol{50}.

\bibitem[Pullin(2000)]{Pullin2000}
{\sc \au{Pullin, Dale~I}} \yr{2000}  \at{A vortex-based model for the subgrid
  flux of a passive scalar}.  \jt{Physics of Fluids}  \bvol{12}~(9),
  \pg{2311--2319}.

\bibitem[Rapp \& Melville(1990)]{Rapp1990}
{\sc \au{Rapp, Ronald~James} \& \au{Melville, W~Kendall}} \yr{1990}
  \at{Laboratory measurements of deep-water breaking waves}.  \jt{Philosophical
  Transactions of the Royal Society of London. Series A, Mathematical and
  Physical Sciences}  \bvol{331}~(1622),  \pg{735--800}.

\bibitem[Reichl \& Deike(2020)]{Reichl2020}
{\sc \au{Reichl, Brandon~G} \& \au{Deike, Luc}} \yr{2020}  \at{Contribution of
  sea-state dependent bubbles to air-sea carbon dioxide fluxes}.
  \jt{Geophysical Research Letters} .

\bibitem[Richter \& Sullivan(2013)]{Richter2013}
{\sc \au{Richter, David~H} \& \au{Sullivan, Peter~P}} \yr{2013}  \at{Sea
  surface drag and the role of spray}.  \jt{Geophysical Research Letters}
  \bvol{40}~(3),  \pg{656--660}.

\bibitem[Risso \& Fabre(1998)]{risso1998oscillations}
{\sc \au{Risso, Fr{\'e}d{\'e}ric} \& \au{Fabre, Jean}} \yr{1998}
  \at{Oscillations and breakup of a bubble immersed in a turbulent field}.
  \jt{Journal of Fluid Mechanics}  \bvol{372},  \pg{323--355}.

\bibitem[Rivi{\`e}re {\em et~al.\/}(2021)Rivi{\`e}re, Mostert, Perrard \&
  Deike]{Riviere2021}
{\sc \au{Rivi{\`e}re, Ali{\'e}nor}, \au{Mostert, Wouter}, \au{Perrard,
  St{\'e}phane} \& \au{Deike, Luc}} \yr{2021}  \at{Sub-hinze scale bubble
  production in turbulent bubble break-up}.  \jt{Journal of Fluid Mechanics}
  \bvol{917}.

\bibitem[Rojas \& Loewen(2007)]{Rojas2007}
{\sc \au{Rojas, G} \& \au{Loewen, MR}} \yr{2007}  \at{Fiber-optic probe
  measurements of void fraction and bubble size distributions beneath breaking
  waves}.  \jt{Experiments in Fluids}  \bvol{43}~(6),  \pg{895--906}.

\bibitem[Romero(2019)]{Romero2019}
{\sc \au{Romero, Leonel}} \yr{2019}  \at{Distribution of surface wave breaking
  fronts}.  \jt{Geophysical Research Letters}  \bvol{46}~(17-18),
  \pg{10463--10474}.

\bibitem[Romero {\em et~al.\/}(2012)Romero, Melville \& Kleiss]{Romero2012}
{\sc \au{Romero, L.}, \au{Melville, W.~K.} \& \au{Kleiss, J.~M.}} \yr{2012}
  \at{Spectral energy dissipation due to surface wave breaking}.  \jt{J. Phys.
  Oceanogr.}  \bvol{42}~(9),  \pg{1421--1444}.

\bibitem[Saket {\em et~al.\/}(2017{\natexlab{{\em a\/}}})Saket, Peirson,
  Banner, Barthelemy \& Allis]{Saket2017}
{\sc \au{Saket, Arvin}, \au{Peirson, William~L}, \au{Banner, Michael~L},
  \au{Barthelemy, Xavier} \& \au{Allis, Michael~J}} \yr{2017{\natexlab{{\em
  a\/}}}}  \at{On the threshold for wave breaking of two-dimensional deep water
  wave groups in the absence and presence of wind}.  \jt{Journal of Fluid
  Mechanics}  \bvol{811},  \pg{642}.

\bibitem[Saket {\em et~al.\/}(2017{\natexlab{{\em b\/}}})Saket, Peirson,
  Banner, Barthelemy \& Allis]{saket2017threshold}
{\sc \au{Saket, Arvin}, \au{Peirson, William~L}, \au{Banner, Michael~L},
  \au{Barthelemy, Xavier} \& \au{Allis, Michael~J}} \yr{2017{\natexlab{{\em
  b\/}}}}  \at{On the threshold for wave breaking of two-dimensional deep water
  wave groups in the absence and presence of wind}.  \jt{Journal of Fluid
  Mechanics}  \bvol{811},  \pg{642}.

\bibitem[Schwendeman \& Thomson(2017)]{Schwendeman2017}
{\sc \au{Schwendeman, M.S.} \& \au{Thomson, J.}} \yr{2017}  \at{Sharp-crested
  breaking surface waves observed from a ship-based stereo video system}.
  \jt{Journal of Physical Oceanography}  \bvol{47}~(4),  \pg{775--792}.

\bibitem[Shi {\em et~al.\/}(2010)Shi, Kirby \& Ma]{Shi2010}
{\sc \au{Shi, F.}, \au{Kirby, J.T.} \& \au{Ma, G.}} \yr{2010}  \at{Modeling
  quiescent phase transport of air bubbles induced by breaking waves}.
  \jt{Ocean Modelling}  \bvol{35}~(1-2),  \pg{105--117}.

\bibitem[Soligo {\em et~al.\/}(2019)Soligo, Roccon \& Soldati]{Soligo2019}
{\sc \au{Soligo, Giovanni}, \au{Roccon, Alessio} \& \au{Soldati, Alfredo}}
  \yr{2019}  \at{Breakage, coalescence and size distribution of
  surfactant-laden droplets in turbulent flow}.  \jt{Journal of Fluid
  Mechanics}  \bvol{881},  \pg{244--282}.

\bibitem[Song \& Sirviente(2004)]{Song2004}
{\sc \au{Song, C.} \& \au{Sirviente, A.~I.}} \yr{2004}  \at{A numerical study
  of breaking waves}.  \jt{Phys. Fluids}  \bvol{16}~(7),  \pg{2649--2667}.

\bibitem[Sreenivasan(1984)]{Sreenivasan1984}
{\sc \au{Sreenivasan, Katepalli~R}} \yr{1984}  \at{On the scaling of the
  turbulence energy dissipation rate}.  \jt{The Physics of fluids}
  \bvol{27}~(5),  \pg{1048--1051}.

\bibitem[Tang {\em et~al.\/}(2017)Tang, Yang, Liu, Dong \& Shen]{Tang2017}
{\sc \au{Tang, Shuai}, \au{Yang, Zixuan}, \au{Liu, Caixi}, \au{Dong, Yu-Hong}
  \& \au{Shen, Lian}} \yr{2017}  \at{Numerical study on the generation and
  transport of spume droplets in wind over breaking waves}.  \jt{Atmosphere}
  \bvol{8}~(12),  \pg{248}.

\bibitem[Terrill {\em et~al.\/}(2001)Terrill, Melville \&
  Stramski]{Terrill2001}
{\sc \au{Terrill, E.J.}, \au{Melville, W.K.} \& \au{Stramski, Dariusz}}
  \yr{2001}  \at{Bubble entrainment by breaking waves and their influence on
  optical scattering in the upper ocean}.  \jt{J. Geophys. Res.}
  \bvol{106}~(C8),  \pg{16815--16823}.

\bibitem[Tian {\em et~al.\/}(2010)Tian, Perlin \& Choi]{Tian2010}
{\sc \au{Tian, Zhigang}, \au{Perlin, Marc} \& \au{Choi, Wooyoung}} \yr{2010}
  \at{Energy dissipation in two-dimensional unsteady plunging breakers and an
  eddy viscosity model}.  \jt{Journal of fluid mechanics}  \bvol{655},
  \pg{217}.

\bibitem[Troitskaya {\em et~al.\/}(2018)Troitskaya, Kandaurov, Ermakova,
  Kozlov, Sergeev \& Zilitinkevich]{troitskaya2018bag}
{\sc \au{Troitskaya, Yu}, \au{Kandaurov, A}, \au{Ermakova, O}, \au{Kozlov, D},
  \au{Sergeev, D} \& \au{Zilitinkevich, Sergei}} \yr{2018}  \at{The “bag
  breakup” spume droplet generation mechanism at high winds. part i: Spray
  generation function}.  \jt{Journal of physical oceanography}  \bvol{48}~(9),
  \pg{2167--2188}.

\bibitem[Tulin \& Waseda(1999)]{Tulin1999}
{\sc \au{Tulin, M.~P.} \& \au{Waseda, T.}} \yr{1999}  \at{Laboratory
  observations of wave group evolution, including breaking effects}.  \jt{J.
  Fluid Mech.}  \bvol{378},  \pg{197--232}.

\bibitem[Vejra{\v{z}}ka {\em et~al.\/}(2018)Vejra{\v{z}}ka, Zedn{\'\i}kov{\'a}
  \& Stanovsk{\`y}]{Vejravzka2018}
{\sc \au{Vejra{\v{z}}ka, J.}, \au{Zedn{\'\i}kov{\'a}, M.} \& \au{Stanovsk{\`y},
  P.}} \yr{2018}  \at{Experiments on breakup of bubbles in a turbulent flow}.
  \jt{AIChE J.}  \bvol{64}~(2),  \pg{740--757}.

\bibitem[Veron(2015)]{Veron2015}
{\sc \au{Veron, Fabrice}} \yr{2015}  \at{Ocean spray}.  \jt{Annual Review of
  Fluid Mechanics}  \bvol{47}~(1),  \pg{507--538}.

\bibitem[Veron {\em et~al.\/}(2012)Veron, Hopkins, Harrison \&
  Mueller]{Veron2012}
{\sc \au{Veron, F}, \au{Hopkins, C}, \au{Harrison, EL} \& \au{Mueller, JA}}
  \yr{2012}  \at{Sea spray spume droplet production in high wind speeds}.
  \jt{Geophysical Research Letters}  \bvol{39}~(16).

\bibitem[Villermaux(2020)]{Villermaux2020}
{\sc \au{Villermaux, Emmanuel}} \yr{2020}  \at{Fragmentation versus cohesion}.
  \jt{Journal of Fluid Mechanics}  \bvol{898}.

\bibitem[Wang {\em et~al.\/}(2016)Wang, Yang \& Stern]{Wang2016}
{\sc \au{Wang, Zhaoyuan}, \au{Yang, Jianming} \& \au{Stern, Frederick}}
  \yr{2016}  \at{High-fidelity simulations of bubble, droplet and spray
  formation in breaking waves}.  \jt{Journal of Fluid Mechanics}  \bvol{792},
  \pg{307–327}.

\bibitem[Watanabe {\em et~al.\/}(2005)Watanabe, Saeki \& Hosking]{Watanabe2005}
{\sc \au{Watanabe, Yasunori}, \au{Saeki, Hiroshi} \& \au{Hosking, Roger~J}}
  \yr{2005}  \at{Three-dimensional vortex structures under breaking waves}.
  \jt{Journal of Fluid Mechanics}  \bvol{545},  \pg{291--328}.

\bibitem[Wu(1979)]{Wu1979}
{\sc \au{Wu, J.}} \yr{1979}  \at{Oceanic whitecaps and sea state}.  \jt{J.
  Phys. Oceanogr.}  \bvol{9}~(5),  \pg{1064--1068}.

\bibitem[Yang {\em et~al.\/}(2018)Yang, Deng \& Shen]{Yang2018}
{\sc \au{Yang, Zixuan}, \au{Deng, Bing-Qing} \& \au{Shen, Lian}} \yr{2018}
  \at{Direct numerical simulation of wind turbulence over breaking waves}.
  \jt{Journal of Fluid Mechanics}  \bvol{850},  \pg{120--155}.

\bibitem[Zhang {\em et~al.\/}(2020)Zhang, Popinet \& Ling]{Zhang2020}
{\sc \au{Zhang, Bo}, \au{Popinet, St{\'e}phane} \& \au{Ling, Yue}} \yr{2020}
  \at{Modeling and detailed numerical simulation of the primary breakup of a
  gasoline surrogate jet under non-evaporative operating conditions}.
  \jt{International Journal of Multiphase Flow}  \bvol{130},  \pg{103362}.

\end{thebibliography}

\end{document}